\documentclass[twocolumn,showpacs,preprintnumbers,amsmath,amssymb,superscriptaddress,floatfix]{revtex4}
\usepackage{graphicx}
\usepackage{dcolumn}
\usepackage{bm}
\usepackage{latexsym}
\usepackage[hang,small,bf]{caption}
\usepackage[subrefformat=parens]{subcaption}
\usepackage{textcase}
\usepackage{color}

\def\nn{\nonumber}

\newcommand{\D}{\partial}
\newcommand{\beqn}{\begin{eqnarray}}
\newcommand{\beq}{\begin{equation}}
\newcommand{\eeqn}{\end{eqnarray}}
\newcommand{\eeq}{\end{equation}}

\newcommand{\tr}{\mathop{\rm Tr}}

\newcommand{\sbra}[1] { \left( #1 \right)}
\newcommand{\mbra}[1] { \left\{ #1 \right\}}

\newcommand{\FS}[1]{\left( #1 \right)}
\newcommand{\BS}[1]{\Bigl( #1 \Bigr)}
\newcommand{\BM}[1]{\Bigl\{ #1 \Bigr\}}
\newcommand{\GM}[1]{\biggl\{ #1 \biggr\}}

\newcommand{\Kanazawa}{\affiliation{Kanazawa University, Kanazawa 920-1192, Japan}}

\begin{document}
\title{Blockspin renormalization-group study of color confinement due to violation of the non-Abelian Bianchi identity}
\author{Tsuneo Suzuki}
\email[e-mail:]{suzuki04@staff.kanazawa-u.ac.jp}
\Kanazawa


\begin{abstract} 
Block-spin transformation of topological defects is applied to the violation of the non-Abelian Bianchi identity (VMABI) on lattice defined as Abelian monopoles. To get rid of lattice artifacts, we introduce 1) smooth gauge fixings such as  the maximal center gauge (MCG), 2) block-spin transformations and 3) the tadpole-improved gauge action. The effective action can be determined by adopting the inverse Monte-Carlo method. The coupling constants $F(i)$ of the effective action depend on the coupling of the lattice action $\beta$ and the number of the blocking step $n$. But it is found that  $F(i)$ satisfy a beautiful scaling, that is,  they are a function of the product $b=na(\beta)$ alone for lattice coupling constants $3.0\le\beta\le3.9$ and the steps of blocking $1\le n\le 12$. The effective action  showing the scaling behavior can be regarded as an almost perfect action corresponding to the continuum limit, since $a\to 0$ as $n\to\infty$ for fixed $b$.  The infrared effective monopole action keeps  the global color invariance when smooth gauges such as MCG keeping the invariance are adopted. The almost perfect action showing the scaling is found to be  independent of the smooth gauges adopted here as naturally expected from the gauge invariance of the continuum theory. Then we compare the results with those obtained 
by   the analytic blocking method of  topological defects from the
continuum, assuming local two-point interactions are dominant as the infrared effective action.  The action is formulated in the
continuum limit while the couplings of these actions can be
derived from simple observables calculated numerically on lattices with a finite lattice spacing. When use is made of Berezinskii-Kosterlitz-Thouless (BKT) transformation, the infrared monopole action can be transformed into that of the string model. Since large $b=na(\beta)$ corresponds to the strong-coupling region in the string model, the physical string tension and the lowest glueball mass can be evaluated \textit{analytically} with the use of the strong-coupling expansion of the string model. The almost perfect action gives us $\sqrt{\sigma}\simeq 1.3\sqrt{\sigma_{phys}}$ for $b\ge 1.0\ \ (\sigma_{phys}^{-1/2})$, whereas the scalar glueball mass is kept to be near $M(0^{++})\sim 3.7\sqrt{\sigma_{phys}}$. In addition, using the effective action composed  of simple 10 quadratic interactions alone, we can almost explain  \textit{analytically} the scaling function of the squared monopole density determined  numerically for large $b$ region $b>1.2\ (\sigma_{phys}^{-1/2})$.   
\end{abstract}

\pacs{11.15.Ha,14.80.Hv,11.10.Wx}

\maketitle

\section{Introduction}
 
 It is shown in the continuum limit that the violation of the non-Abelian Bianchi identities (VNABI) $J_{\mu}$  is equal to Abelian-like monopole currents $k_{\mu}$ defined by the violation of the Abelian-like Bianchi identities~\cite{Suzuki:2014wya, SIB201711}. Although VNABI is an adjoint operator satisfying the covariant conservation rule $D_{\mu}J_{\mu}=0$, it  gives us, at the same time, the Abelian-like conservation rule $\partial_{\mu}J_{\mu}=0$. There are $N^2-1$ conserved magnetic charges in the case of color $SU(N)$.  The charge of each component of VNABI is quantized \`{a} la Dirac. The color invariant eigenvalue $\lambda_\mu$ of VNABI also satisfies the Abelian conservation rule $\partial_\mu\lambda_\mu=0$ and the magnetic charge of the eigenvalue is also quantized \`{a} la Dirac. If the color invariant eigenvalue make condensation in the QCD vacuum, each color component of the non-Abelian electric field $E^a$ is squeezed by the corresponding color component of the sorenoidal current $J^a_{\mu}$. Then only the color singlets alone can survive as a physical state and non-Abelian color confinement is realized.

 To prove if such a new confinement scheme is realized in nature, studies in the framework of pure $SU(2)$ lattice gauge theories have been done as a simple model of QCD~\cite{SIB201711}.  An Abelian-like definition of a monopole following DeGrand-Toussaint~\cite{DeGrand:1980eq} is adopted as a lattice version of VNABI, since the Dirac quantization condition of the magnetic charge is taken into account on lattice. In Ref~\cite{SIB201711}, the continuum limit of the lattice VNABI density is studied by introducing various techniques of smoothing the thermalized vacuum which is contaminated by lattice artifacts originally. With these improvements, beautiful and convincing scaling behaviors are seen when we plot the density $\rho(a(\beta),n)$ versus $b=a(\beta)$, where  $\rho(a(\beta),n)=\sum_{s,\mu}\sqrt{\sum_{a=1}^3  (K_{\mu}^a(s))^2}/(4\sqrt{3}Vb^3)$,  $K_{\mu}^a(s)$ is an $n$ blocked monopole in the color direction $a$, $n$ is the number
of blocking steps, V is the four-dimensional lattice volume and $b=na(\beta)$ is the lattice  spacing
of the blocked lattice. A single universal curve $\rho(b)$ is found from $n=1$ up to $n=12$,
which suggests that $\rho(a(\beta),n)$ is a function of $b=na(\beta)$ alone.  The scaling means that the lattice definition of VNABI has the continuum limit.

 The monopole dominance and the dual Meissner effect of the new scheme were studied already several years ago without any gauge fixing~\cite{Suzuki:2009xy} by making use of huge number of thermalized vacua produced by random gauge transformations. The monopole dominance of the string tension was shown beautifully. The dual Meissner effect with respect to each color electric field was shown also beautifully by the   Abelian monopole in the corresponding color direction. 
 
Now  in this paper we perform the blockspin renormalization-group study of lattice $SU(2)$ gauge theory and try to get the infrared effective VNABI action by introducing a blockspin transformation of lattice VNABI (Abelian monopoles). Since lattice VNABI is defined as Abelian monopoles following Degrand-Toussaint~\cite{DeGrand:1980eq}, the renormalization-group study is similar to the previous works done  in maximally Abelian (MA) gauge~\cite{Ivanenko:1991wt,Shiba:1994db,Kato:1998ur,Chernodub:2000ax}. However here we mainly  adopt global color-invariant maximal center gauge (MCG)~\cite{DelDebbio:1996mh,DelDebbio:1998uu} as a gauge smoothing the lattice vacuum, although comparison of the results in other smooth gauges is discussed. Beautiful scaling and gauge-independent behaviors are found to exist, not only with respect to the monopole density done in Ref.~\cite{SIB201711}, but also with respect to the effective monopole action. 

 After numerically deriving the infrared effective action with the simple assumption of two-point monopole interactions alone, we try to get the monopole action in the continuum limit by applying the method called blocking from the continuum~\cite{ref:BFC}. When use is made of Berezinskii-Kosterlitz-Thouless (BKT) transformation, the infrared monopole action can be transformed into the string model action.
Since large $b=na(\beta)$ corresponds to the strong-coupling region in the string model, the string tension and the lowest glueball mass can be evaluated \textit{analytically}  with the use of the strong-coupling expansion. The almost perfect action gives us $\sqrt{\sigma}\simeq 1.3\sqrt{\sigma_{phys}}$ for $b\ge 1.0\ \ (\sigma_{phys}^{-1/2})$, whereas the lowest scalar glueball mass is kept to be near $M(0^{++})\sim 3.7\sqrt{\sigma}$~\cite{Lucini2004}. Finally, we try to explain the scaling behavior of the monopole density observed in Ref.~\cite{SIB201711} starting from the obtained effective monopole action composed of 10 quadratic interactions alone. Since the square-root operator is difficult to evaluate, we adopt the squared monopole density  $R(b)=\sum_{s,\mu}(\sum_{a=1}^3(K_{\mu}^a(s))^2)/(4Vb^3)$. $R(b)$ is found numerically to be a function of 
$b=na(\beta)$ alone. It is interesting to see the numerically determined scaling behavior of  $R(b)$ can almost be reproduced analytically by the simple monopole action for $b>1.2\ (\sigma_{phys}^{-1/2})$, although there remains around 30\% discrepancy due mainly to the choice of simplest 10 quadratic monopole interactions alone.

\section{The effective monopole action and the blockspin transformation of lattice monopoles}
The method to derive the monopole action is the following:

\begin{itemize}
\item[1] 
We generate $SU(2)$ link fields $\mbra{U(s,\mu)}$ using the
tadpole-improved action  \cite{Alford:1995hw} for SU(2) gluodynamics:
\beq
    S =   \beta \sum_{pl} S_{pl}
       - {\beta \over 20 u_0^2} \sum_{rt} S_{rt}
\label{eq:improved_action}
\eeq
where $S_{pl}$ and $S_{rt}$ denote plaquette and $1 \times 2$ rectangular loop terms in the action,
\beq
S_{pl,rt}\ = \ {1\over 2}{\rm Tr}(1-U_{pl,rt}) \, ,
\label{eq:terms}
\eeq
the parameter $u_0$ is the {\it input} tadpole improvement
factor taken here equal to the fourth root of the average plaquette
$P=\langle \frac{1}{2} {\mathrm tr} U_{pl} \rangle$.
We consider $48^4$ ($24^4$) hyper-cubic lattice for $\beta=3.0\sim 3.9$
(for $\beta=3.0\sim 3.7$). For details of the vacuum generation using the tadpole-improved action, see Ref.~\cite{SIB201711}.
\item[2] Monopole loops in the thermalized vacuum produced from the above improved action (\ref{eq:improved_action}) still contain large amount of lattice artifacts.
Hence we adopt a gauge-fixing technique smoothing the vacuum, although any gauge-fixing is not necessary for smooth continuum configurations.
The first smooth gauge is the maximal center gauge~\cite{DelDebbio:1996mh,DelDebbio:1998uu} which is usually discussed in the framework of the center vortex idea. We adopt the so-called direct maximal center gauge which requires maximization of the quantity
\begin{eqnarray}
R=\sum_{s,\mu}(\tr U(s,\mu))^2 \label{eq:MCG}
\end{eqnarray}
with respect to local gauge transformations. Here $U(s, \mu)$ is a lattice gauge field. The above condition  fixes the gauge up to
$Z(2)$ gauge transformation and can be considered as the Landau gauge for the adjoint representation. In our simulations, we choose simulated
annealing algorithm as the gauge-fixing method which is known to be  powerful for finding the global maximum. For details, see the reference~\cite{Bornyakov:2000ig}.
 
  For comparison, we also consider the direct Lalacian center gauge(DLCG)~\cite{Faber:2001zs}, Maximal Abelian Wilson loop (AWL) gauge~\cite{SIB201711} and Maximally Abelian (MA) plus $U1$ Landau gauge(MAU1)~\cite{Kronfeld:1987ri,Kronfeld:1987vd,SIB201711,Bali:1996dm}. 
\item[3]
Next we perform an abelian projection in the above smooth gauges
to separate abelian link variables. We explain how to extract the Abelian fields and the color-magnetic
monopoles from the thermalized non-Abelian SU(2) link variables~\cite{Suzuki:2009xy},
\begin{equation}
U(s,\mu) = U^0(s,\mu)+i\vec{\sigma}\cdot\vec{U}(s,\mu),
\end{equation}
where $\vec{\sigma}=(\sigma^{1},\sigma^{2},\sigma^{3})$ 
is the Pauli matrix.
Abelian link variables in one of the color directions, 
for example, in the $\sigma^1$ direction are defined as 
\begin{equation}
u_{\mu}(s) = \cos \theta_{\mu}(s) +i\sigma^1\sin \theta_{\mu}(s) \; ,
\end{equation}
where
\begin{equation}
\theta^1_{\mu}(s) 
= \arctan\bigl(\frac{U^1(s,\mu)}{U^0(s,\mu)}\bigr) \; 
\end{equation}
corresponds to the Abelian field.

\item[4] 
Monopole currents can be defined from abelian plaquette
variables $\theta_{\mu\nu}^a (s)$ following DeGrand and Toussaint~\cite{DeGrand:1980eq}.
The abelian plaquette variables are written by
\begin{eqnarray*}
\theta_{\mu\nu}^a(s)&\equiv&\theta_{\mu}^a(s)+\theta_{\nu}^a(s+\hat{\mu})
                   -\theta_{\mu}^a(s+\hat{\nu})-\theta_{\nu}^a(s),\\
& & (-4\pi< \theta_{\mu\nu}^a(s) < 4\pi).
\end{eqnarray*}
It is decomposed into two terms:
\begin{eqnarray*}
\theta_{\mu\nu}^a(s)&\equiv&\bar\theta_{\mu\nu}^a(s)+2\pi{n}_{\mu\nu}^a(s),\\
&&(-\pi \le \bar\theta_{\mu\nu}^a(s) < \pi).
\end{eqnarray*}
Here, $\bar\theta_{\mu\nu}^a(s)$ is interpreted as the electro-magnetic
flux with color $a$ through the plaquette and
the integer $n_{\mu\nu}^a (s)$ corresponds to the number of Dirac string
penetrating the plaquette.
One can define quantized conserved monopole currents
\begin{eqnarray}
k_\mu^a(s)=\frac{1}{2}\epsilon_{\mu\nu\rho\sigma}\partial_\nu
                           n_{\rho\sigma}^a(s+\hat{\mu}), 
\label{eqn:km}
\end{eqnarray}
where $\partial$ denotes the forward difference on the 
lattice.
The monopole currents satisfy a conservation law
$\partial^{\prime}_{\mu}k_\mu^a(s) = 0$ by definition,
where $\partial^{\prime}$ denotes the backward difference on the 
lattice.
\item[5]
We consider a set of independent and local monopole interactions 
which are summed up over
the whole  lattice. We denote each operator as ${\cal S}_i[k]$.
Then
the monopole action can be written as a linear combination of these
operators:
\begin{eqnarray}
 {\cal S}[k] = \sum_{i} F(i) {\cal S}_i[k], \label{eq:monoact}
\end{eqnarray}
where $F(i)$ are coupling constants.

The effective monopole action is defined  as follows:
\begin{eqnarray*}
e^{-{\cal S}[k]}&=&\int DU(s,\mu)e^{-S(U)}\\
&&\times \prod_a \delta(k_\mu^a(s)-\frac{1}{2}\epsilon_{\mu\nu\rho\sigma}\partial_\nu n_{\rho\sigma}^a(s+\hat{\mu})),
\end{eqnarray*}
where $S(U)$ is the gauge-field action (\ref{eq:improved_action}).

We determine the monopole action~(\ref{eq:monoact}), that is, the set of couplings $F(i)$ from the monopole current 
ensemble $\mbra{k_{\mu}^a(s)}$ with the aid of an inverse 
Monte-Carlo method first developed by Swendsen~\cite{swendsen} 
and extended to closed monopole currents by Shiba and Suzuki 
~\cite{Shiba:1994db}. The details of the inverse Monte-Carlo method are reviewed in Appendix\ref{APD:swendson}. See also the previous paper~\cite{Kato:1998ur}. 

Practically, we have to restrict the number of interaction terms. 
It is natural to assume that monopoles which are far apart do not 
interact strongly and to consider only short-ranged local interactions 
of monopoles. The form of actions adopted here
 are shown in Appendix~\ref{APD:action} and in Appendix~\ref{APD:comparison}.
Some comments are in order:
\begin{itemize}
  \item Contrary to previous studies in MA gauge, there are three colored Abelian monopoles here. Due to the possible interactions between  
gauge fields and monopoles, there may appear interactions between different 
colored monopoles. When we consider here only effective actions of Abelian 
monopoles, such induced interactions between monopoles of different colors become jnevitablly non-local.  Also no two-point color-mixed interactions appear. 
\item We adopt only monopole interactions which are local and have no color mixing, since stable convergence could not be obtained with introduction of color-mixed four and six-point local interactions. 
\item Actually, we study here in details assuming two-point monopole interactions alone, although some four and six point interactions without any color mixing are studied for comparison. For the discussions concerning the set of monopole interactios, see  Appendix~\ref{APD:comparison}. 
\item All possible types of interactions are not independent due to
the conservation law of the monopole current. So we get rid of
almost all perpendicular interactions by the use of 
the conservation rule~\cite{Shiba:1994db, Chernodub:2000ax}.
\end{itemize}

\item[6] 
We perform a blockspin transformation in terms of 
the monopole currents on the dual lattice 
to investigate the renormalization flow in the IR region. 
We adopt $n=1,2,3,4,6,8,12$ extended conserved monopole currents
as an $n$ blocked operator~\cite{Ivanenko:1991wt}:
\begin{eqnarray}
K_\mu(s^{(n)})&=&
  \sum_{i,j,l=0}^{n-1}
    k_\mu(s(n,i,j,l))\nn\\ 
&\equiv&{\cal B}_{k_\mu}(s^{(n)}),
\label{pfac:2}
\end{eqnarray}
where $s(n,i,j,l)\equiv ns^{(n)}+(n-1)\hat{\mu}+i\hat{\nu}+j\hat{\rho}+l\hat{\sigma}$.
The renormalized lattice spacing is $b=na(\beta)$ and the continuum limit is 
taken as the limit $n \to \infty$ for a fixed physical length $b$.

We determine the effective monopole action from the blocked 
monopole current ensemble $\mbra{K_{\mu}(s^{(n)})}$.
Then one can obtain the renormalization group flow in the coupling constant space.

\item[7]
The physical length $b=n a(\beta)$ is taken in unit of the physical 
string tension $\sigma_{phys}^{-1/2}$. We evaluate 
the string tension $\sigma_{lat}$ from the monopole part of the abelian
Wilson loops for each $\beta$ since the error bars are small in 
this case. 
The lattice spacing $a(\beta)$ is given
by the relation $a(\beta)=\sqrt{\sigma_{lat}/\sigma_{phys}}$.
Note that 
$b=1.0\ \sigma_{phys}^{-1/2}$ corresponds to $0.45fm$, when we 
assume $\sigma_{phys}\cong (440 MeV)^2$.
\end{itemize}

\begin{figure}[htb]
\caption{The coupling constants of the self and the two nearest-neighbor interactions in the effective monopole action versus $b=na(\beta)$  in MCG on $48^4$. }
\label{figF123_MCG_48}
  \begin{minipage}[b]{0.9\linewidth}
    \centering
    \includegraphics[width=8cm,height=6.cm]{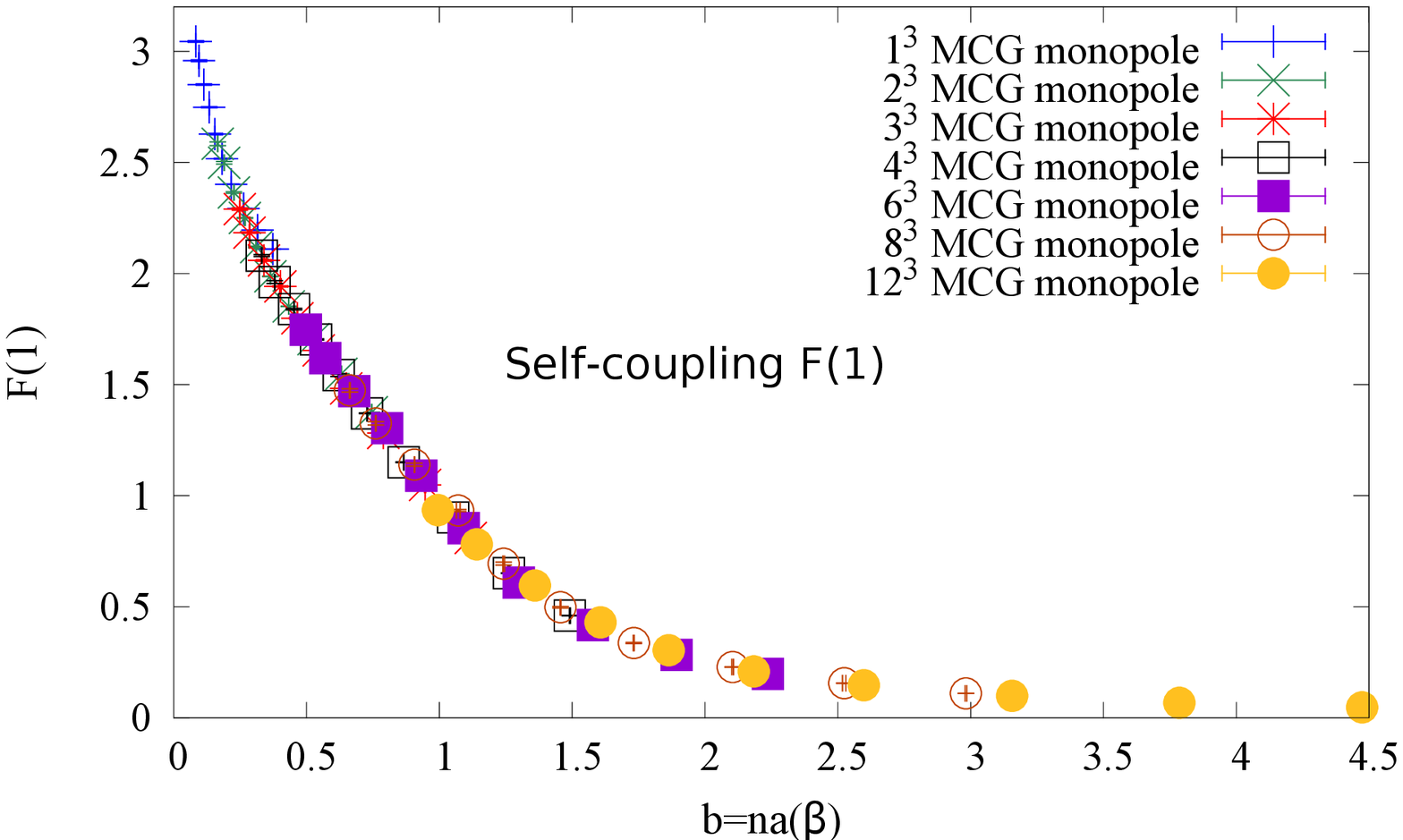}
  \end{minipage}
  \begin{minipage}[b]{0.9\linewidth}
    \centering
    \includegraphics[width=8cm,height=6.cm]{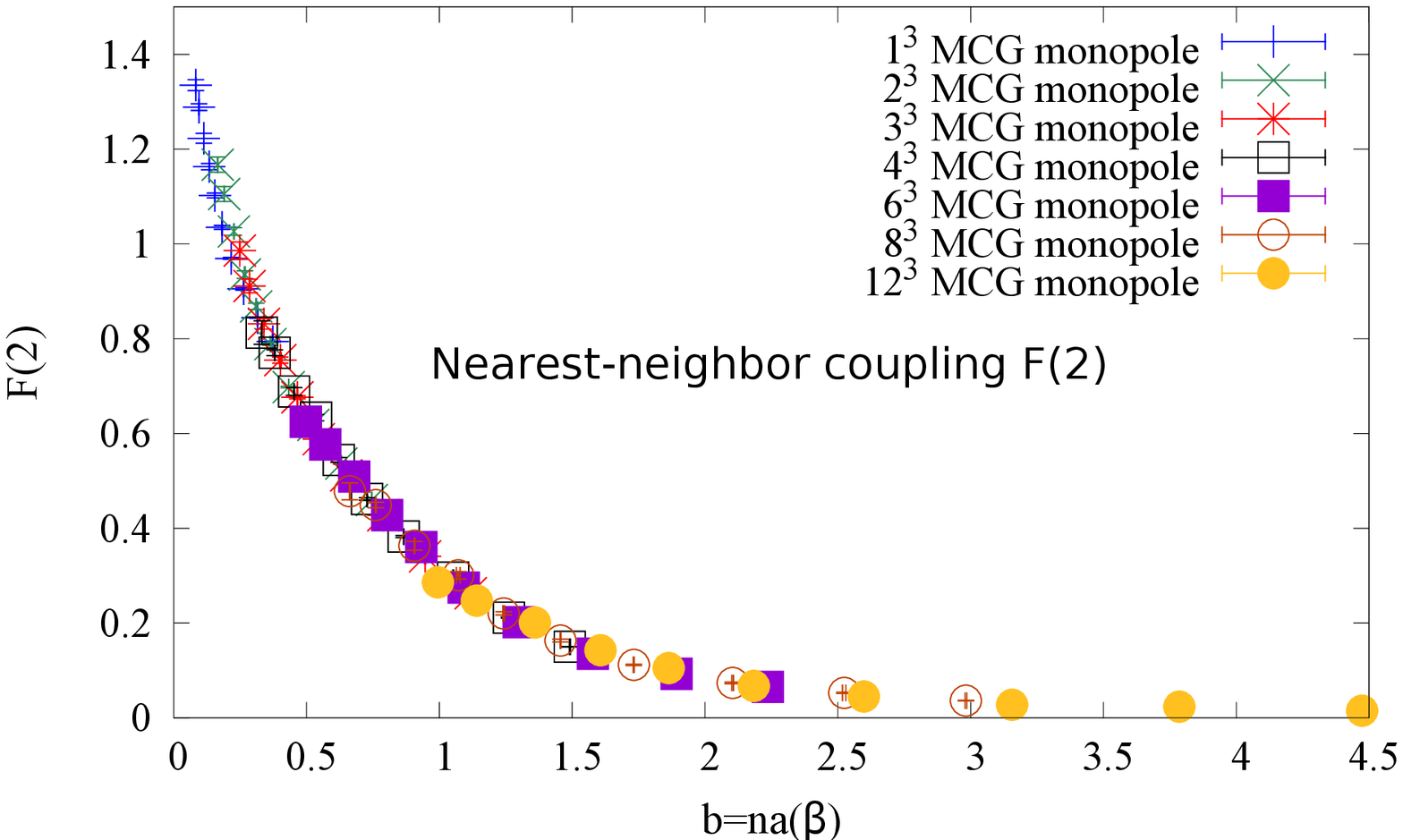}
  \end{minipage}
  \begin{minipage}[b]{0.9\linewidth}
    \centering
    \includegraphics[width=8cm,height=6.cm]{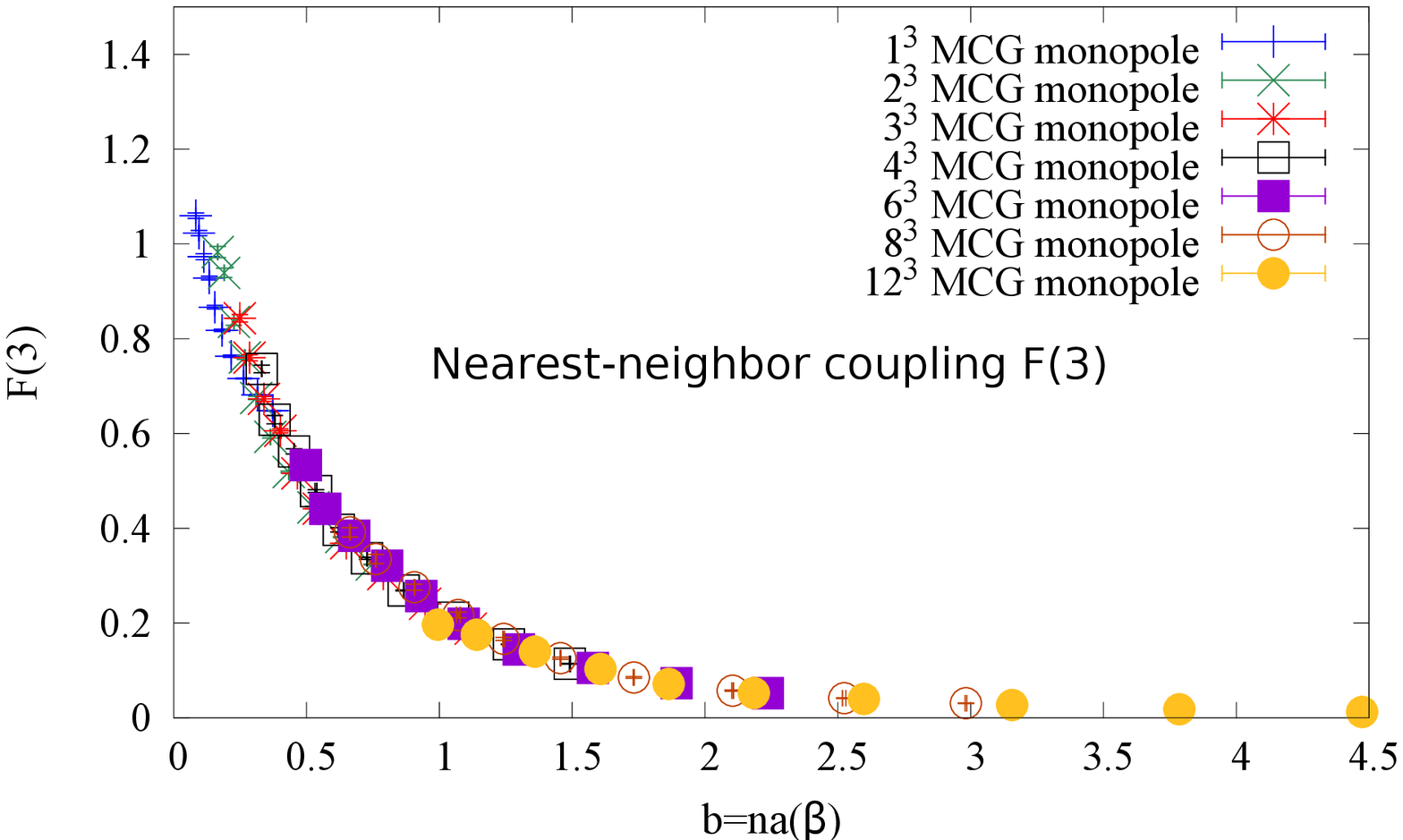}
  \end{minipage}
\end{figure}
\begin{figure}[htb]
\caption{The coupling constants of the two next to the nearest-neighbor interactions in the effective monopole action versus $b=na(\beta)$  in MCG on $48^4$. }
\label{figF45_MCG_48}
  \begin{minipage}[b]{0.9\linewidth}
    \centering
    \includegraphics[width=8cm,height=6.cm]{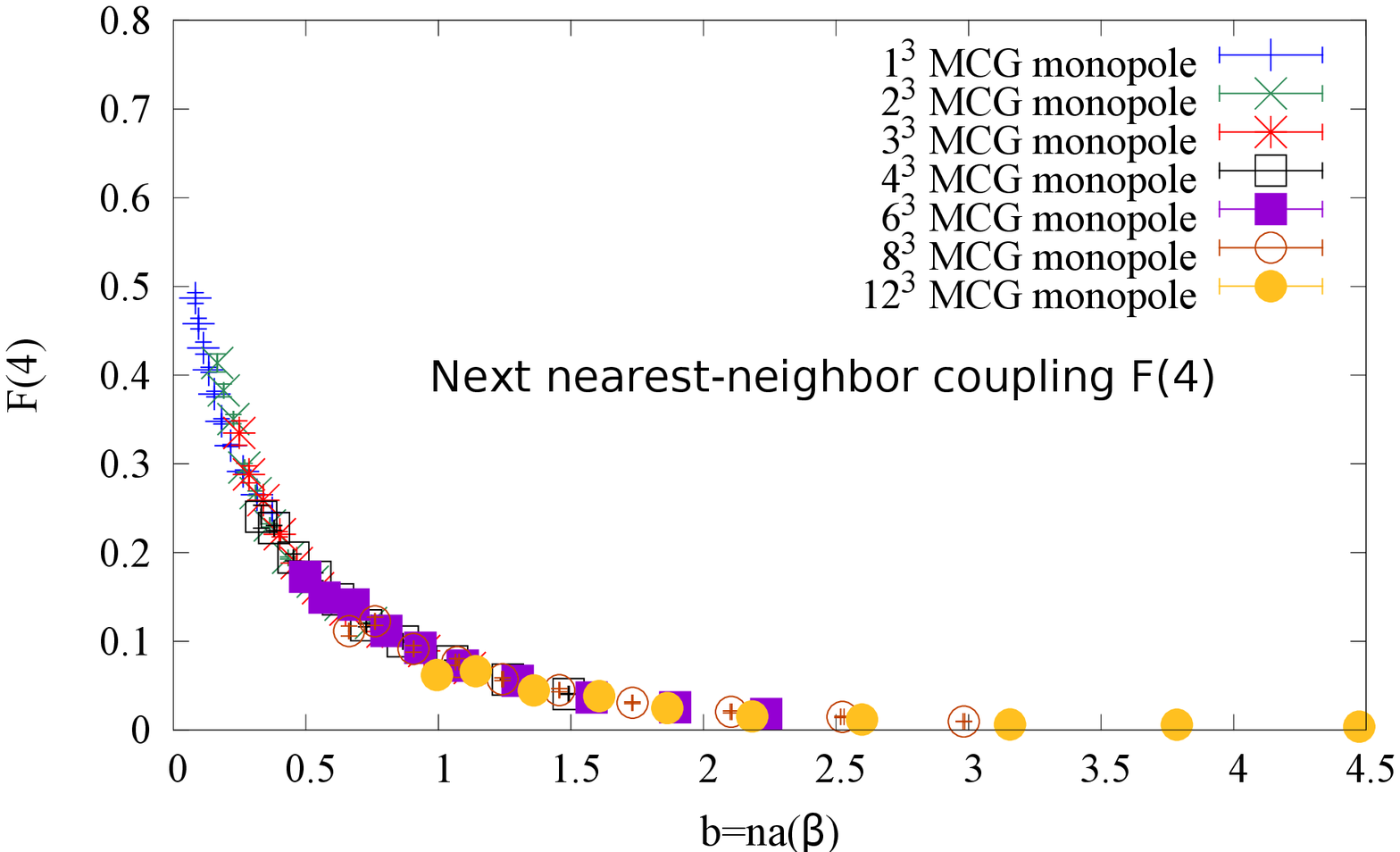}
      \end{minipage}
  \begin{minipage}[b]{0.9\linewidth}
    \centering
    \includegraphics[width=8cm,height=6.cm]{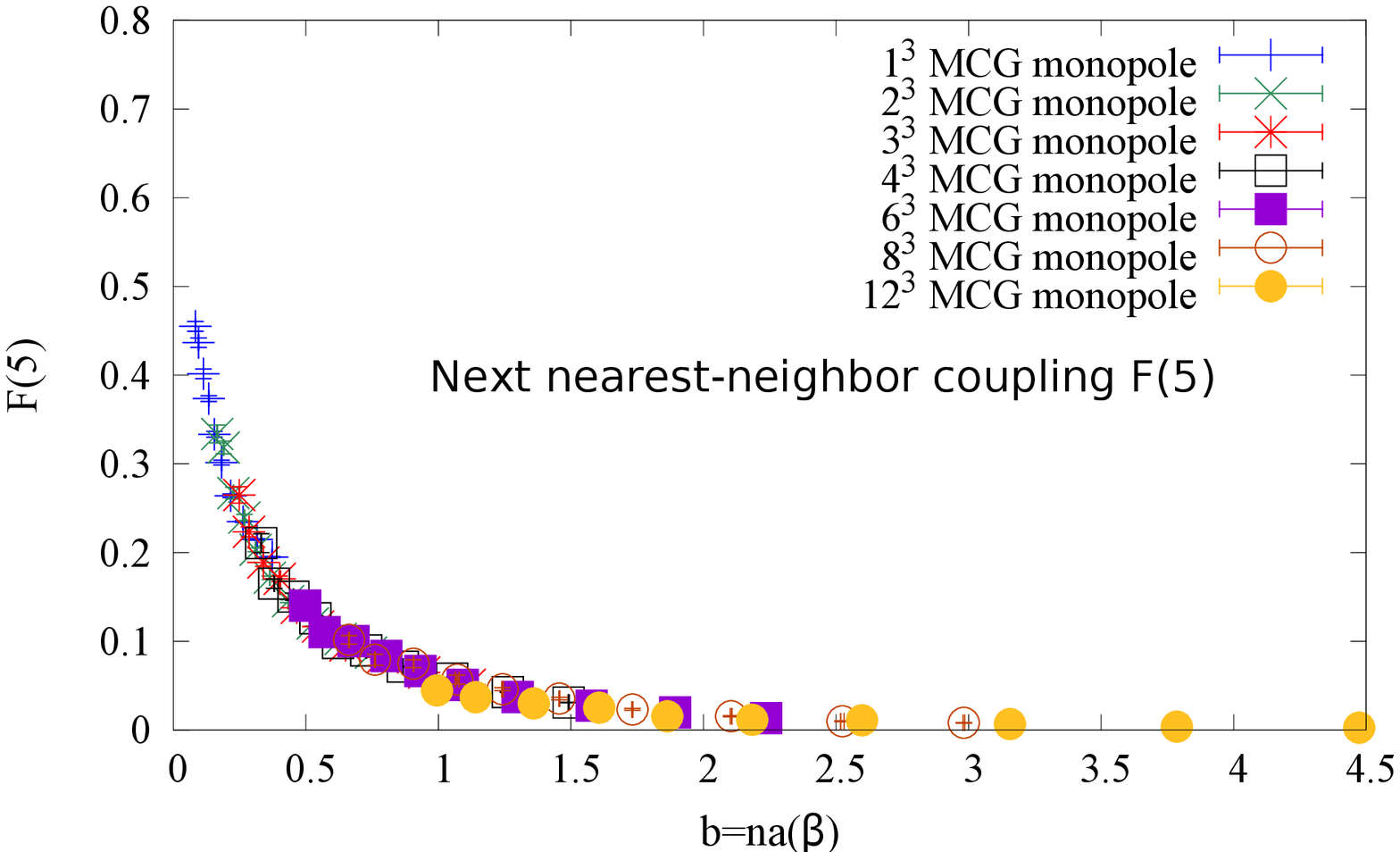}
  \end{minipage}
\end{figure}
\begin{figure}[htb]
\caption{The renoramlization-group flow projected onto the two-dimensional coupling constant planes in MCG on $48^4$. }
\label{fig_flowF1_234}
  \begin{minipage}[b]{0.9\linewidth}
    \centering
    \includegraphics[width=8cm,height=6.cm]{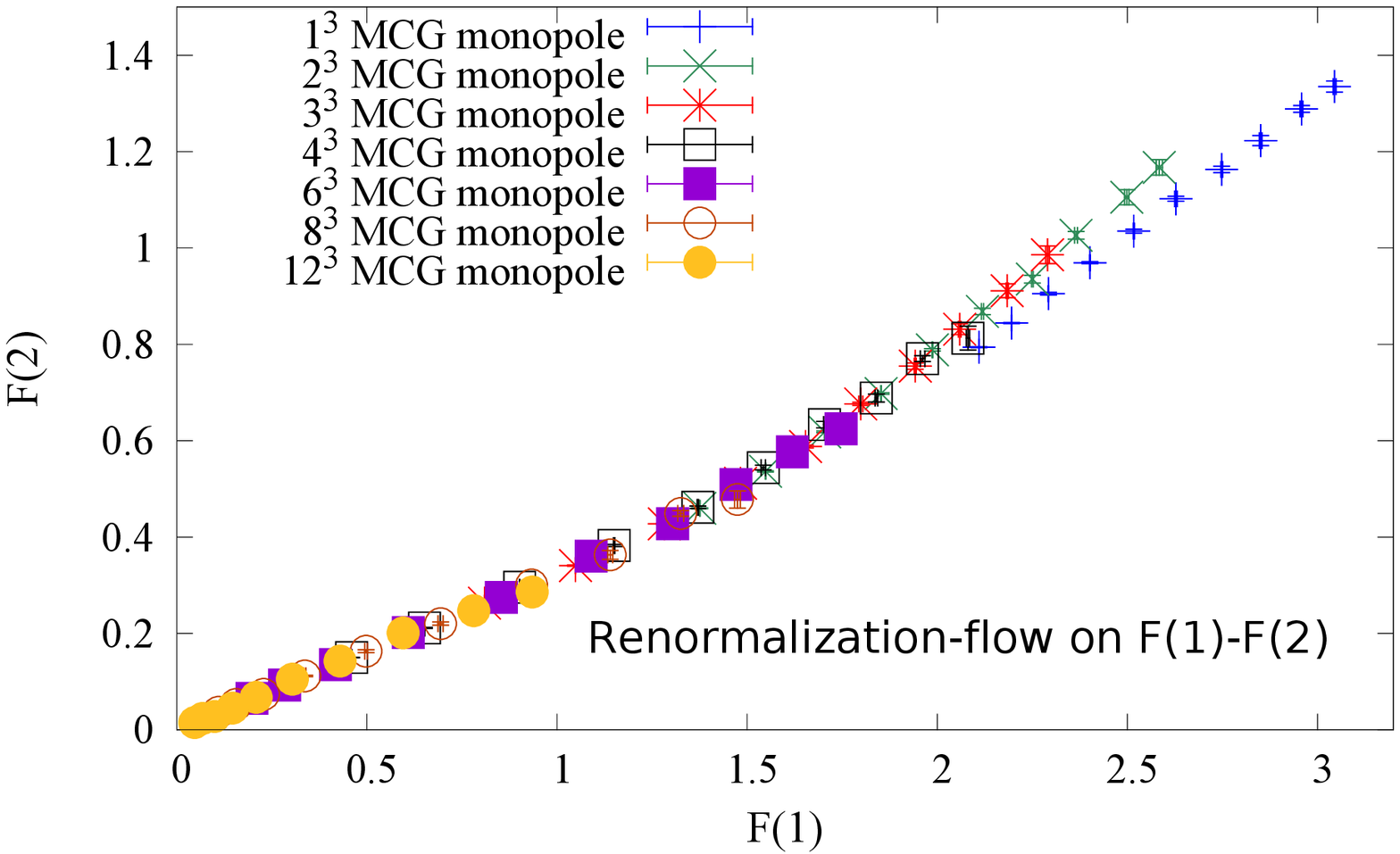}
  \end{minipage}
  \begin{minipage}[b]{0.9\linewidth}
    \centering
    \includegraphics[width=8cm,height=6.cm]{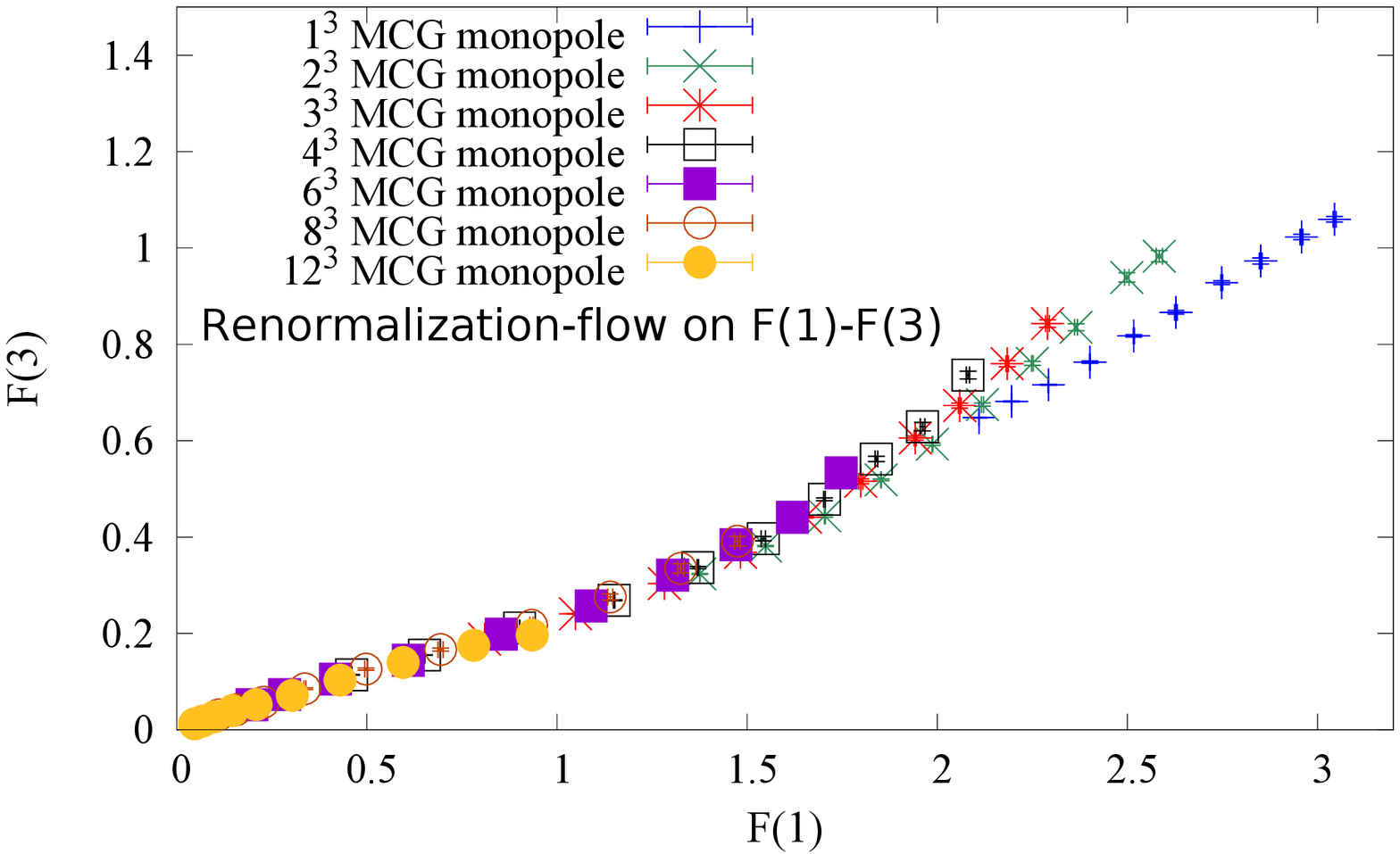}
  \end{minipage}
  \begin{minipage}[b]{0.9\linewidth}
    \centering
    \includegraphics[width=8cm,height=6.cm]{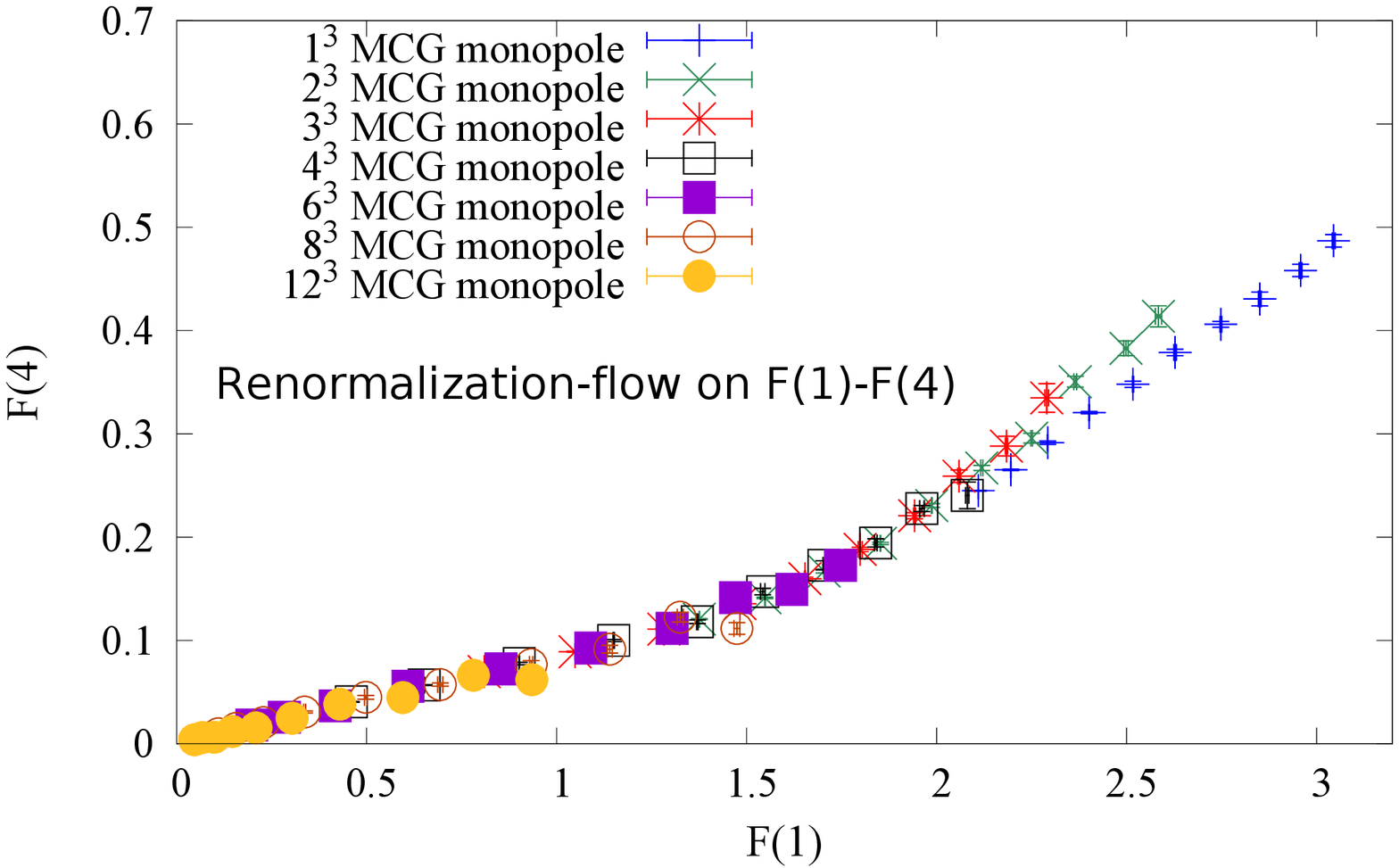}
  \end{minipage}
\end{figure}
\begin{figure}[htb]
\caption{The renoramlization-group flow projected onto the two-dimensional coupling constant planes in MCG on $48^4$. }
\label{fig_flow_F15_F23}
  \begin{minipage}[b]{0.9\linewidth}
    \centering
    \includegraphics[width=8cm,height=6.cm]{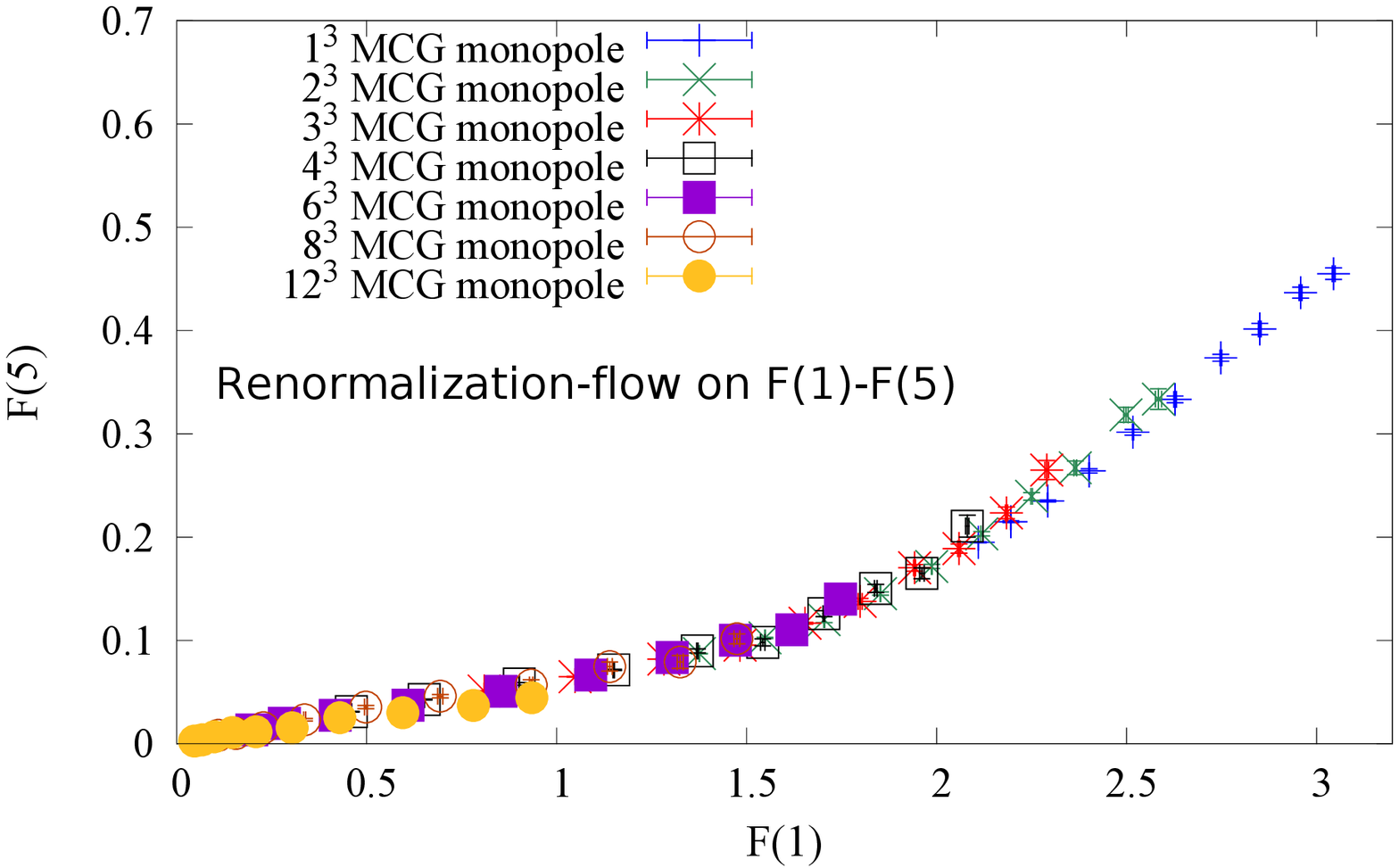}
  \end{minipage}
  \begin{minipage}[b]{0.9\linewidth}
    \centering
    \includegraphics[width=8cm,height=6.cm]{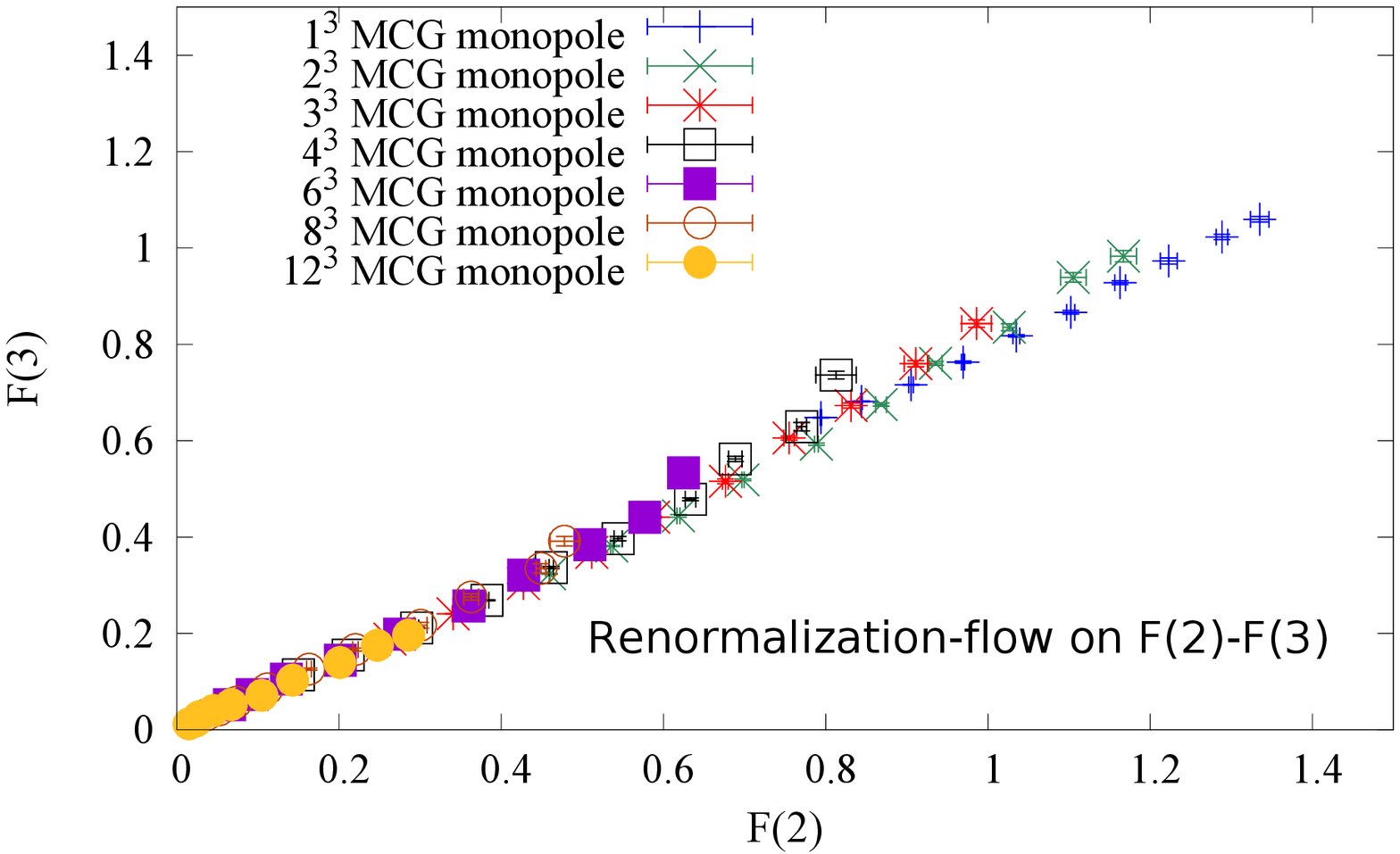}
  \end{minipage}
\end{figure}
\begin{figure}[htb]
\caption{Volume dependence of the infrared effective monopole action in MCG on $24^4$ and $48^4$. The coupling constants of the self $F(1)$ and the next next nearest-neighbor interactions $F(6)$ are shown as examples.}
\label{fig_volume}
  \begin{minipage}[b]{0.9\linewidth}
    \centering
    \includegraphics[width=8cm,height=6.cm]{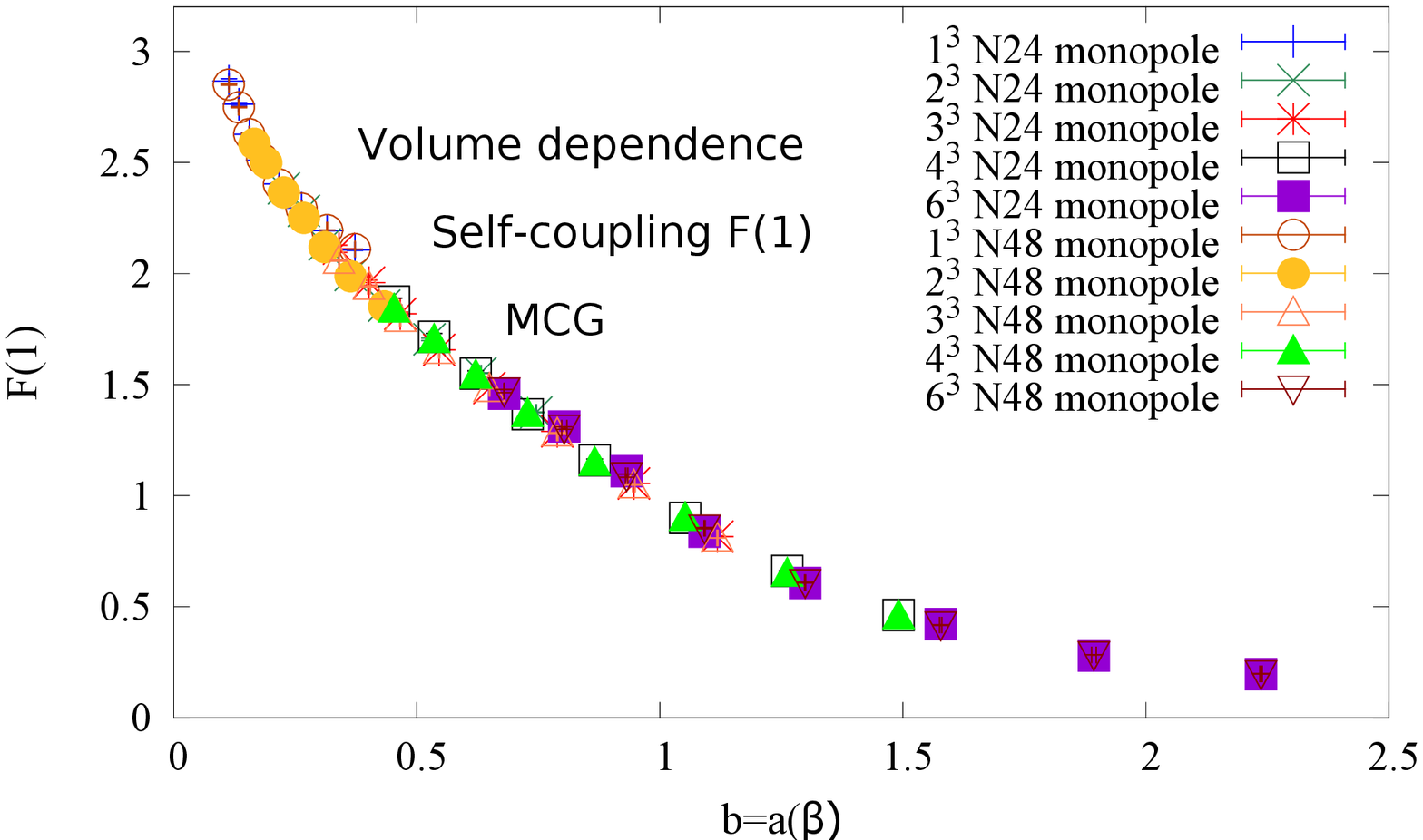}
    \subcaption{F(1)}
  \end{minipage}
  \begin{minipage}[b]{0.9\linewidth}
    \centering
    \includegraphics[width=8cm,height=6.cm]{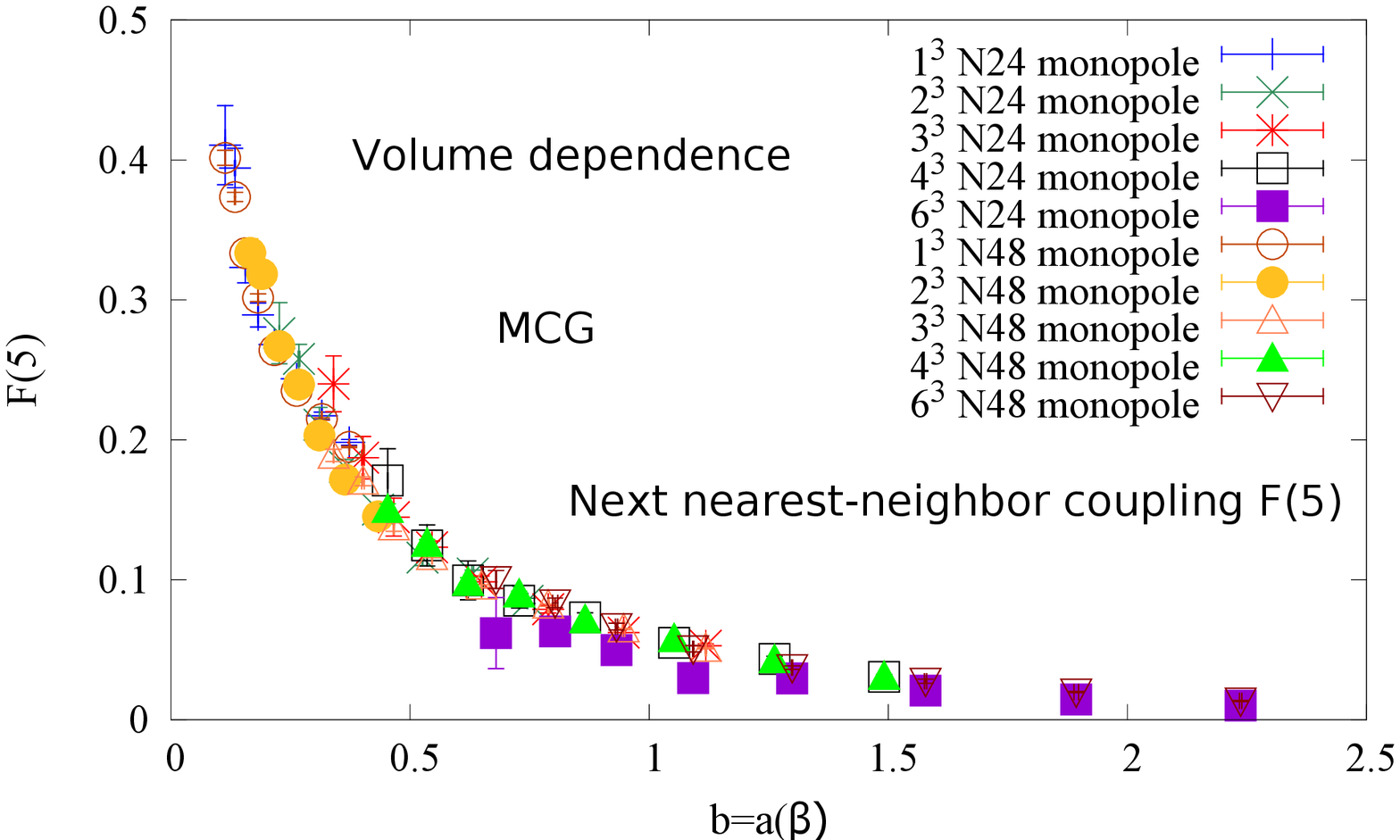}
    \subcaption{F(6)}
  \end{minipage}
\end{figure}
\begin{figure}[htb]
\caption{The self coupling of the infrared effective monopole action in NGF case in comparison with that in MCG case.}
\label{F1_NGF_MCG}  
    \centering
    \includegraphics[width=8cm,height=6.cm]{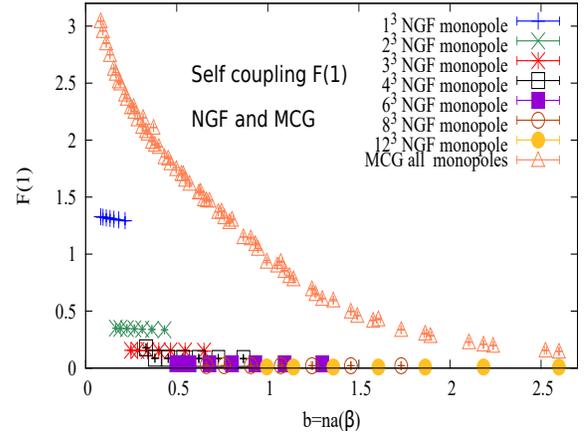}
\end{figure}
\begin{figure}[htb]
\caption{The infrared effective monopole action in DLCG and MCG on $24^4$. The coupling constants of the self and the next next nearest-neighbor interactions are shown as an example.}
\label{fig_DLCG}
  \begin{minipage}[b]{0.9\linewidth}
    \centering
    \includegraphics[width=8cm,height=6.cm]{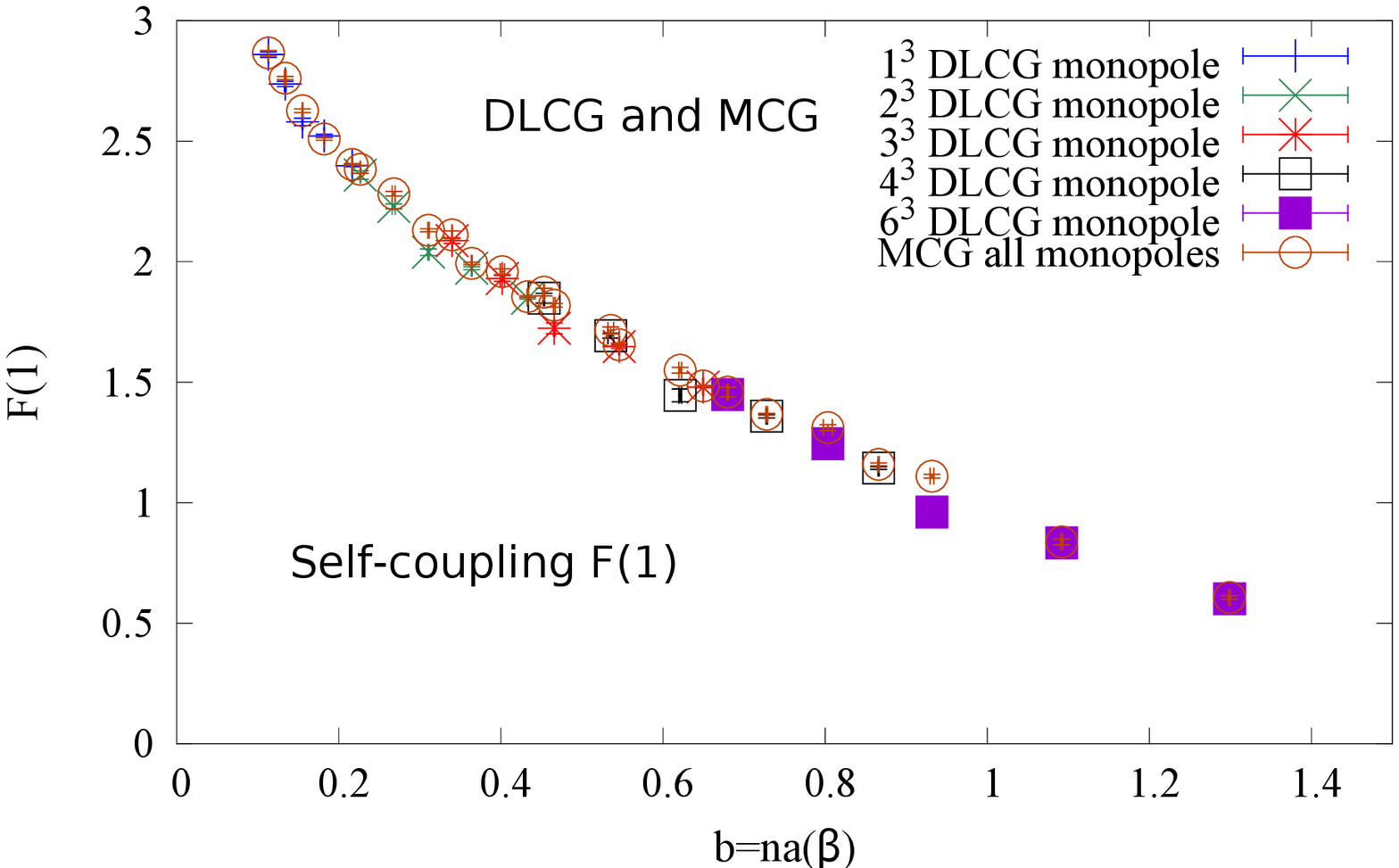}
  \end{minipage}
  \begin{minipage}[b]{0.9\linewidth}
    \centering
    \includegraphics[width=8cm,height=6.cm]{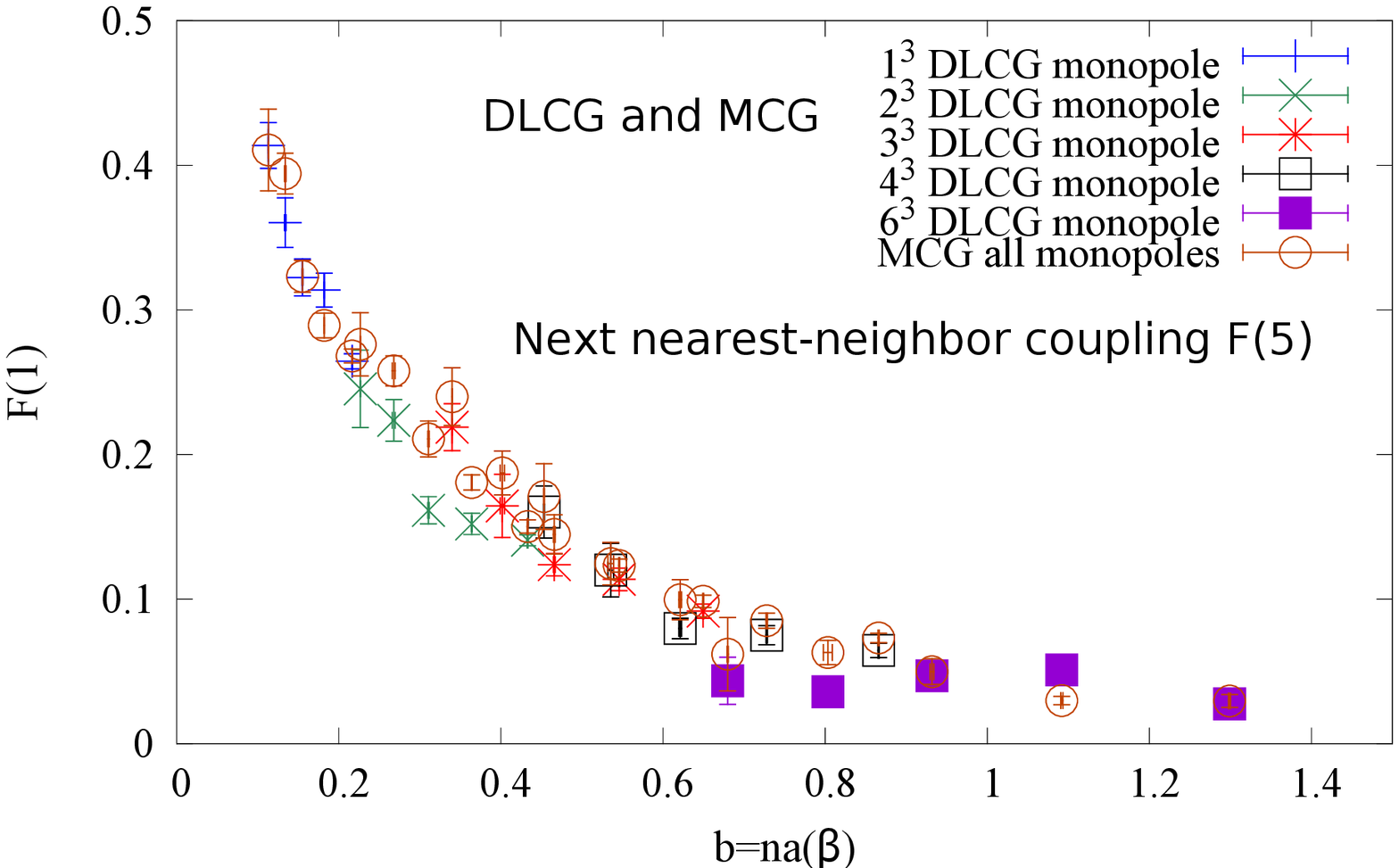}
  \end{minipage}
\end{figure}
\begin{figure}[htb]
\caption{The infrared effective monopole action in AWL and MCG on $48^4$. The coupling constants of the self  and the next nearest-neighbor interactions  are shown as an example.}
\label{fig_AWL}
  \begin{minipage}[b]{0.9\linewidth}
    \centering
    \includegraphics[width=8cm,height=6.cm]{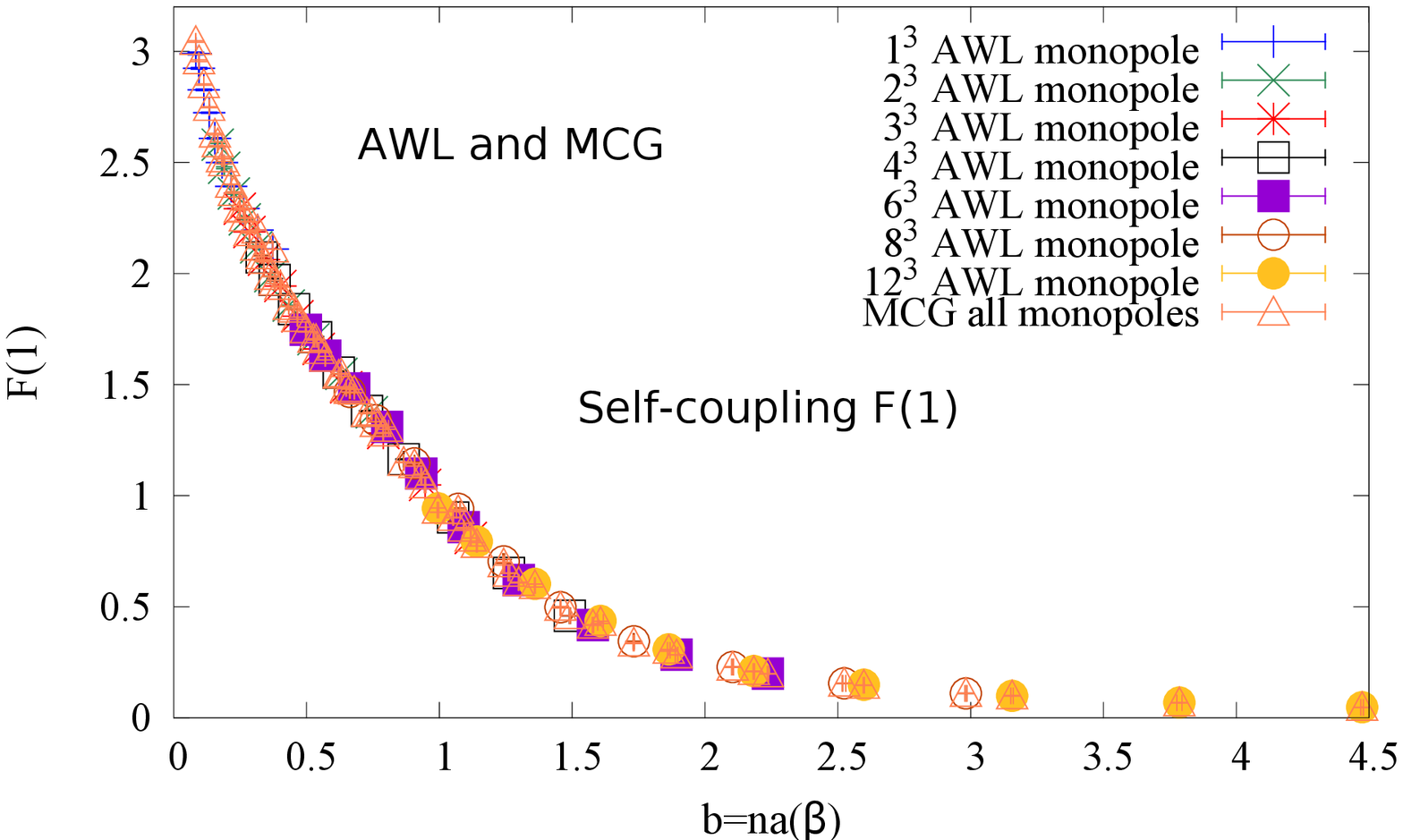}
  \end{minipage}
  \begin{minipage}[b]{0.9\linewidth}
    \centering
    \includegraphics[width=8cm,height=6.cm]{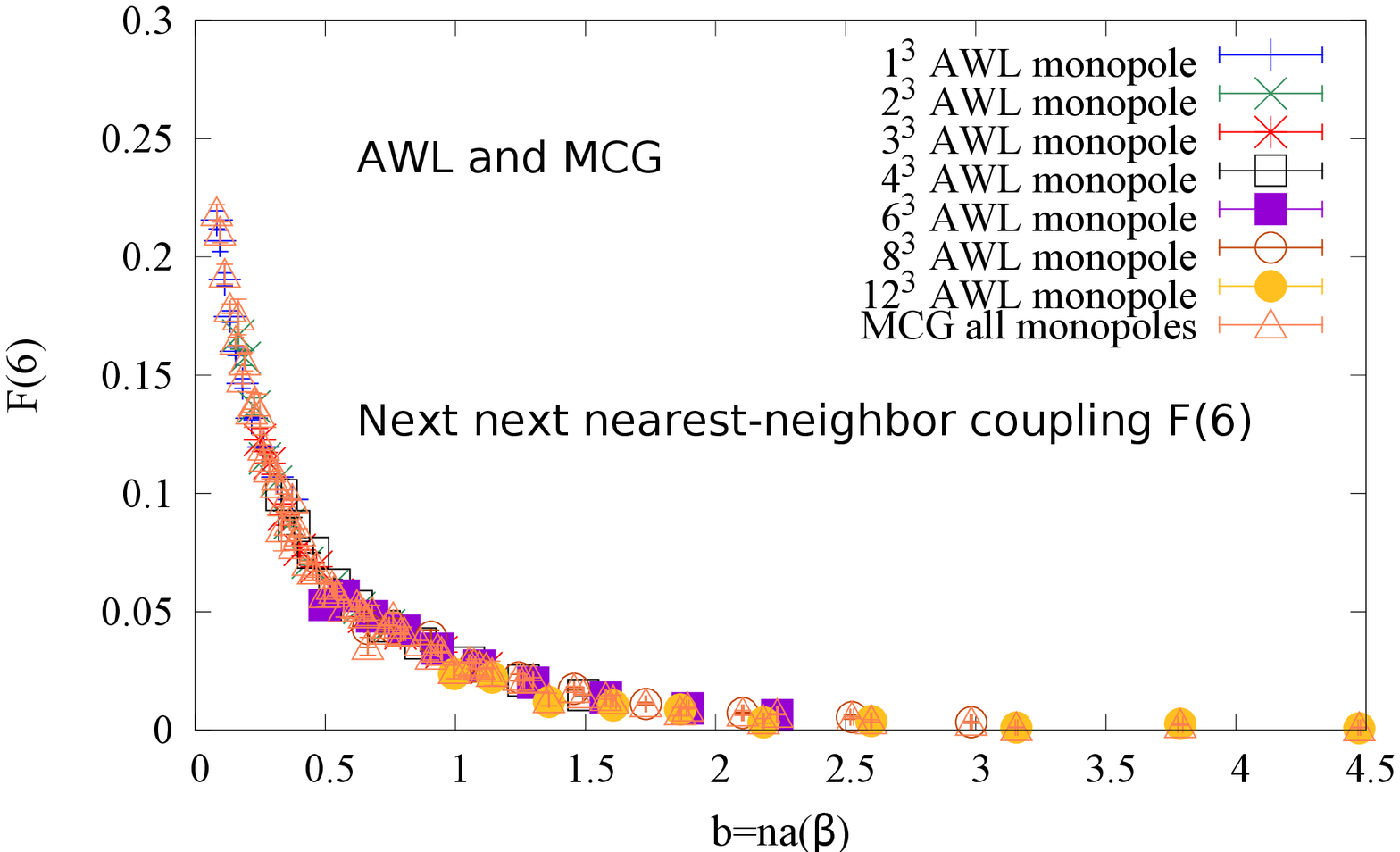}
  \end{minipage}
\end{figure}
\begin{figure}[htb]
\caption{The coupling constants of the self and the two nearest-neighbor interactions in the effective monopole action versus $b=na(\beta)$  in MAU1 on $48^4$. }
\label{figMAU1_F123}
  \begin{minipage}[b]{0.9\linewidth}
    \centering
    \includegraphics[width=8cm,height=6.cm]{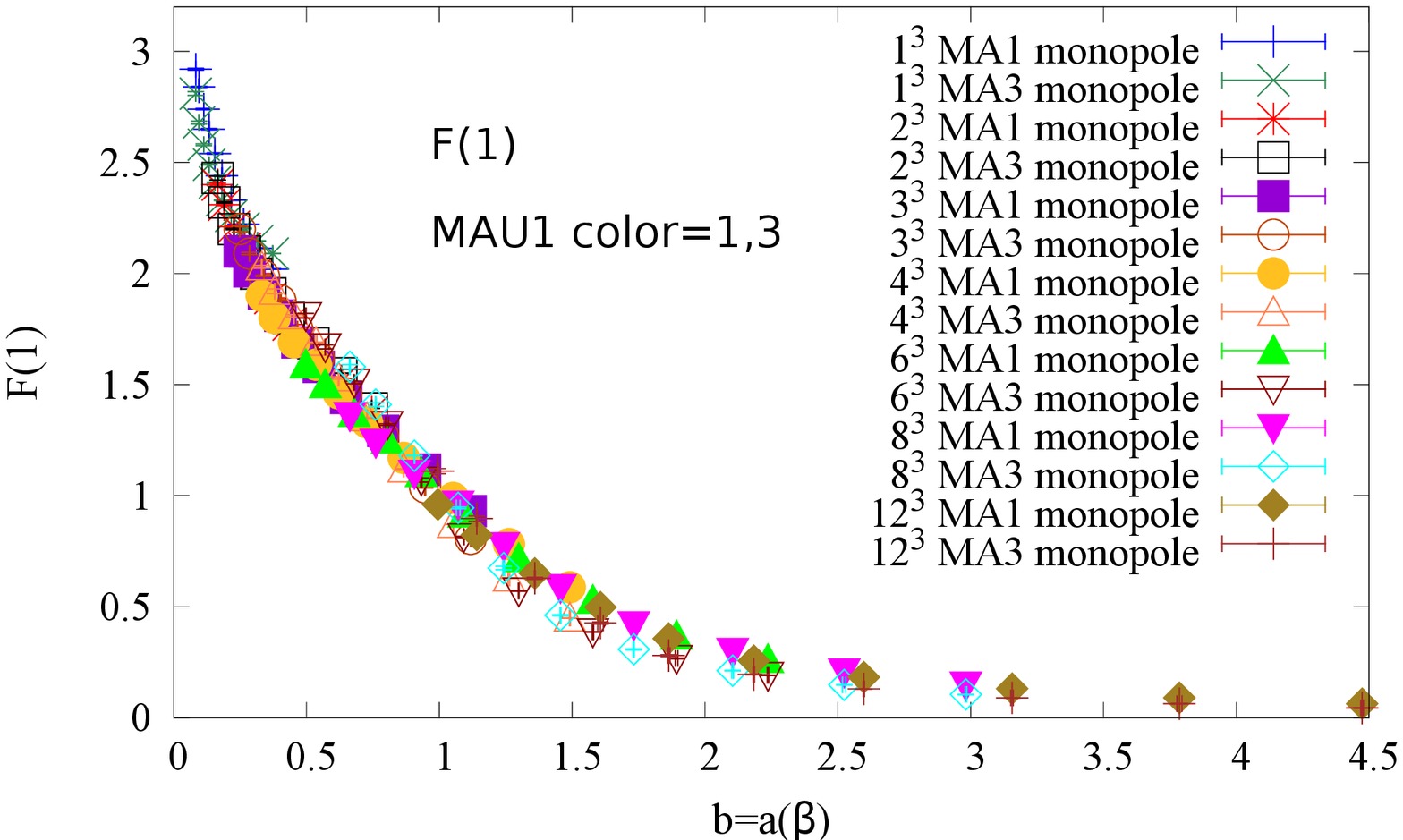}
  \end{minipage}
  \begin{minipage}[b]{0.9\linewidth}
    \centering
    \includegraphics[width=8cm,height=6.cm]{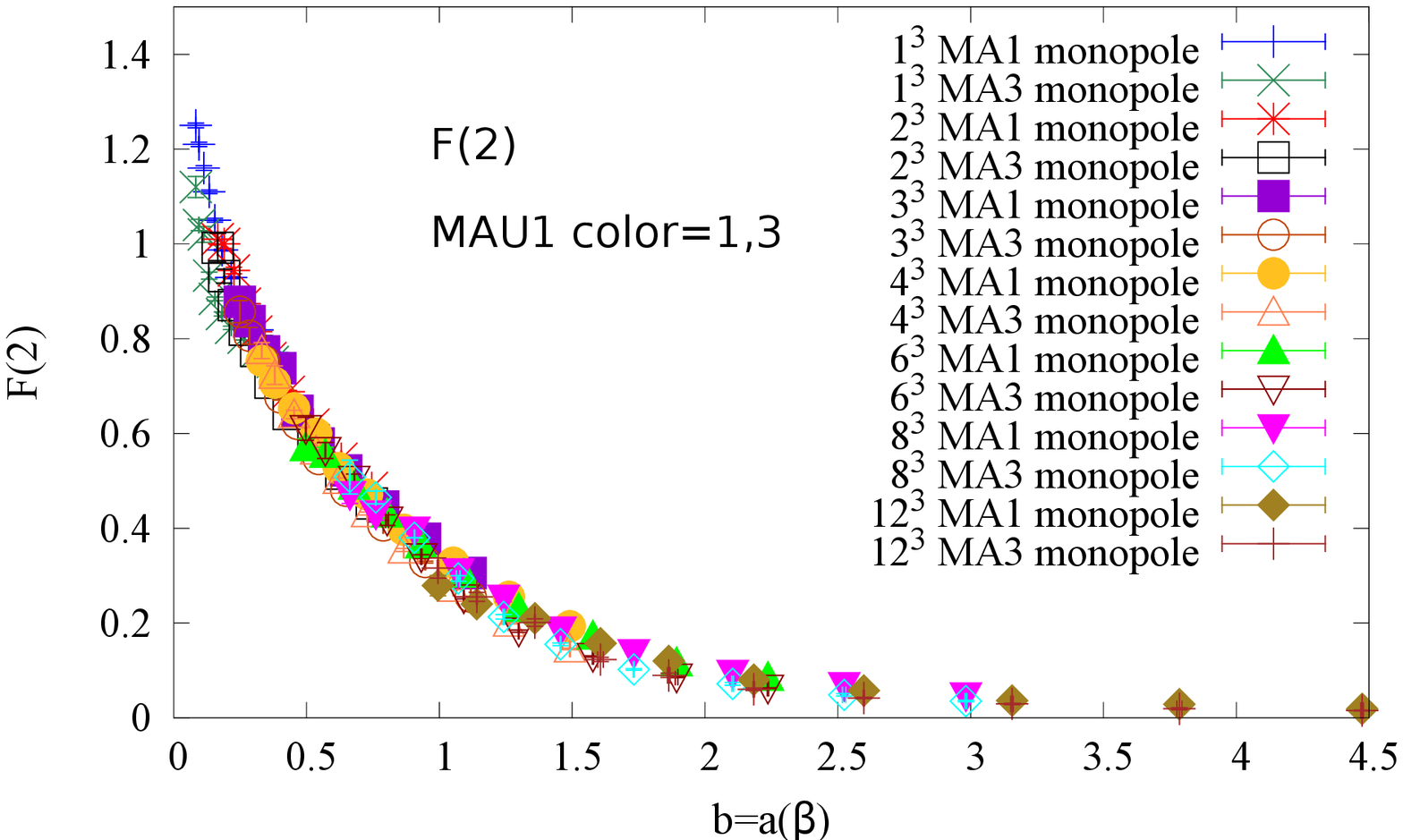}
  \end{minipage}
  \begin{minipage}[b]{0.9\linewidth}
    \centering
    \includegraphics[width=8cm,height=6.cm]{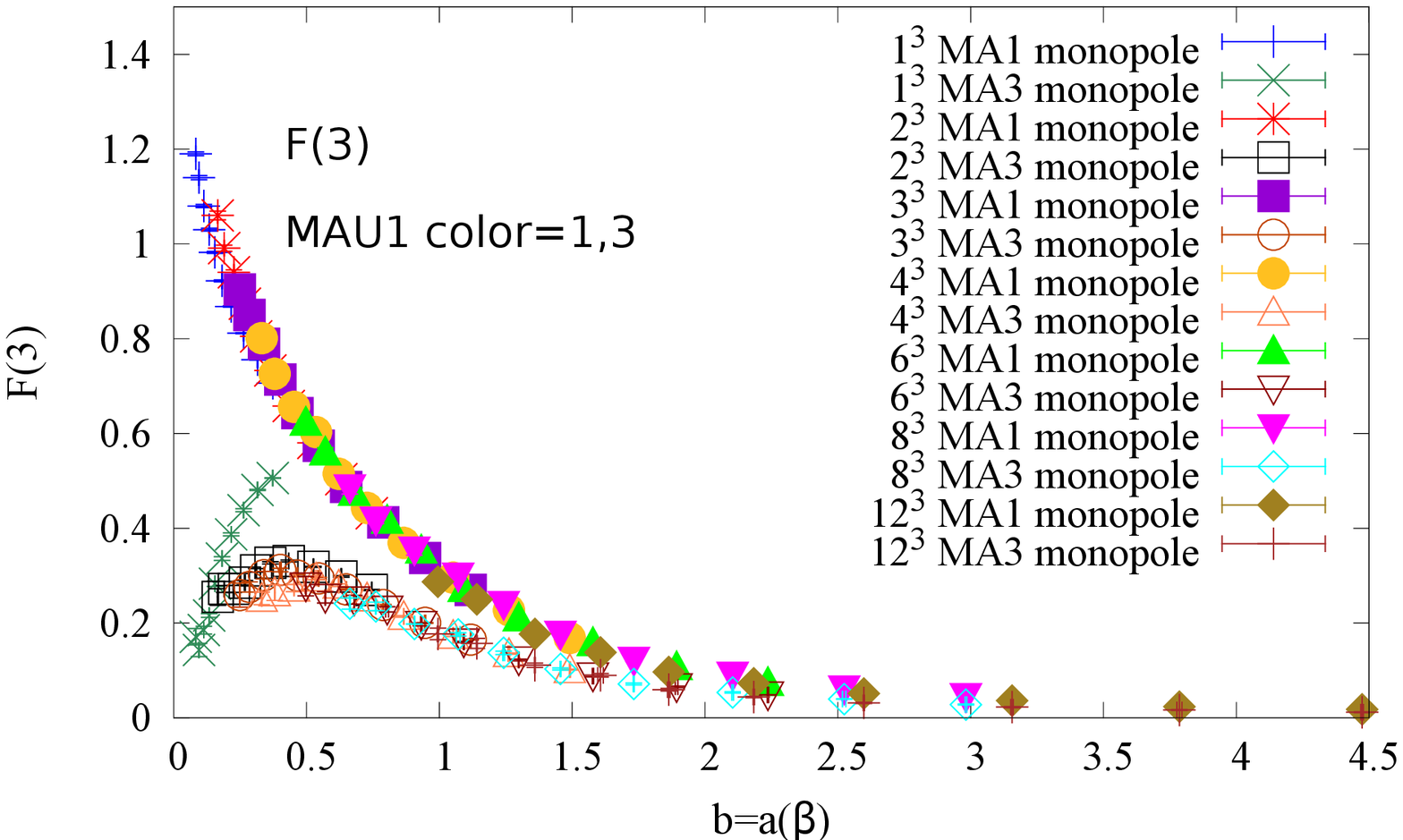}
  \end{minipage}
\end{figure}
\begin{figure}[htb]
\caption{The coupling constants of the self and the nearest-neighbor interactions in the effective monopole action versus $b=na(\beta)$  in MAU1 and MCG on $48^4$. The sum of each coupling constants with respect to three color components are shown. }
\label{figMAU1_MCG_F123}
  \begin{minipage}[b]{0.9\linewidth}
    \centering
    \includegraphics[width=8cm,height=6.cm]{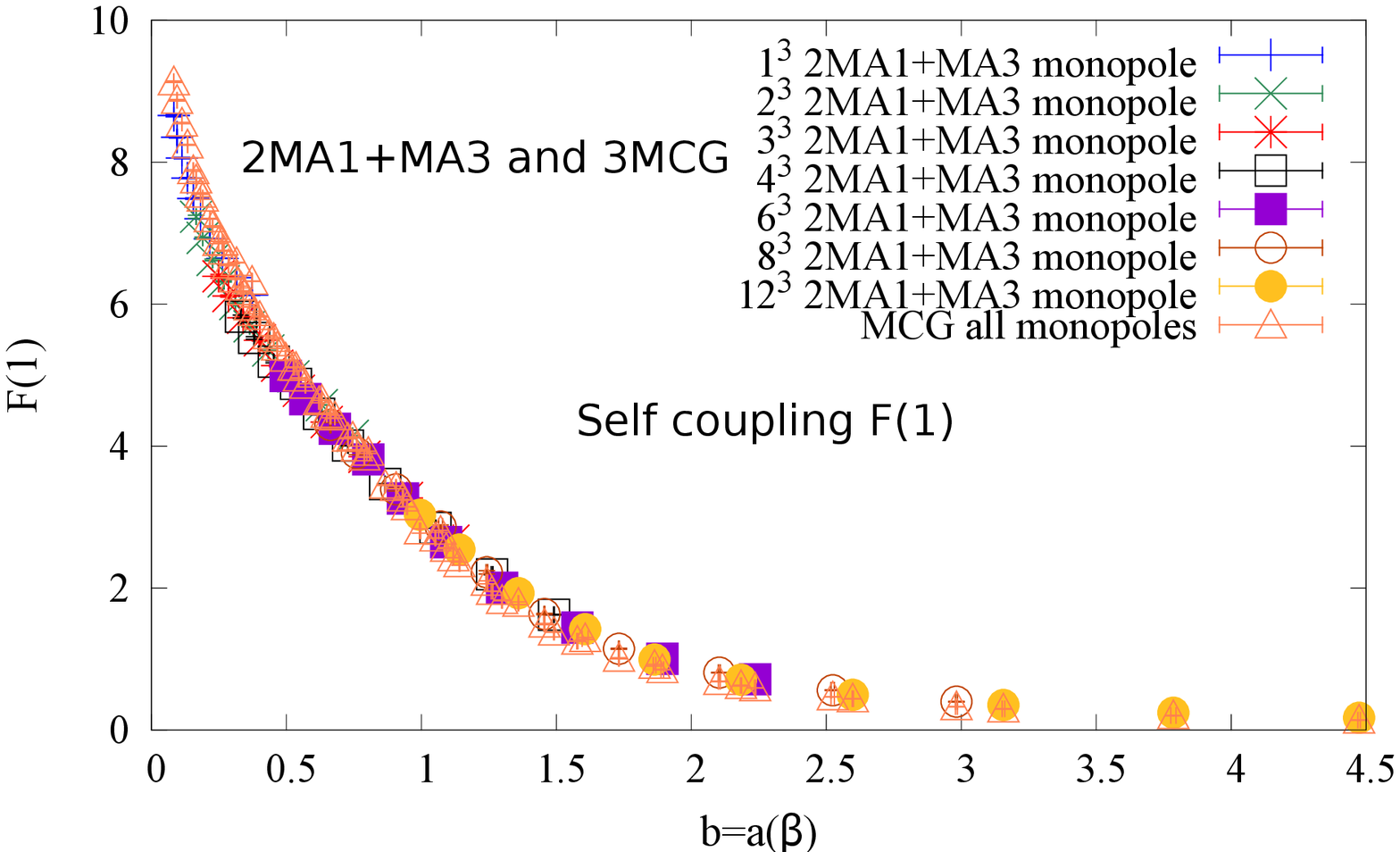}
  \end{minipage}
  \begin{minipage}[b]{0.9\linewidth}
    \centering
    \includegraphics[width=8cm,height=6.cm]{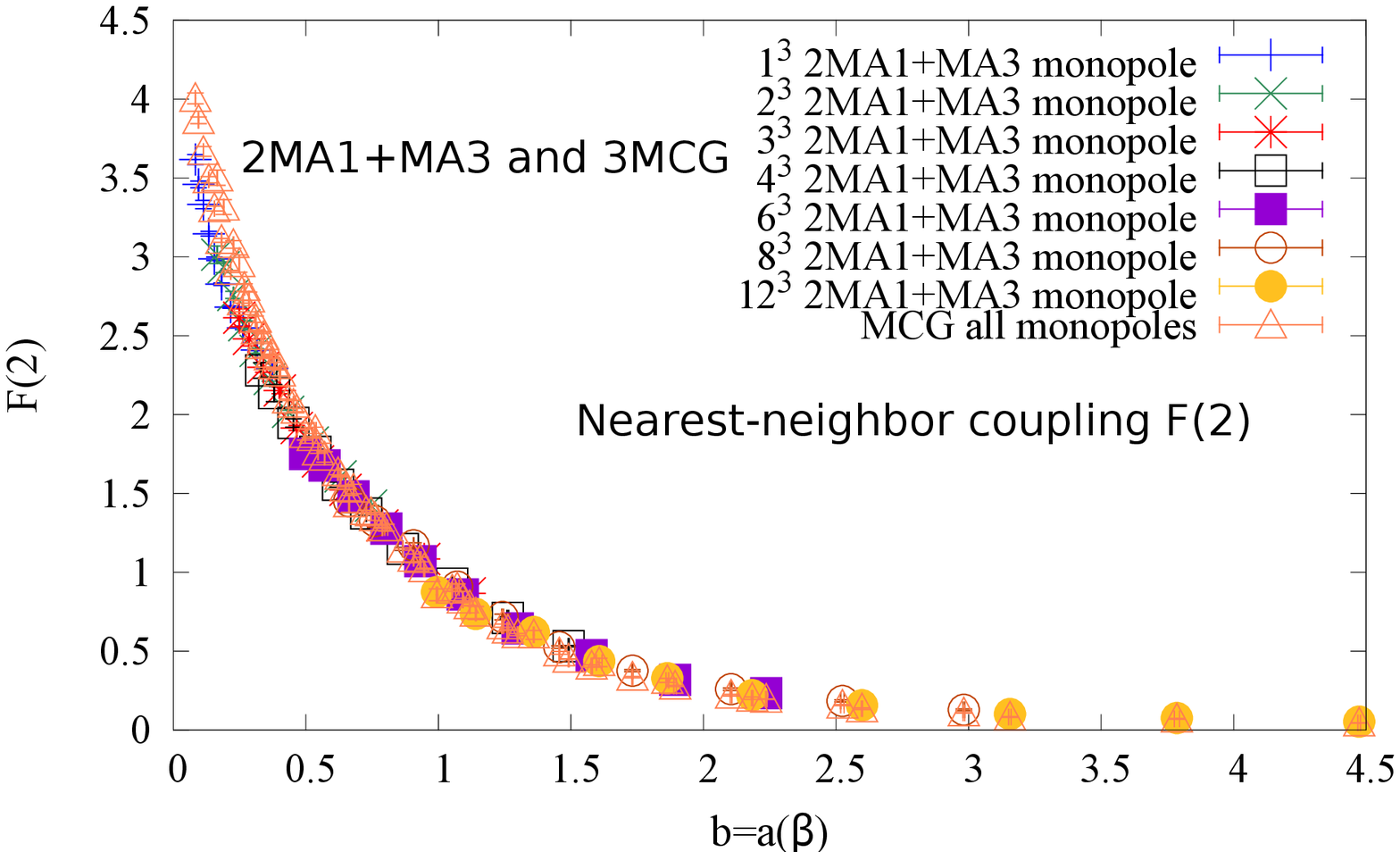}
  \end{minipage}
  \begin{minipage}[b]{0.9\linewidth}
    \centering
    \includegraphics[width=8cm,height=6.cm]{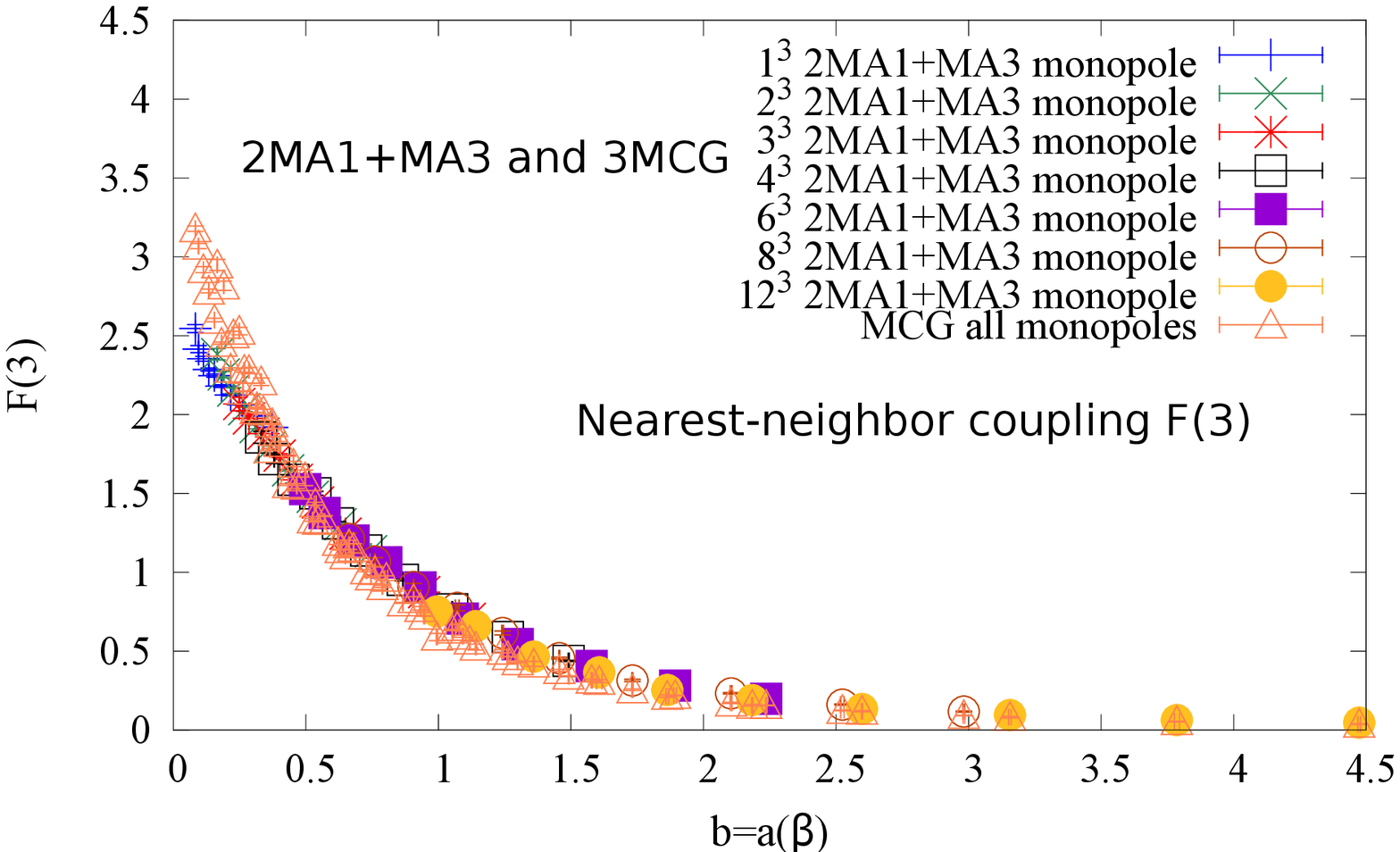}
  \end{minipage}
\end{figure}
\begin{figure}[htb]
\caption{The coupling constants of the two next to the nearest-neighbor interactions in the effective monopole action versus $b=na(\beta)$  in MAU1 and MCG on $48^4$. The sum of each coupling constants with respect to three color components are shown. }
\label{figMAU1_MCG_F45}
  \begin{minipage}[b]{0.9\linewidth}
    \centering
    \includegraphics[width=9cm,height=6.cm]{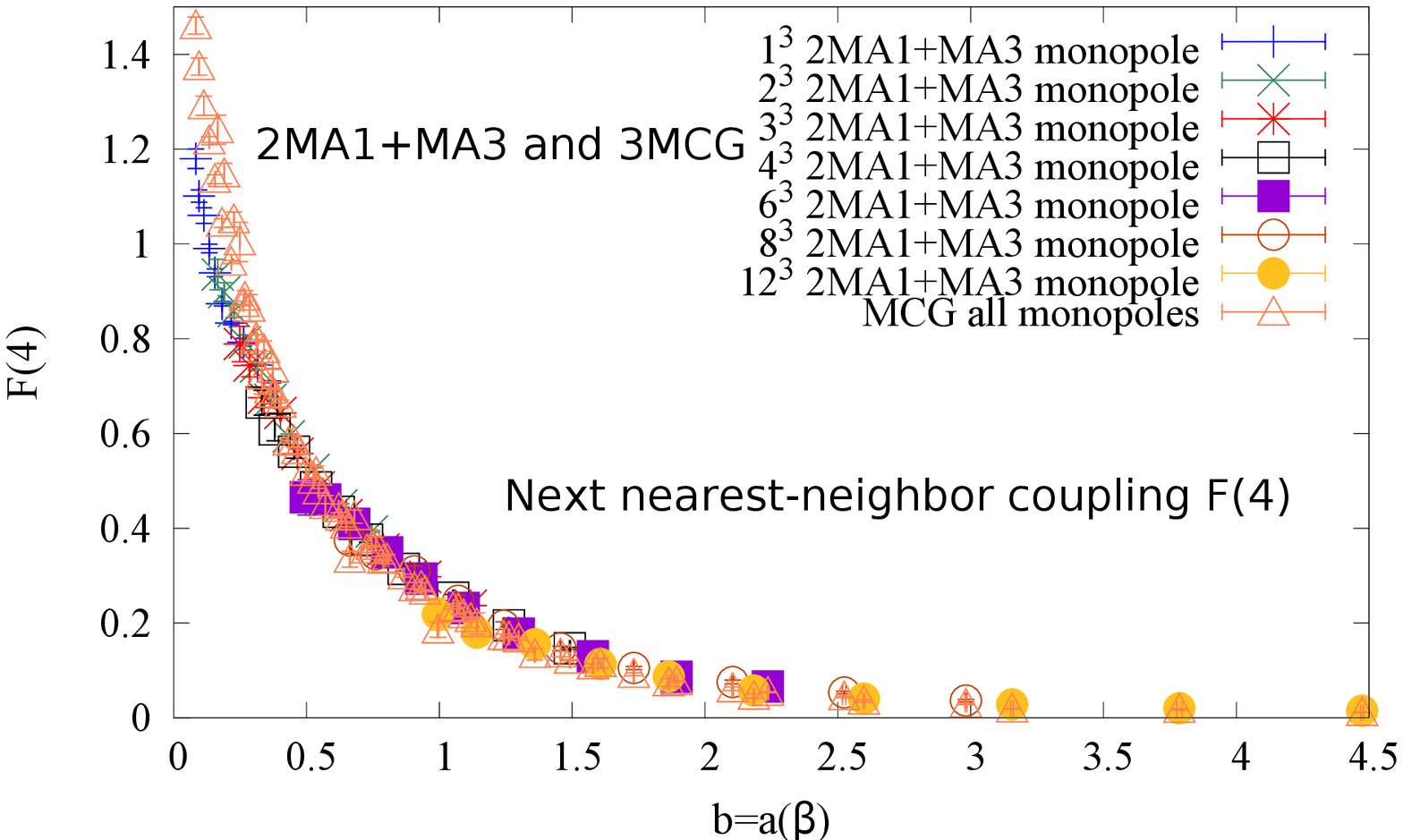}
  \end{minipage}
  \begin{minipage}[b]{0.9\linewidth}
    \centering
    \includegraphics[width=9cm,height=6.cm]{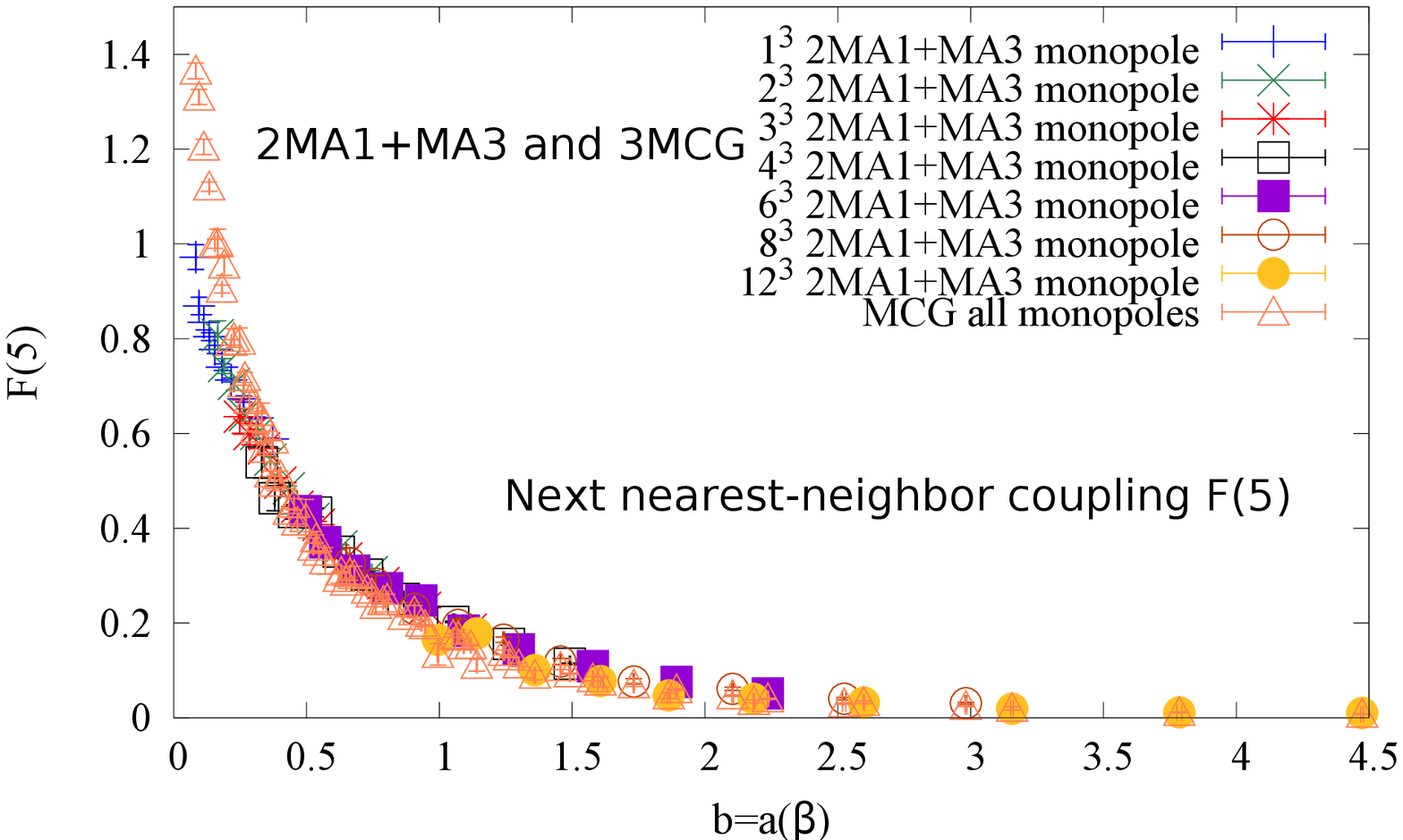}
  \end{minipage}
\end{figure}

\section{Numerical results}
As discussed in Appendix~\ref{APD:action} and Appendix~\ref{APD:comparison}, in the main part of this work, we adopt 10 short-ranged quadratic interactions alone as the form of the effective monopole action for simplicity and also for the comparison with the analytic blocking from the continuum limit. 

\subsection{Results in MCG gauge on $48^4$ lattice}
The $10$ coupling constants $F(i)\ (i=1\sim 10)$ of quadratic interactions are fixed very beautifully for lattice coupling constants $3.0\le\beta\le3.9$ and the steps of blocking $1\le n\le 12$. Remarkablly they are all expressed by a function of 
$b=na(\beta)$ alone, although they originally depend on two parameters $\beta$ and $n$. Namely the scaling is satisfied and the continuum limit is obtained when $n\to\infty$ for fixed $b=na(\beta)$. The obtained action can be considered as the projection of the perfect action onto the 10 quadratic coupling constant plane. These behaviors are shown for the first $5$ dominant couplings in Fig.\ref{figF123_MCG_48} and Fig.\ref{figF45_MCG_48}. These data are actually much more beautiful than those obtained in previous  works in MA gauge considering the third color component alone~\cite{Chernodub:2000ax}.
\subsection{Renormalization group flow diagrams}
The perfect monopole action draws a unique trajectory in the multi-dimensional coupling-constant space. To see if such a behavior is realized in our case, we plot the renormalization group flow line of our data projected onto some two-dimensional coupling-constant planes in Fig.\ref{fig_flowF1_234} and Fig.\ref{fig_flow_F15_F23}. Except the case for small $b=na(\beta)$ regions especially with $n=1$ case, the  unique trajectory is seen clearly. The behaviors are again much more beautiful than those obtained previously in MA gauge~\cite{Chernodub:2000ax}.

\subsection{Volume dependence in MCG gauge}
Volume dependence is checked in comparison with the data on $24^4$ and $48^4$ lattices in MCG gauge. Fig.~\ref{fig_volume} shows examples of the most dominant self-coupling coupling $F(1)$ and the coupling of the  next nearest-neighbor interaction $F(5)$. Volume dependence is seen to be small, although the error bars of the data on $24^4$ become naturally larger due to the boundary effect when the couplings at larger distances are considered. 

\subsection{Smooth gauge dependence}
The above results are all obtained in MCG gauge. Before studying other smooth gauges, we show the result without any gauge-fixing. In this case, the vacuum is contaminated by dirty artifatcs. Nevertheless, the infrared effective monopole action is determined.  Fig.\ref{F1_NGF_MCG} shows an example of the coupling of the self-interaction $F(1)$ in comparison with that in MCG gauge. One can see that scaling is not seen at all in NGF case.

\subsubsection{DLCG gauge}
The direct Lapacian center gauge(DLCG) is a gauge used to study the center vortex~\cite{Faber:2001zs} as MCG. Since DLCG gauge fixing needs more maschine time, we take data on smaller $24^4$ lattice only. The results are shown in comparison with those in MCG in Fig.\ref{fig_DLCG} with respect to the self-coupling $F(1)$ and the  next nearest-neighbor coupling $F(5)$ as an example. Both data are almost equal for the $b=na(\beta)$ regions considered, although small deviations are seen in the $F(5)$ case having the finite-size effects
on small $24^4$ lattice.

\subsubsection{AWL gauge}
The third smooth gauge is the maximally Abelian Wilson loop (AWL) gauge~\cite{Suzuki:1996ax,SIB201711}, where Abelian $1\times 1$ Wilson loop is maximized as much as possible. The data in AWL is shown in
Fig.\ref{fig_AWL} along with those in MCG with respect to the self-coupling $F(1)$ and the next next nearest-neighbor coupling $F(6)$ as an example.
The scaling is found very clearly and the both data are almost the same
even with respect to  $F(6)$ on $48^4$ lattice.

\subsubsection{MAU1 gauge}
Now let us compare MCG and MAU1 gauges, the latter of which is the combination of the maximally Abelian(MA) gauge-fixing~\cite{Kronfeld:1987vd} and Landau gauge fixing with respect to the remaining $U(1)$~\cite{Bali:1996dm}. In MAU1, the global isospin invariance is broken and the effective action $S(k^3)$ is different 
from those of the off-diagonal monopole currents $S(k^1)$ and $S(k^2)$.
See Fig.\ref{figMAU1_F123} as an example. With respect to $F(1)$ and $F(2)$, the isospin breaking is not so big, but large deviation is observed with respect to $F(3)$.

However, if the effective actions in both MAU1 and MCG are on the renormalized trajectory corresponding to the continuum limit, the total sum of the monopole actions in three color directions in MAU1 should be equivalent to the sum of three monopole actions in MCG gauge.   It is very interesting to see from Fig.\ref{figMAU1_MCG_F123} and Fig.\ref{figMAU1_MCG_F45} that
the expectation is realized. Actually except for small $b=na(\beta)$ regions, the gauge-invariance is seen clearly. 

\subsection{Summary of studies in smooth gauges}
From the above data in various gauges, one can conclude that if scaling behaviors are obtained and the effective monopole action is on the renormalized trajectory with the introduction of some smooth gauge fixing, the trajectory obtained becomes universal naturally.
In fact, the renormalized trajectory represents the effective action in the continuum limit and gauge dependence should not exist in the continuum. It is exciting to see that this natural expectation is realized actually at least for larger $b$ regions $b\ge 0.5 ~(\sigma_{phys}^{-1/2})$.

\begin{table*}[htbp]
\caption{Best parameters fitted}
\label{tabl:best_para}
\begin{center}
\begin{tabular}{|c|c|c|c|c|c|c|c|c|c|}
\hline
$b=na(\beta)$ & 0.5 & 1 & 1.5 & 2 & 2.5 & 3 & 3.5 & 4 & 4.5 \\
\hline
$\kappa$ & 0.117504 & 0.470017 & 1.057538 & 1.880067 & 2.937605 & 4.230151 & 5.757705 & 7.520268 & 9.51784 \\
\hline
$m_1$ & 9 & 18 & 27 & 36 & 45 & 54 & 63 & 72 & 81 \\
\hline
$m_2$ & 0.9 & 1.8 & 2.7 & 3.6 & 4.5 & 5.4 & 6.3 & 7.2 & 8.1 \\
\hline
$\bar{\alpha}$ & 8.682261 & 2.170565 & 0.964696 & 0.542641 & 0.34729 & 0.241174 & 0.177189 & 0.13566 & 0.107188 \\
\hline
$\bar{\beta}$ & 6.963001 & 6.963001 & 6.963001 & 6.963001 & 6.963001 & 6.963001 & 6.963001 & 6.963001 & 6.963001 \\
\hline
$\bar{\gamma}$ & 1.06e-01 & 6.63e-03 & 1.31e-03 & 4.15e-04 & 1.70e-04 & 8.19e-05 & 4.42e-05 & 2.59e-05 & 1.62e-05 \\
\hline
\end{tabular}
\end{center}
\end{table*}
\begin{figure}[htb]
\caption{Comparison of the coupling constants of the self and the nearest-neighbor interactions in the effective monopole action between numerical MCG data and theoretical values derived from the almost perfect action. }
\label{figF123MCG_perfect}
  \begin{minipage}[b]{0.9\linewidth}
    \centering
    \includegraphics[width=8cm,height=6.cm]{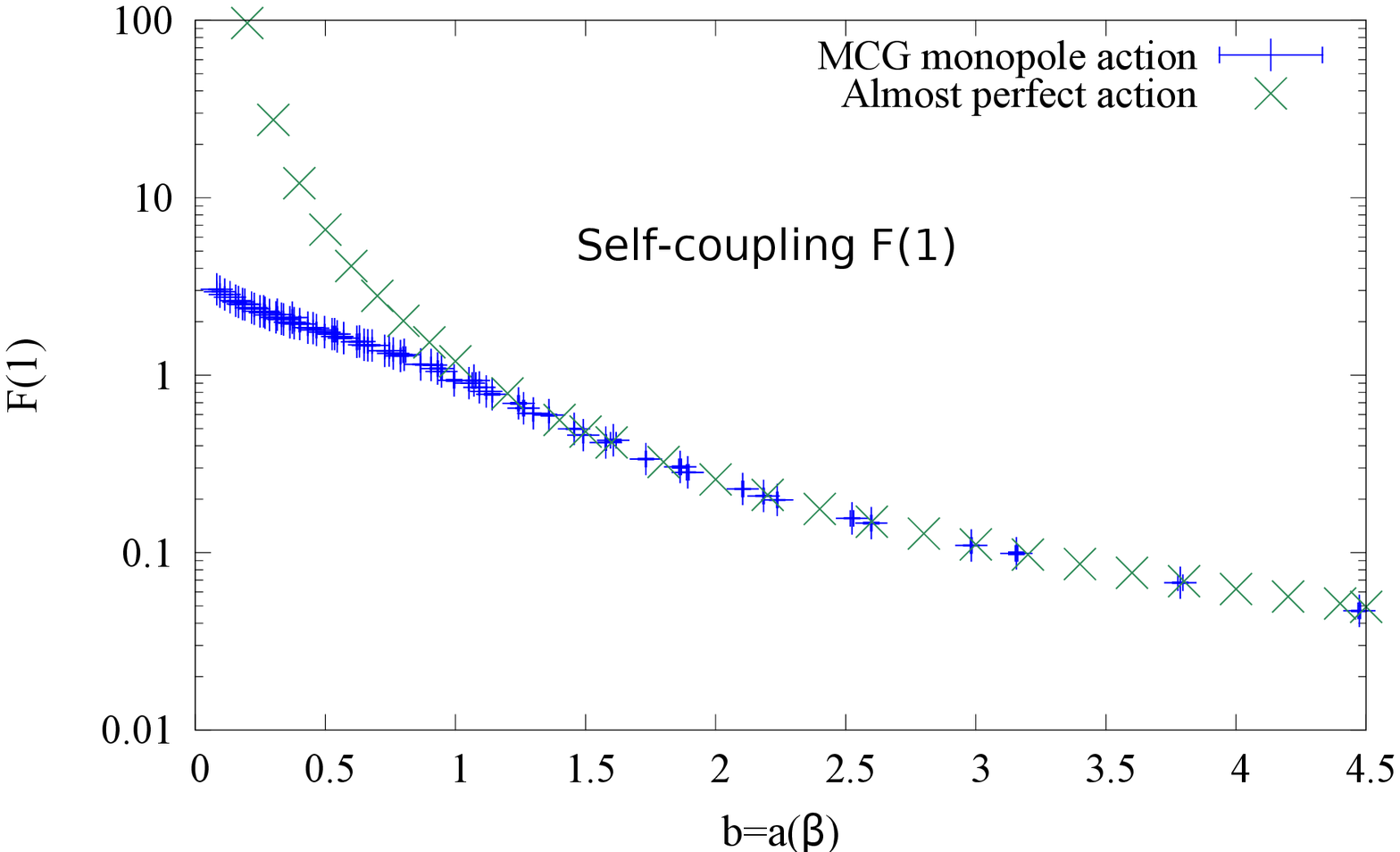}
  \end{minipage}
  \begin{minipage}[b]{0.9\linewidth}
    \centering
    \includegraphics[width=8cm,height=6.cm]{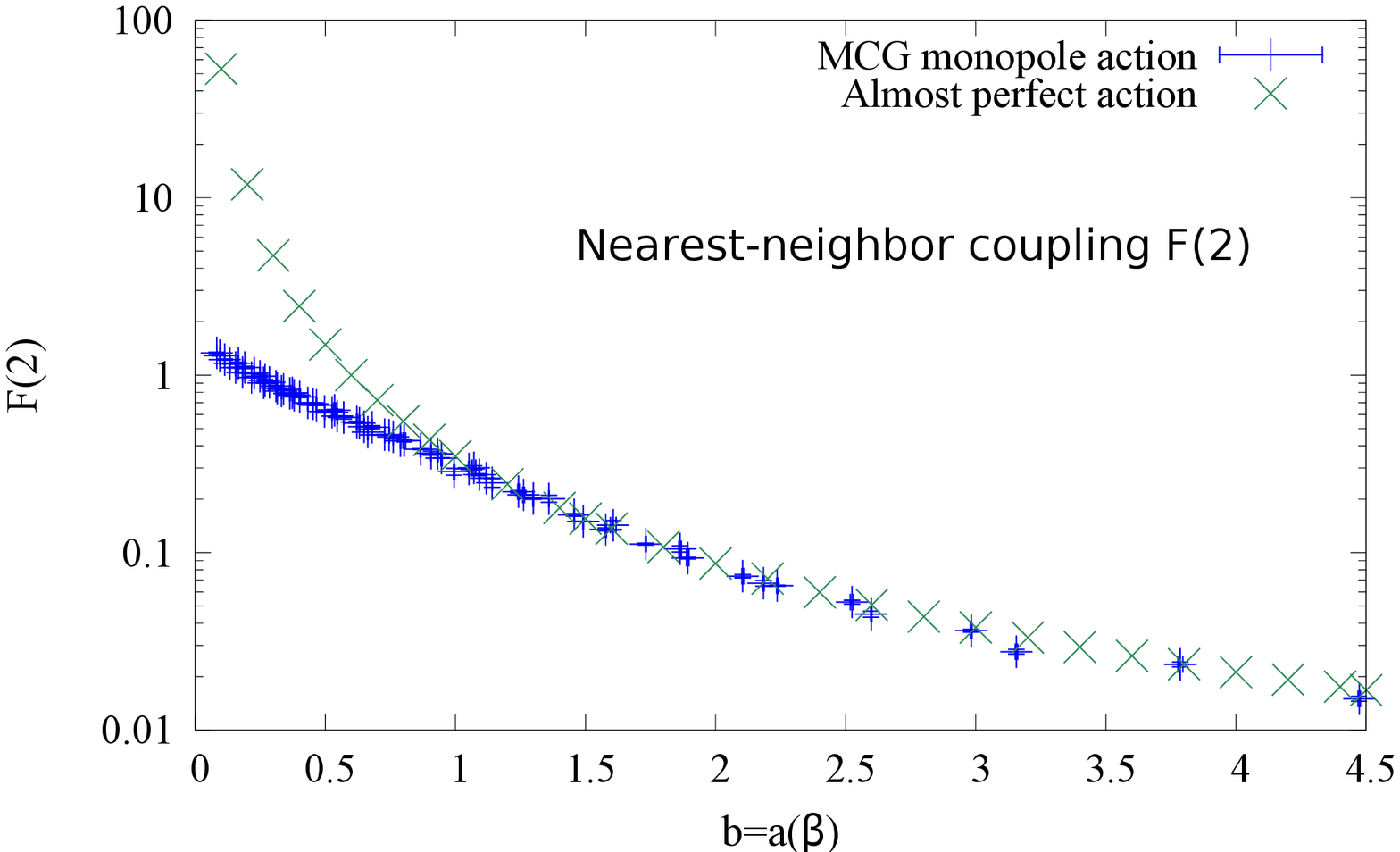}
  \end{minipage}
  \begin{minipage}[b]{0.9\linewidth}
    \centering
    \includegraphics[width=8cm,height=6.cm]{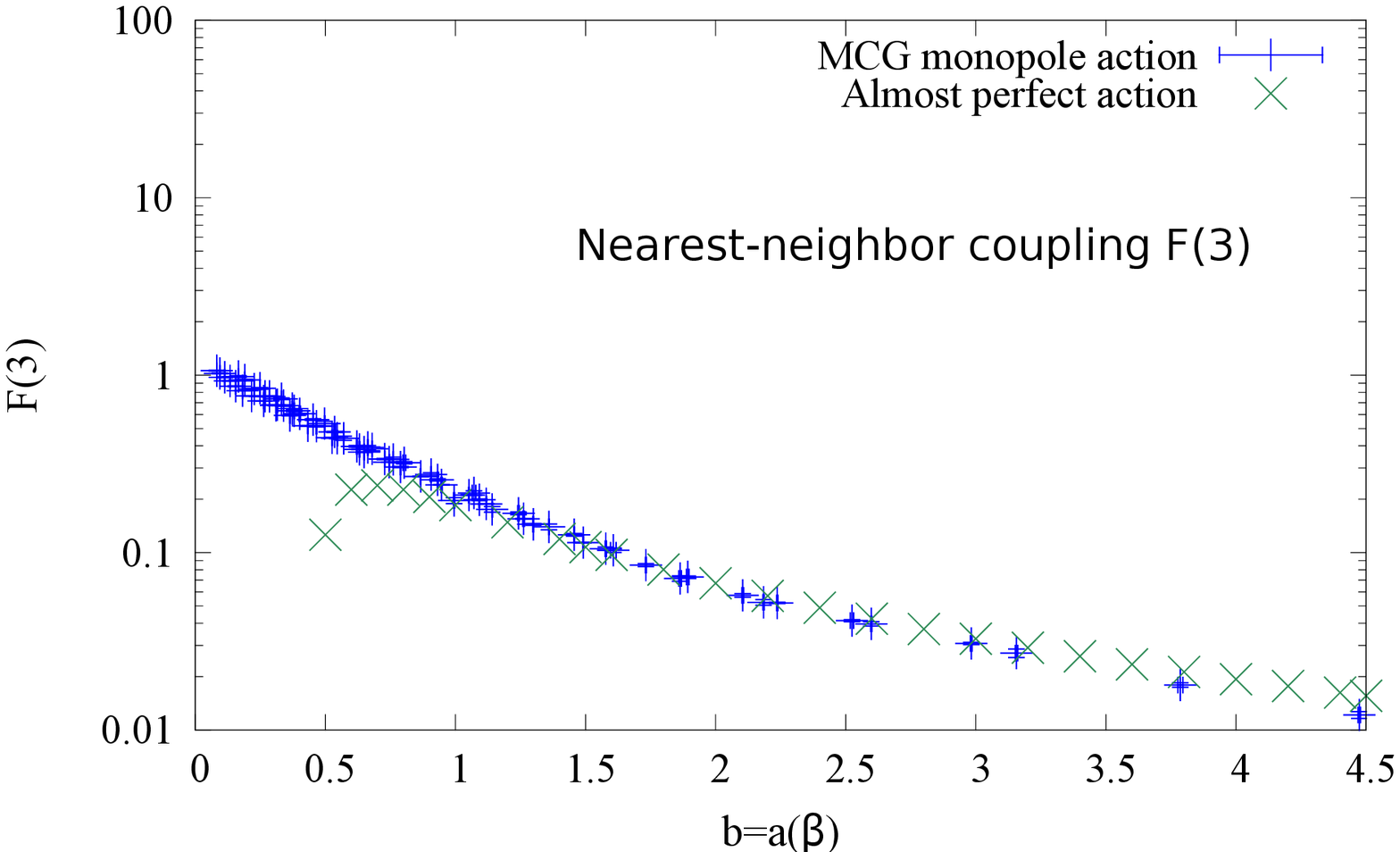}
  \end{minipage}
\end{figure}
\begin{figure}[htb]
\caption{Comparison of the coupling constants of the next nearest-neighbor interactions in the effective monopole action between numerical MCG data and theoretical values derived from the almost perfect action.}
\label{figF45MCG_perfect}
  \begin{minipage}[b]{0.9\linewidth}
    \centering
    \includegraphics[width=8cm,height=6.cm]{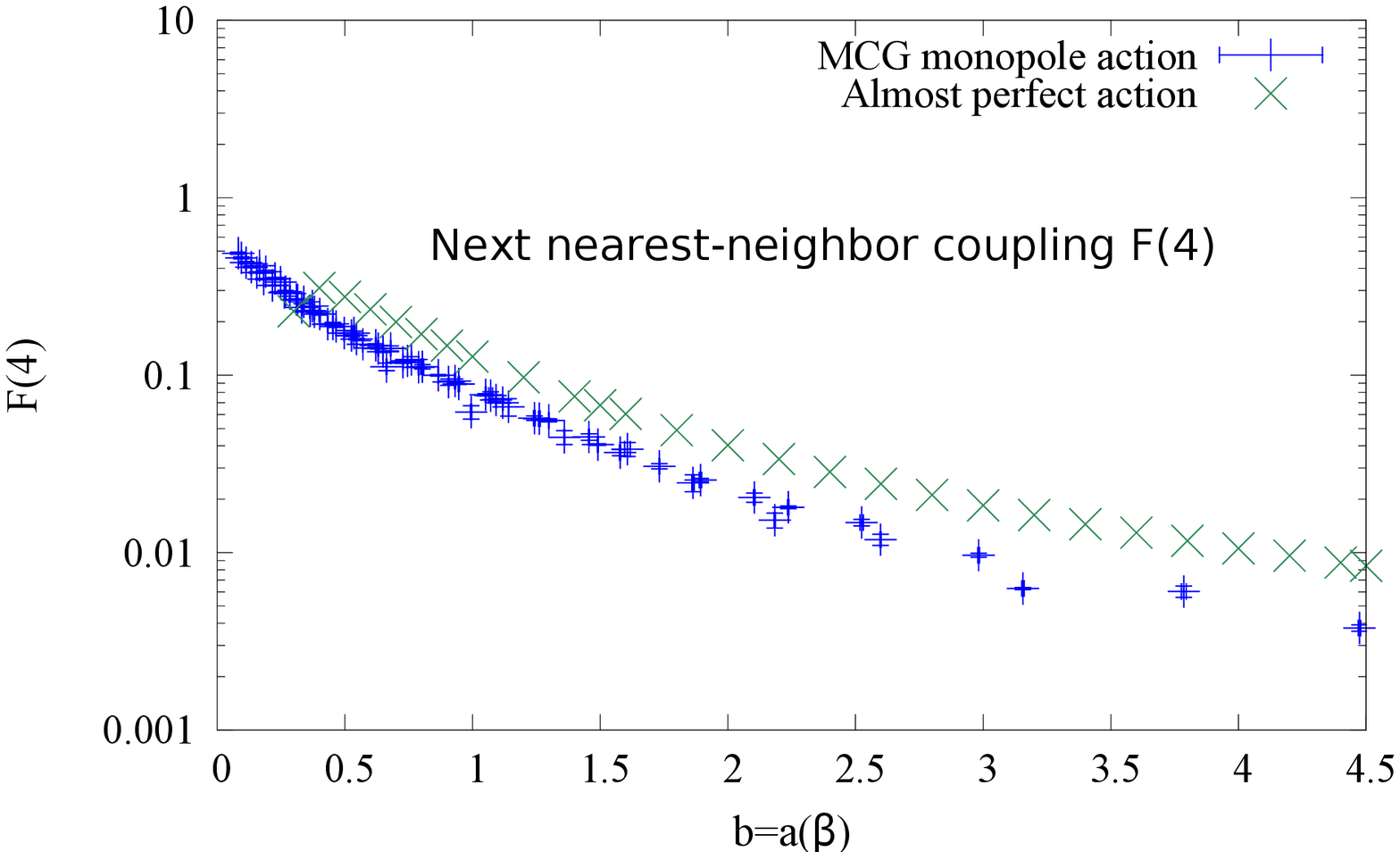}
  \end{minipage}
  \begin{minipage}[b]{0.9\linewidth}
    \centering
    \includegraphics[width=8cm,height=6.cm]{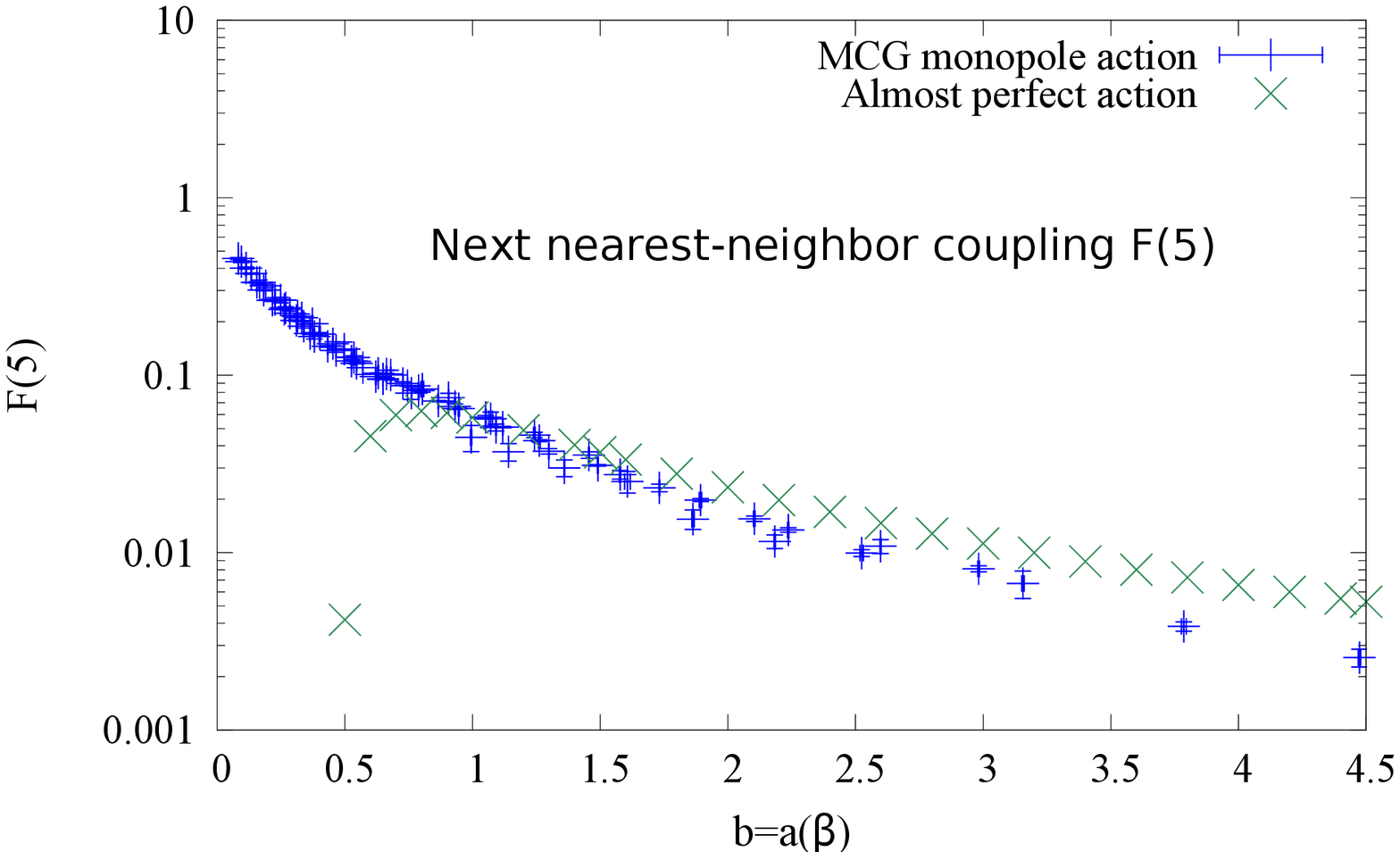}
  \end{minipage}
\end{figure}

\section{Blocking from the continuum limit}
The infrared effective action determined above numerically shows a clear scaling, that is, a function of $b=na(\beta)$ alone and it can be regarded as an action in the continuum limit. But it is an action still formulated on a lattice with the finite lattice spacing $b=na(\beta)$. Hence
various symmetries such as rotational invariance of physical quantities in the continuum limit is difficult to observe, since the action itself does not satisfy, say, the rotational invariance.  
One has to consider a perfect operator in addition to a perfect action on $b$ lattice in order to reproduce a symmetry such as rotational invariance in the continuum limit~\cite{Fujimoto:1999ci,Chernodub:2000ax}. For example, a simple Wilson loop on a plane does not reproduce the rotational-invariant static potential on the $b$ lattice.

It is highly desirable to get a perfect action formulated in the continuum space-time which reproduce the same physics at the scale $b$ as those obtained by the above perfect action formulated on the $b$ lattice.
If such a perfect action in the continuum space-time is given, the rotational invariance of physical quantities is naturally 
reproduced with simple operators such as a simple Wilson loop, since the action also respects the invariance.

If the infrared effective monopole action is quadratic, it is possible to perform analytically the blocking from the continuum  and to get the infrared monopole action formulated on a coarse $b=na(\beta)$ lattice~\cite{Fujimoto:1999ci,Chernodub:2000ax}. Perfect operators are also 
obtained.
This is similar to the method developed by Bietenholz and Wiese~\cite{ref:BFC}.

We review the above old works~\cite{Fujimoto:1999ci,Chernodub:2000ax}
shortly. Let us start from the following action composed of quadratic 
interactions between magnetic monopole currents.  
It is formulated on an infinite lattice 
with very small lattice spacing $a$:
\begin{eqnarray}
S[k]=\sum_{s,s',\mu}k_\mu(s)D_0(s-s')k_\mu(s').\label{eqn.S}
\end{eqnarray}
Here we omit the color index.
Since we are starting from the region very near to the continuum limit,
it is natural to assume the direction independence of $D_0(s-s')$.
Also we adopt only parallel interactions, since we can avoid 
perpendicular interactions from  short-distant terms
using the current conservation.
Moreover, for simplicity, we adopt only the first three Laurent
expansions, i.e., Coulomb, self and nearest-neighbor 
interactions. Explicitly, $D_0(s-s')$ is expressed as 
$\bar{\alpha}\delta_{s,s'}+\bar{\beta}\Delta_L^{-1}(s-s')+\bar{\gamma}\Delta_L(s-s')$ where $\bar{\alpha},\ \bar{\beta}$ and $\bar{\gamma}$ are free parameters.
Here $\Delta_L(s-s')=-\sum_\mu\partial_\mu\partial'_\mu\delta_{s,s'}$.  Including more complicated quadratic interactions is not difficult.

When we define an operator on the fine $a$ lattice, 
we can find a perfect operator along the projected flow 
in the $a\to 0$ limit for fixed $b$.
We assume the perfect operator on the projected space 
as an approximation of the correct
operator for the action ${\cal S}[k]$ on the coarse $b$ lattice.

Let us start from
\begin{eqnarray}
 \langle W_m({\cal C}) \rangle
&=&  \sum_{k_{\mu}(s)=-\infty
         \atop{\partial^{\prime}_{\mu}k_{\mu}(s)=0}}^{\infty}
  \exp\{-\sum_{s,s',\mu} k_{\mu}(s)D_0(s-s')k_{\mu}(s')\nn \\
 &&      +2\pi i\sum_{s,\mu} N_{\mu}(s)k_{\mu}(s)\} \nonumber \\
&\times&  \prod_{s^{(n)},\mu}\delta\bigg(K_{\mu}(s^{(n)})-{\cal B}_{k_\mu}(s^{(n)})\bigg)
      /{\cal Z}[k],
\label{pfac:3}
\end{eqnarray}
where ${\cal B}_{k_\mu}(s^{(n)})\equiv \sum_{i,j,l=0}^{n-1}k_\mu(s(n,i,j,l)) ~($\ref{pfac:2}). Note that the monopole contribution to the static potential is given by the term in Eq.(\ref{pfac:3})
\begin{eqnarray}
W_m({\cal C})&=&
\exp\bigg( 2\pi i\sum_{s,\mu}N_{\mu}(s)k_{\mu}(s) \bigg),
\nn\\
N_{\mu}(s)&=&\sum_{s'}\Delta_L^{-1}(s-s')\frac{1}{2}
\epsilon_{\mu\alpha\beta\gamma}\partial_{\alpha}
S^J_{\beta\gamma}(s'+\hat{\mu}), \label{eq:pla}
\end{eqnarray}
where $S^J_{\beta\gamma}(s'+\hat{\mu})$ is a plaquette variable satisfying 
$\partial'_{\beta}S^J_{\beta\gamma}(s)=J_{\gamma}(s)$ and the coordinate 
displacement $\hat{\mu}$ is due to the interaction between dual
variables. Here $J_{\mu}(s)$ is an Abelian integer-charged electric current corresponding to an Abelian Wilson loop. See Ref.~\cite{Chernodub:2000ax}.

The cutoff effect of the operator (\ref{pfac:3}) is $O(a)$ by definition.
This $\delta$-function renormalization group transformation can be
done analytically. Taking the continuum limit $a\to 0$, $n\to \infty$
(with $b=na$ is fixed) finally, we obtain the expectation
value of the operator on the coarse lattice with spacing $b=na(\beta)$
~\cite{Fujimoto:1999ci}: 
\begin{eqnarray}
\langle W_m({\cal C}) \rangle
&=&
  \exp\Biggl\{
    - \pi^2 \int_{-\infty}^{\infty}\!\! d^4xd^4y
    \sum_{\mu}N_{\mu}(x)\nn\\
&\times&  D_0^{-1}(x-y)N_{\mu}(y)
    + \pi^2 b^8\!\!\!\! \sum_{s^{(n)},s^{(n)'}\atop{\mu,\nu}}
    \!\!\!\!
    B_{\mu}(bs^{(n)})\nn\\
&\times& D_{\mu\nu}(bs^{(n)}-bs^{(n)'})
    B_{\nu}(bs^{(n)'})
  \Biggr\}
\nonumber \\
&&
\times \!\!\!\!
\sum_{b^3K_\mu(bs)=-\infty\atop\partial'_\mu K_\mu=0}^{\infty}
\!\!\!\!\!\!
  \exp\Biggl\{
    - S[K_{\mu}(s^{(n)})]
\nonumber \\
&&
    +2 \pi i b^8\!\!\!\! \sum_{s^{(n)},s^{(n)'}\atop{\mu,\nu}}
    \!\!\!\!
     B_{\mu}(bs^{(n)})
       D_{\mu\nu}(bs^{(n)}-bs^{(n)'})\nn\\
&\times&       
     K_{\nu}(bs^{(n)'})
  \Biggr\}
  \Bigg/
  \!\!\!\!
  \sum_{b^3K_\mu(bs)=-\infty\atop\partial'_\mu K_\mu=0}^{\infty}
  \!\!\!\!\!\! Z[K,0],
\label{opwil:1}
\end{eqnarray}
where
\begin{eqnarray}
B_\mu(bs^{(n)}) &\equiv&
\lim_{a\to 0 \atop{n\to\infty}}
  a^8\sum_{s,s',\nu}
    \Pi_{{\neg}\mu}(bs^{(n)}-as)\nn\\
&\times& \left\{
  \delta_{\mu\nu}
  -\frac{\partial_{\mu}\partial'_{\nu}}{\sum_{\rho}\partial_{\rho}
  \partial'_{\rho}}
\right\}\nn\\
&\times&D_0^{-1}(as-as')
N_{\nu}(as'),
\label{opwil:9}
\\
\Pi_{\neg\mu}(bs^{n}-as)&\equiv&
\frac{1}{n^3}
  \delta\left( nas_\mu^{(n)}+(n-1)a-as_\mu \right)\nonumber \\
  &\times&
  \prod_{i(\ne \mu)}\left(
    \sum_{I=0}^{n-1}\delta\left( nas_i^{(n)}+Ia-as_i \right)
  \right).\nn
\end{eqnarray}
$S[K_{\mu}(s^{(n)})]$ denotes the effective action defined on the coarse
lattice:
\begin{eqnarray}
  S[K_{\mu}(s^{(n)})] &=&
  b^8 \sum_{s^{(n)},s^{(n)'}}\sum_{\mu,\nu}K_{\mu}(bs^{(n)})\nn\\
 &\times& D_{\mu\nu}(bs^{(n)}-bs^{(n)'})K_{\nu}(bs^{(n)'}).
\label{pfac:5}
\end{eqnarray}
Since we take the continuum limit analytically, the operator (\ref{opwil:1})
does not have no cutoff effect. For clarity, we have recovered the scale factor $a$ and $b$ in (\ref{opwil:1}), (\ref{opwil:9}) and (\ref{pfac:5}).

The momentum representation of $D_{\mu\nu}(bs^{(n)}-bs^{(n)'})$
takes the form
\begin{eqnarray}
D_{\mu\nu}(p)=
  A_{\mu\nu}^{GF-1}(p) 
  - \frac{1}{\lambda}
      \frac{\hat{p_\mu}\hat{p_\nu}}{(\hat{p}^2)^2}e^{i(p_\mu-p_\nu)/2},
\label{fit:2}
\end{eqnarray}
where $\hat{p_\mu}=2\sin(p_{\mu}/2)$ and  $A_{\mu\nu}^{GF^{-1}}(p)$ is the gauge-fixed 
inverse of the following operator
\begin{eqnarray}
A'_{\mu\nu}(p)\!&\equiv&\!
\left(\prod_{i=1}^4\sum_{l_i=-\infty}^{\infty} \right)
\!\!
\Biggl\{
  \!D_0^{-1}(p+2\pi l)
  \Biggl[
    \delta_{\mu\nu}\!\nn\\
&& -\frac{(p+2\pi l)_\mu(p+2\pi l)_\nu}{\sum_i(p
+2\pi l)_i^2}
  \Biggr]\nn\\
&\times&  
  \frac{(p+2\pi l)_\mu(p+2\pi l)_\nu}{\prod_i(p+2\pi l)_i^2}
\Biggr\}\frac{\left(\prod_{i=1}^4\hat{p}_i \right)^2}
{\hat{p}_\mu\hat{p}_\nu}.
\label{fit:11}
\end{eqnarray}
The explicit form of $D_{\mu\nu}(p)$ is written in Ref.~\cite{Fujimoto:1999ci}.
Performing the BKT transformation explained in Appendix B of Ref.~\cite{Chernodub:2000ax} on the coarse
lattice, we can get the loop operator for the static potential 
in the framework of the string model:
\begin{eqnarray}
\langle W_m({\cal C}) \rangle
&=& \langle W_m({\cal C}) \rangle_{cl} \nn\\
&\times& \frac{1}{Z}
  \!\!\!\!
  \sum_{\sigma_{\mu\nu}(s)=-\infty
        \atop{\partial_{[\alpha}\sigma_{\mu\nu]}(s)=0}}^{\infty}
  \!\!\!\!
  \exp
  \Bigg\{
    -\pi^2\sum_{ s,s'\atop{ \mu\neq\alpha\atop{ \nu\neq\beta } } }
    \sigma_{\mu\alpha}(s)\partial_{\alpha}\partial_{\beta}'\nn\\
 &\times&  D_{\mu\nu}^{-1}(s-s_1)\Delta_L^{-2}(s_1-s')
 \sigma_{\nu\beta}(s')\nonumber \\
&-2\pi^2&\sum_{s,s'\atop{\mu,\nu}}\sigma_{\mu\nu}(s)\partial_{\mu}
    \Delta_L^{-1}(s-s')B_{\nu}(s') 
  \Bigg\},
\label{opwil:4}
\end{eqnarray}
where $\sigma_{\nu\mu}(s) \equiv \partial_{[ \mu}s_{\nu ]}$ is the
closed string variable satisfying the conservation rule
\begin{eqnarray}
\partial_{[\alpha}\sigma_{\mu\nu]}=
  \partial_{\alpha}\sigma_{\mu\nu}+\partial_{\mu}\sigma_{\nu\alpha}
  +\partial_{\nu}\sigma_{\alpha\mu}=0.\label{conv}
\end{eqnarray}
 The classical part $\langle W_m({\cal C}) \rangle_{cl}$ is defined by
\begin{eqnarray}
\langle W_m({\cal C}) \rangle_{cl}
&=&\exp\Bigg\{
    -\pi^2 \int_{-\infty}^{\infty}\!\!\!\!d^4xd^4y
    \sum_{\mu}N_{\mu}(x)\nonumber\\
 &\times& D_0^{-1}(x-y)N_{\mu}(y)\Bigg\}.
\label{opwil:5}
\end{eqnarray}
\begin{figure}[htb]
\caption{Strong-coupling calculations of the Wilson loops}
\label{fig:strong}
    \centering
    \includegraphics[width=7cm,height=2.cm]{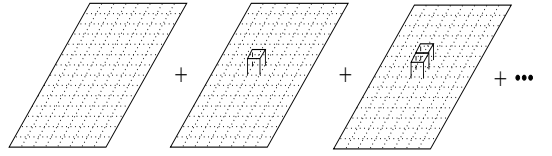}
\end{figure}
\begin{figure}[htb]
\caption{Comparison of monopole density from MCG numrical data and that from the perfect action}
\label{dens_MCG_perfect}
    \centering
    \includegraphics[width=8cm,height=7cm]{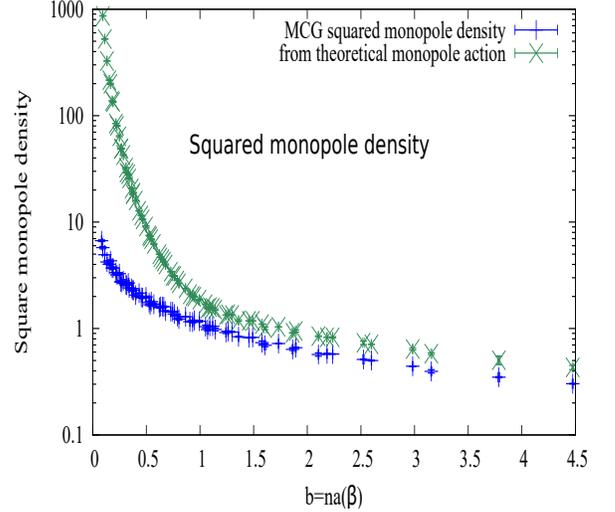}
\end{figure}

\section{Analytic evaluation of non-perturbative quantities}
\subsection{Parameter fitting}
To derive non-perturbative physical quantites analytically, we have to fix first the propagator $D_0(s)$ in (\ref{pfac:3}) of the continuum limit. It can be done by comparing $D_{\mu\nu}^{-1}(s-s')$ in Eq.(\ref{pfac:5}) with 
the set of coupling constants $F(i)\ \ (i=1\sim 10)$ of the monopole action determined numerically
in Eq.(\ref{eq:monoact}).
 
$D_0(s-s')$ in the monopole action (\ref{pfac:3}) is assumed to be 
$\bar{\alpha}\delta_{s,s'}+\bar{\beta}\Delta_L^{-1}(s-s')+\bar{\gamma}\Delta_L(s-s')$.
We can consider more general quadratic interactions, but as we see later,
this choice is almost sufficient to derive the IR region of SU(2) gluodynamics.

The inverse operator of $D_0(p)=\bar{\alpha}+\bar{\beta}/p^2+
\bar{\gamma} p^2$ 
takes the form
\begin{eqnarray}
D_0^{-1}(p)=
\kappa
\left(
  \frac{m_1^2}{p^2+m_1^2} - \frac{m_2^2}{p^2+m_2^2}
\right),
\label{fit:1}
\end{eqnarray}
where the new parameters $\kappa$, $m_1$ and $m_2$ satisfy
$\kappa (m_1^2-m_2^2)=\bar{\gamma}^{-1}, m_1^2+m_2^2=\bar{\alpha}/\bar{\gamma}, m_1^2m_2^2=\bar{\beta}/\bar{\gamma}$.

Substituting Eq.(\ref{fit:1}) into Eq.(\ref{fit:11}) and
performing the First Fourier transform(FFT) on a momentum lattice for the 
several input values $\kappa$, $m_1$ and $m_2$ we calculate $D_{\mu\nu}(p)$\footnote{In most calculations, we have adopted  $10^4$ momentum lattice with a cutoff $lmax=20$ with respect to the sum over $l$ in
(\ref{fit:11}). To check the reliablitiy of the cutoff parameters, a case with $lmax=40$ on $16^4$ momentum lattice has been done, but the difference is found to be less than $10\%$ with respect to all coupling constants,
although the computer time costs more than 100 times more.}.

To be noted, the three parameters as a function of $b=na(\beta)$ can not be uniquely determined. 
As shown later, $m_1$ ($m_2$) corresponds to the inverse of the coherence (penetration) length. Moreover $m_2/b$ is found to correpond to the mass of the lowest scalar glueball. Hence we assume
\begin{itemize}
  \item $m_1/m_2=10$ for all $b=na(\beta)$ regions.
  \item $m_2/b\sim 1.8$ correponding to $M_{0^{++}}\sim 3.7\sqrt{\sigma_{phys}}$.
  \item The string tension calculated analytically is as near as possible to the physical string tension $\sigma_{phys}$ and shows scaling, namely $\sigma/\sigma_{phys}$ is constant for all $b=na(\beta)$ regions considered.  
\end{itemize}
Table \ref{tabl:best_para} shows the results of the best fit.

\subsection{Comparison of the couplings from numerical analyses and theoretical calculations}
Now let us show the coupling constants determined by the analytical blocking method using the above best-fit parameters in Fig.\ref{figF123MCG_perfect} and Fig.\ref{figF45MCG_perfect}. As seen from these figures, the fit is nice for $b=na(\beta)\ge 1.0$, although the deviation becomes larger at smaller $b$ regions, especially for the couplings at larger distace. Note that the logscale is adopted in the $y$ axis.

\subsection{The string tension (1)}
Let us evaluate the string tension using the perfect operator 
(\ref{opwil:4})~\cite{Fujimoto:1999ci}.
The plaquette variable $S^J_{\alpha\beta}$ in Eq.(\ref{eq:pla}) for the 
static potential $V(Ib,0,0)$ is expressed by
\begin{eqnarray}
S^J_{\alpha\beta}(z)
&=&
  \delta_{\alpha 1}\delta_{\beta 4}\delta(z_{2})\delta(z_{3})
  \theta(z_{1})\nn\\
  &\times&\theta(Ib-z_{1})
  \theta(z_{4})\theta(Tb-z_{4})\label{a_loop_op}.
\end{eqnarray}
We have seen that the monopole action on 
the dual lattice 
is in the weak coupling region for large $b$.
Then the string
model on the original lattice is in the strong coupling region. 
Therefore, we evaluate Eq.(\ref{opwil:4}) by the 
strong coupling expansion.
The method can be shown diagrammatically in Figure~\ref{fig:strong}.

As explicitly evaluated in Ref.~\cite{Fujimoto:1999ci}, 
the dominant classical part of the string tension coming 
from Eq.~(\ref{opwil:5}) is
\begin{eqnarray}
\sigma_{cl}=\frac{\pi\kappa}{2b^2} \ln\frac{m_1}{m_2}.
\label{sigma_cl}
\end{eqnarray} 
This is consistent with the analytical results~\cite{suzu89} in Type-2
superconductor.
The two constants $m_1$ and $m_2$ may be regarded as the coherence and
the penetration lengths.

The ratio $\sqrt{\sigma_{cl}/\sigma_{phys}}$ using the optimal values $\kappa$, 
$m_1$ and $m_2$ given in Table~\ref{tabl:best_para} becomes a bit higher, namely about 1.3 for all $b$ regions considered. As shown previously~\cite{Fujimoto:1999ci}, quantum fluctuations are so small to recover the difference. 
This is due mainly to that the assumption of 10 quadratic monopole couplings alone is too simple.

Note that the rotational invariance of the static potential is maintained by the calculation using the classical part as naturally expected from the perfect action. For example, the variable $S_{\alpha\beta}$ for the static potential
$V(Ib,Ib,0)$ is given by
\begin{eqnarray}
S_{\alpha\beta}(z)
&=&
  \Bigl(
    \delta_{\alpha 1}\delta_{\beta 4}+\delta_{\alpha 2}\delta_{\beta 4}
  \Bigr)
  \delta(z_{3})\theta(z_{4})\theta(Tb-z_{4})
\nonumber \\
&&
\times
  \theta(z_{1})\theta(Ib-z_{1})
  \theta(z_{2})\theta(Ib-z_{2})
  \delta(z_{1}-z_{2})\nn.
\end{eqnarray}
The static potential $V(Ib,Ib,0)$ can be written as
\begin{eqnarray}
V(Ib,Ib,0) &=& \frac{\sqrt{2}\pi\kappa Ib}{2} \ln\frac{m_1}{m_2}. 
\end{eqnarray}
The potentials from the classical part take only the linear form and the 
rotational invariance is recovered completely even for the 
nearest $I=1$ sites. 

\begin{table}[htbp]
 \caption{$\sqrt{\sigma_{cl}/\sigma_{phys}}$ evaluated from the effective action on the $b$ lattice at three typical $b$ values. Errorbars of at most a few \% order exist but  are not shown explicitly.}\label{sigma_L_D}
 \begin{center}
  \begin{tabular}{|c|c|c|c|}
    \hline
    $b$   &$\beta$  & $n$ &$\sqrt{\sigma_{cl}/\sigma_{phys}}$  \\
    \hline
   1.4912  &3.0&4&     1.25 \\
    \hline
  2.9824  &3.0&8&      1.25 \\
    \hline
  4.4736  &3.0&12&      1.31 \\
    \hline
  \end{tabular}
 \end{center}
\end{table}

\subsection{The string tension (2)}
In the above calculation of the string tension, we have started from the source term corresponding to the loop operator (\ref{a_loop_op}) for the static potential of the fine $a$ lattice and have constructed the operator on the coarse $b$ lattice by making the block-spin transformation.
But as shown in Ref.~\cite{Chernodub:2000ax}, the same string tension for the flat on-axis Wilson loop can be obtained for $I,T\to\infty$ when we consider a naive Wilson loop operator on the coarse $b$ lattice. In this method, we can evaluate the string tension directly by the numerical data of the  coupling constants of the the effective monopole action.

Consider the source term on the $1-4$ plane of the coarse $b$ lattice:
\begin{eqnarray}
\bar{S}_{14}(s)=\delta(s_2)\delta(s_3)\theta(s_1)\theta(I-s_1)\theta(s_4)\theta(T-s_4).
\label{b_loop_op}
\end{eqnarray}
Define
\begin{eqnarray*}
\bar{N}_{\mu}(s,\bar{S})=\sum_{s'}\Delta_L^{-1}(s-s')\frac{1}{2}\epsilon_{\mu\alpha\beta\gamma}
\partial_{\alpha}\bar{S}_{\beta\gamma}(s'+\hat{\mu})
\end{eqnarray*}
Then the classical part of the static potential is written as 
\begin{eqnarray}
\langle W_m({\cal C}) \rangle_{cl}=e^{-\pi^2\sum_{s,s'\atop\mu,\nu}\bar{N}_{\mu}(s)
D_{\mu\nu}^{-1}(s-s')\bar{N}_{\nu}(s')},\label{b_Wc}
\end{eqnarray}
where $D_{\mu\nu}^{-1}(s-s')$ is the inverse of the propagator of the effective action on the coarse lattice. Since only the parallel interactions are considered here, the momentum representation of the inverse propagator becomes $D_{\mu\nu}^{-1}(k)=\delta_{\mu\nu}D^{-1}(k)$.
Then the exponent $X({\cal C})$ of (\ref{b_Wc}) is written in the momentum representation as
\begin{eqnarray}
X({\cal C})&=&-4\pi^2\int_{-\pi}^{\pi}\frac{d^4k}{(2\pi)^4}\Delta_L^{-2}(k)
[\sin^2(\frac{k_1}{2})D_{22}^{-1}(k)\nn\\
&&+ \sin^2(\frac{k_2}{2})D_{11}^{-1}(k)]\bar{S}_{14}(k)
\bar{S}_{14}(-k).
\end{eqnarray}
This can be calculated easily when we take the limit $I,T\to\infty$ as
\begin{eqnarray}
X({\cal C})&=&
-\frac{IT\pi^2}{4}\int_{-\pi}^{\pi}\frac{d^2k}{(2\pi)^2}
\frac{1}{(\sin^2(\frac{k_1}{2})+\sin^2(\frac{k_2}{2})}\nn\\
&&[\sin^2(\frac{k_1}{2})D_{22}^{-1}(k)+ \sin^2(\frac{k_2}{2})D_{11}^{-1}(k)]\label{X_c}.
\end{eqnarray}
Using the 10 quadratic coupling constant, we get for example 
\begin{eqnarray}
D_{11}(k_1,k_2,\vec{0})&=&4[f_1+f_2\cos(k_1)+f_3(2+\cos(k_2))\nn\\
&+&f_4\cos(k_1)(2+\cos(k_2))+f_5(1+2\cos(k_2))\nn\\
&+&f_6\cos(k_1)(1+2\cos(k_2))+f_7\cos(k_2)\nn\\
&+&f_8\cos(2k_2)
+f_9\cos(k_1+k_2)\nn\\
&+&f_{10}(2+\cos(2k_2)]\nn.
\end{eqnarray}
Then (\ref{X_c}) can be evaluated using FFT calculations in the momentum space when use is made of numerical 10 coupling constants. The results are shown for typical three $b$ values in Table~\ref{sigma_L_D}. Again, the ratio $\sqrt{\sigma_{cl}/\sigma_{phys}}$ is around 30\% larger at these $b$ values.
Hence we see that better agreement can not be gotten with the simple 10 quadratic monopole inteactions alone.

\subsection{The lowest scalar glueball mass}
We consider here the following U(1) singlet and Weyl invariant
operator
\begin{eqnarray}
  \Psi (t) = L^{-3/2}\sum_{\vec{x}} 
                   Re \sbra{\Psi_{12} + \Psi_{23} 
                       + \Psi_{31}}(\vec{x},t) \label{Psi}
\end{eqnarray} 
on the $a$-lattice at timeslice $t$.
Here $\Psi_{ij}(\vec{x},t)$ is an $na \times na$ abelian Wilson loop
 and $L$ stands for the linear size of the lattice.
One can check easily that this operator carries $0^{++}$ quantum number
~\cite{Montovay}.
Then we evaluate the connected two point correlation function of $\Psi$
by  using the string model just as done in the case of the calculations 
of the string tension.
It turns out that the quantum correction is also negligibly small for large $b$.
Refer to the paper~\cite{Chernodub:2000ax} for details. Assuming the lowest mass gap obtained by the $\Psi$ operator (\ref{Psi}) for finite $b$ is the scalar glueball mass, we get the lowest scalar glueball mass as $M_{0^{++}}=2m_2$. In the best-fit parameters listed in Table~\ref{tabl:best_para}, we have fixed $m_2$ so to reproduce $M_{0^{++}}/\sigma_{phys}\sim 3.7$ which is consistent with the direct calculations done in Ref.~\cite{Lucini2004}.

\subsection{Monopole density distribution}
As shown in our previous work~\cite{SIB201711}, the monopole density
\begin{eqnarray}
r(b)\equiv\frac{\rho}{b^3}=\frac{1}{4\sqrt{3}Vb^3}\sum_{s,\mu}\sqrt{\sum_{a}(K_{\mu}^a(s))^2}
\label{eq:monoden}
\end{eqnarray}
  shows beautiful scaling behaviors in smooth gauges such as MCG, where $V$ is the lattice volume.
Namely the monopole density (\ref{eq:monoden}) can be written in terms of a unique function $r(b)$ of $b=na(\beta)$. But in the paper~\cite{SIB201711}, the meaning of $r(b)$ has  not been clarified.

Now we have derived the infrared effective monopole action showing also beautiful scaling. It is interesting to evaluate the monopole density from the effective action analytically. Since the square-root operator is rather difficult to evaluate analytically, we consider the squared monopole density defined as 
\begin{eqnarray}
R(b)\equiv\frac{1}{4Vb^3}\sum_{s,\mu}(\sum_{a}(K_{\mu}^a(s))^2)
\label{eq:square_monoden}
\end{eqnarray}

The effective monopole action on the coarse lattice is written as (\ref{pfac:5}). Then the squared monopole density (\ref{eq:square_monoden}) can be expressed by evaluating the partition function
\begin{eqnarray}
Z(\eta)&=&
  \sum_{K_{\mu}=-\infty \atop{\partial^{\prime}_{\mu}K_{\mu}=0}}^{\infty}
  \exp
  \Biggl\{
    - \sum_{s,s'\atop{\mu,\nu}}
    K_{\mu}(s)D_{\mu\nu}(s-s')K_{\nu}(s')\nn\\ 
&&    + i \sum_{s\atop{\mu}}
    \eta_{\mu}(s)K_{\mu}(s)
 \Biggr\}
\nonumber\\
&=&
  \int_{-\infty}^{+\infty}\!\!\!\!{\cal D}F_{\mu}(s) 
  \int_{-\pi}^{+\pi}\!\!\!\!{\cal D}\phi(s)
  \!\!\!\!\sum_{K_{\mu}(s)=-\infty}^{\infty}\!\!\!\!
  \delta(F_{\mu}(s)-K_{\mu}(s))\nn\\
&&\exp
  \Bigg\{
    -\sum_{s,s'\atop{\mu,\nu}}
    F_{\mu}(s)D_{\mu\nu}(s-s')F_{\nu}(s')
\nonumber \\
&&\qquad +i\sum_{s,\mu}
    F_{\mu}(s)
    \Big[
      \partial_{\mu}\phi(s) +  \eta_{\mu}(s)
    \Big]
  \Bigg\} ,\nonumber \\
&=&
  \int_{-\pi}^{+\pi}\!\!\!\!{\cal D}\phi(s)
  \!\!\!\!\sum_{l_{\mu}(s)=-\infty}^{\infty}\!\!\!\! \nn\\
&&  \exp
  \Bigg\{
    -\frac{1}{4}\sum_{s,s'\atop{\mu,\nu}}
    \Big[
      \partial_{\mu}\phi(s) + 2\pi l_{\mu}(s)+  \eta_{\mu}(s)
    \Big]
\nonumber\\
&&  D_{\mu\nu}^{-1}(s-s')
    \Big[
      \partial_{\nu}\phi(s') + 2\pi l_{\nu}(s')+  \eta_{\nu}(s')
    \Big]
  \Bigg\}.\label{z_eta}
\end{eqnarray}
Performing BKT transformation and Hodge decomposition, we obtain
\begin{eqnarray}
l_{\mu}(s)
&=&
  s_{\mu}(s) + \partial_{\mu}r(s)
\nonumber \\
&=&
  \partial_{\mu}
  \Big\{
    -\sum_{s'} \Delta^{-1}_{L\;s,s'}
    \partial'_{\nu} s_{\nu}(s') + r_{\mu}(s')
  \Big\}
\nonumber \\
&&\qquad
  +\sum_{s'} \partial'_{\nu}\Delta^{-1}_{L\;s,s'}\sigma_{\nu\mu}(s'),
\end{eqnarray} 
where $\sigma_{\nu\mu}(s) \equiv \partial_{[ \mu}s_{\nu ]}$ is the
closed string variable satisfying the conservation rule (\ref{conv}).
The compact field $\phi(s)$ is absorbed into a non-compact 
field $\phi_{NC}(s)$. 
Integrating out the auxiliary non-compact field,
we see 
\begin{eqnarray}
Z(\eta)
&=&
  \!\!\!\!
  \sum_{\sigma_{\mu\nu}(s)=-\infty
        \atop{\partial_{[\alpha}\sigma_{\mu\nu]}(s)=0}}^{\infty}
  \!\!\!\!
  \exp
  \Bigg\{-S(\sigma)-\sum_{\mu, s}X_{\mu}(s)\eta_{\mu}(s)\nn\\
&&    - \frac{1}{4}\sum_{s,s'\atop{\mu,\nu}}
    \eta_{\mu}(s)D_{\mu\nu}^{-1}(s-s')\eta_{\nu}(s')
  \Bigg\},\label{z_eta2} \\
 S(\sigma)&=& 
    \pi^2\sum_{ s,s'\atop{ \mu\neq\alpha\atop{ \nu\neq\beta } } }
      \sigma_{\mu\alpha}(s)\partial_{\alpha}\partial_{\beta}'
      D_{\mu\nu}^{-1}(s-s_1)\nn\\
  &\times&   \Delta_L^{-2}(s_1-s')\sigma_{\nu\beta}(s')
\nonumber \\
X_{\mu}(s)&=&
  \pi\sum_{s',s''\atop{\nu,\alpha}}
    \sigma_{\nu\alpha}(s)\partial_{\nu}
    \Delta_L^{-1}(s'-s'')D_{\alpha\mu}^{-1}(s''-s).\nn
\end{eqnarray}

Then the squared monopole density (\ref{eq:square_monoden}) is evaluated as 
\begin{eqnarray}
R(b)&=&
  -\frac{1}{4Vb^3 Z(0)}\frac{\delta^2}{\delta\eta_{\mu}^2(s)}Z(\eta)|_{\eta=0}\nn\\
  &=&\frac{3}{2b^3}D_{ii}^{-1}(0)-Q(b)\label{z_eta3}\\
 Q(b)&=&\frac{1}{4Vb^3 Z(0)}\nn\\
 &\times&  \sum_{\sigma_{\mu\nu}(s)=-\infty
        \atop{\partial_{[\alpha}\sigma_{\mu\nu]}(s)=0}}^{\infty}
  \!\!\!\!
  \exp(-S(\sigma))\sum_{\mu,s}X_{\mu}(s)^2, 
\label{den_perf}
\end{eqnarray}
where $D_{ii}^{-1}(0)$ denotes the self-copling term of the inverse of the propagator $D_{\mu\nu}(s-s')$ in (\ref{pfac:5}).

The  quantum part $Q(b)$ (\ref{den_perf}) is expected to be small for large $b$ strong-coupling regions and hence we evaluate the first part  in (\ref{z_eta3}) alone.
The self-coupling term $D_{ii}^{(-1)}(0,0,0,0)$ is calculated explicitly in Eq.(\ref{D_in}) of Appendix~\ref{dinv}. 

The squared density   $R(b)$ is plotted in Fig.\ref{dens_MCG_perfect} in comparison with that calculated numerically with the help of the MCG data obtained in Ref.~\cite{SIB201711}.   One can see from Fig.\ref{dens_MCG_perfect} a rough agreement for $b=na(\beta)> 1.2 ~(\sigma_{phys}^{-1/2})$. The difference may comes again from the simple assumption of 10 quadratic interactions alone adopted here. Anyway, the features are new found in the global color-invariant smooth gauge like in MCG.

\subsection{Discussions about the disagreement between analytical calculations and numerical data}
As shown above, we have obtained around 30\% larger theoretical values with respect to both the string tension and the  monopole density. 
Let us discuss the disagreement, comparing the forms of the effective 
monopole action. First of all,
the assumption of adopting quadratic interactions alone leads us to the type-2 dual superconductor as seen from (\ref{sigma_cl}). But as found numerically in the previous paper~\cite{Suzuki:2009xy}, the dual Meissner effect shows that the confined vacuum is near the border between the type-1 and the type-2 dual superconductor. Hence only from this fact, the assumption that the action form composed of simple quadratic interactions alone is insufficient.
To be noted that both the string tension and the monopole density depend on the 
inverse of the propagator of the effective monopole action on the coarse $b$ lattice as seen from (\ref{X_c}) and (\ref{z_eta3}). The self-coupling term is dominant in the propagator and so let us compare the self-coupling term starting from (1) the simplest 10 quadratic ineraction case and (2) the 27 quadratic plus higher four- and six-point interactions case.
See an example shown in Table~\ref{tbl:comparison} for $\beta=3.2,\ n=4\ (b=1.054 (\sigma_{phys})^{-1/2})$.

Since analytic calculations including four- and six-point interactions are too
difficult to perform exactly as discussed in Ref.\cite{Chernodub:1999xf}, we adopt a simple mean-field assumption
using the averaged monopole density $RQ$ evaluated from the numerical squared monopole density $R(b)$, i.e., $RQ=<(K_{\mu}^a)^2>\equiv R(b)/3$. Then using the form of four- and six-point interactions defined in  Table~\ref{tbl:higher}, we get the effective self-coupling term of 
the case (2) as 
\begin{eqnarray}
F(1)_{effective}=F(1)+\frac{32RQ}{3}F(28)+\frac{128RQ^2}{3}F(29)\nn.
\end{eqnarray}

In the typical example shown in Table~\ref{tbl:comparison} where $R(b=1.052)=1.04\ (\sigma_{phys}^{-1/2})$, we get $F(1)=0.902$ in the case (1), whereas in the case (2)
\begin{eqnarray*}
F(1)_{effective}&=&1.56-0.0455*32*1.04/3\nn\\
&+& 0.00123*128*1.04^2/3=1.112.
\end{eqnarray*}
This is 23\% larger than that of $F(1)$ of the simple 10 quadratic case (1).
Hence the above 30\% discrepancies are most probablly due to the too simple assumption of 10 quadratic monopole actions alone.

\begin{acknowledgments}
The numerical simulations of this work were done using  computer clusters HPC and SX-ACE at RCNP of Osaka University. The author would like to thank RCNP for their support of computer facilities. 
The vacuum configurations in MCG used here are the same used in Ref.~\cite{SIB201711} and were generated by Dr.Vitaly Bornyakov. The author acknowledges very much
Vitaly's contribution. He would like to thank also Dr.Shouji Fujimoto and Dr.Katsuya Ishiguro for illuminating discussions. 
\end{acknowledgments}

\appendix

\begin{widetext}

\section{The inverse Monte-Carlo method\label{APD:swendson}}
The effective monopole actions $S(k)$ is derived following the Swendsen's method
~\cite{swendsen,Shiba:1994db}.
The effective monopole action $S(k)$ is assumed to be a sum of independent Lorentz invariant monopole currents interactions summed over all space-time links.
Define these operators adopted as $S_i[k]$. Then
$S[k]=\sum_i F(i) S_i[k]$,
where $F(i)$ are coupling constants which should be determined by the Swendsen 
method.

Let us consider the expectation value of an operator ${\cal O}_a[k]$:
\beqn
\langle{\cal O}_a[k]\rangle&=&\frac{(\prod_{s,\mu}\sum_{k_\mu(s)=-\infty}^{\infty})
                    (\prod_{s}\delta_{\partial'_\mu k_\mu(s),0})
                   {\cal O}_a[k]\exp(-\sum_i F(i) S_i [k])}{\prod_{s,\mu}\sum_{k_\mu(s)=-\infty}^{\infty}\exp\FS{-\sum_i F(i) S_i [k]}}
.\label{eqn:sw0}
\eeqn
Now notice one plaquette $(s',\hat{\mu}',\hat{\nu}')$ on the dual lattice and the monopole currents around the plaquette:
\beqn
\{ k_{\mu'}(s'),\ k_{\nu'}(s'+\hat{\mu}')
   ,\ k_{\mu'}(s'+\hat{\nu}'),\ k_{\nu'}(s')\}
\label{cur:pla}
\eeqn
Define a part of the monopole action containing the currents (\ref{cur:pla})  as $\tilde{S}[k]$. Then we get:
\beqn
\lefteqn{
\BS{{\prod_{s,\mu}}{\sum_{k_\mu(s)=-\infty}^{\infty}}}
\BS{{\prod_{s}}\delta_{\partial'_\mu k_\mu(s),0}}
         {\cal O}_a[k]\exp\BM{-\sum_i F(i) S_i [k]}}\nn\\
&=& \BS{{\prod_{s,\mu}}^\prime
        {\sum_{k_\mu(s)=-\infty}^{\infty}}}
    \BS{{\prod_{s}}^\prime\delta_{\partial'_\mu k_\mu(s),0}}
    \exp\BM{-\sum_i F(i) \BS{S_i [k]-\tilde{S}_i [k]}}\nn\\
&&  \GM{\sum_{k_{\mu'}(s')=-\infty}^{\infty}
        \sum_{k_{\nu'}(s'+\hat{\mu}')=-\infty}^{\infty}
        \sum_{k_{\mu'}(s'+\hat{\nu}')=-\infty}^{\infty}
        \sum_{k_{\nu'}(s')=-\infty}^{\infty}
        \delta_{\D_{\mu}'k_{\mu}(s'),0}
        \delta_{\D_{\mu}'k_{\mu}(s'+\hat{\mu}'),0}
        \delta_{\D_{\mu}'k_{\mu}(s'+\hat{\nu}'),0}
        \delta_{\D_{\mu}'k_{\mu}(s'+\hat{\mu}'+\hat{\nu}'),0}\nn\\
&&  \quad{\cal O}_a[k]\exp\BM{-\sum_i F(i) \tilde{S}_i [k]}},\label{eqn:sw1}
\eeqn
where ${\prod}^\prime$ means the product excluding the sites and the links in the plaquette considered. Using the current conservation rule, we can rewrite one  $\delta$ function among four $\delta$ functions around the plaquette as
\beqn
\delta_{\D_{\mu}'k_{\mu}(s')+\D_{\mu}'k_{\mu}(s'+\hat{\mu}')
        +\D_{\mu}'k_{\mu}(s'+\hat{\nu}')
        +\D_{\mu}'k_{\mu}(s'+\hat{\mu}'+\hat{\nu}'),0}.
\eeqn
Now let us note that the $\delta$ function does not contain 
any monopole currents (\ref{cur:pla}). Then we get
\beqn
\mbox{(\ref{eqn:sw1})}
&=&\BS{{\prod_{s,\mu}}^\prime\sum_{k_\mu(s)=-\infty}^{\infty}}
    \BS{{\prod_{s}}^\prime\delta_{\partial'_\mu k_\mu(s),0}}
     \delta_{\D_{\mu}'k_{\mu}(s')+\D_{\mu}'k_{\mu}(s'+\hat{\mu}')
        +\D_{\mu}'k_{\mu}(s'+\hat{\nu}')
        +\D_{\mu}'k_{\mu}(s'+\hat{\mu}'+\hat{\nu}'),0}\nn\\
&&  \GM{\BS{\sum \delta}_{\hat{k}}
        {\cal O}_a[\hat{k},\{k\}']
        \exp\BM{-\sum_i F(i) \tilde{S}_i [\hat{k},\{k\}']}}
        \exp\BM{-\sum_i F(i) \BS{S_i [k]-\tilde{S}_i [k]}},\label{eqn:sw2}
\eeqn
where $\{k\}'$ denotes the monopole currents excluding those on the plaquette
(\ref{cur:pla}) and 
$\BS{\sum \delta}_{\hat{k}}$ is given by 
\beqn
\BS{\sum \delta}_{\hat{k}}\equiv
        \sum_{\hat{k}_{\mu'}(s')=-\infty}^{\infty}
        \sum_{\hat{k}_{\nu'}(s'+\hat{\mu}')=-\infty}^{\infty}
        \sum_{\hat{k}_{\mu'}(s'+\hat{\nu}')=-\infty}^{\infty}
        \sum_{\hat{k}_{\nu'}(s')=-\infty}^{\infty}
        \delta_{\D_{\mu}'\hat{k}_{\mu}(s'),0}
        \delta_{\D_{\mu}'\hat{k}_{\mu}(s'+\hat{\mu}'),0}
        \delta_{\D_{\mu}'\hat{k}_{\mu}(s'+\hat{\nu}'),0}.
\eeqn
 
Now define a new operator $\hat{\cal O}_a[\{k\}']$ as
\beqn
\hat{\cal O}_a[\{k\}']=
\frac{\BS{\sum \delta}_{\hat{k}}
        {\cal O}_a[\hat{k},\{k\}']
        \exp\BM{-\sum_i F(i) \tilde{S}_i [\hat{k},\{k\}']}}
     {\BS{\sum \delta}_{\hat{k}}
        \exp\BM{-\sum_i F(i) \tilde{S}_i [\hat{k},\{k\}']}},
\eeqn
we get
\beqn
\lefteqn{
\BS{{\prod_{s,\mu}}{\sum_{k_\mu(s)=-\infty}^{\infty}}}
\BS{{\prod_{s}}\delta_{\partial'_\mu k_\mu(s),0}}
         {\cal O}_a[k]\exp\BM{-\sum_i F(i) S_i [k]}}\nn\\
&=&\BS{{\prod_{s,\mu}}\sum_{k_\mu(s)=-\infty}^{\infty}}
    \BS{{\prod_{s}}\delta_{\partial'_\mu k_\mu(s),0}}
        \hat{\cal O}_a[\{k\}']
        \exp\BM{-\sum_i F(i) S_i [k]}.\label{eqn:sw3}
\eeqn

Now consider further $\hat{\cal O}_a[\{k\}']$. Noting that the monopole current 
conservation holds good on every site in Eq.(\ref{eqn:sw2}), we see
\beqn
\D_{\mu}'\hat{k}_{\mu}(s')&=&\D_{\mu}' k_{\mu}(s')
  +\hat{k}_{\mu'}(s')+\hat{k}_{\nu'}(s')
  -k_{\mu'}(s')-k_{\nu'}(s')\nn\\
  &=&\hat{k}_{\mu'}(s')+\hat{k}_{\nu'}(s')
  -k_{\mu'}(s')-k_{\nu'}(s').
\eeqn
and
\beqn
&&\D_{\mu}'\hat{k}_{\mu}(s'+\hat{\mu'})=
   \hat{k}_{\nu'}(s'+\hat{\mu'})-\hat{k}_{\mu'}(s')
  -k_{\nu'}(s'+\hat{\mu'})+k_{\mu'}(s'),\\
&&\D_{\mu}'\hat{k}_{\mu}(s')+\D_{\mu}'\hat{k}_{\mu}(s'+\hat{\nu'})=
   \hat{k}_{\mu'}(s'+\hat{\nu'})+\hat{k}_{\mu'}(s')
  -k_{\mu'}(s'+\hat{\nu'})-k_{\mu'}(s').
\eeqn
Also using a relation
\beqn
\sum_{M=-\infty}^{\infty}\delta_{\hat{k}_{\mu'}(s'),
                                 k_{\mu'}(s')+M}=1,
\eeqn
where $M$ is an integer, we get
\beqn
\BS{\sum \delta}_{\hat{k}}F[\hat{k},\{k\}']
&=&     \sum_{M=-\infty}^{\infty}
        \sum_{\hat{k}_{\mu'}(s')=-\infty}^{\infty}
        \sum_{\hat{k}_{\nu'}(s'+\hat{\mu}')=-\infty}^{\infty}
        \sum_{\hat{k}_{\mu'}(s'+\hat{\nu}')=-\infty}^{\infty}
        \sum_{\hat{k}_{\nu'}(s')=-\infty}^{\infty}\nn\\
&&      \delta_{\hat{k}_{\mu'}(s'),k_{\mu'}(s')+M}
        \delta_{\hat{k}_{\nu'}(s'+\hat{\mu'}),k_{\nu'}(s'+\hat{\mu'})+M}
        \delta_{\hat{k}_{\mu'}(s'+\hat{\nu'}),k_{\mu'}(s'+\hat{\nu'})-M}
        \delta_{\hat{k}_{\nu'}(s'),k_{\nu'}(s')-M}\nn\\
&&      F[\hat{k}_{\mu'}(s'),\hat{k}_{\nu'}(s'+\hat{\mu}')
          ,\hat{k}_{\mu'}(s'+\hat{\nu}'),\hat{k}_{\nu'}(s'),\{k\}']\nn\\
&=&     \sum_{M=-\infty}^{\infty}
        F[k_{\mu'}(s')+M,k_{\nu'}(s'+\hat{\mu'})+M,
          k_{\mu'}(s'+\hat{\nu'})-M,k_{\nu'}(s')-M
          ,\{k\}'],\nn\\
\eeqn
where $F[\hat{k},\{k\}']$ is  any function of $k$.

The value of the lattice monopole current defined by DeGrand and Toussaint~\cite{DeGrand:1980eq} is restricted to the region [$-2$,$+2$], so that the type-2 $n$ extended monopole defined by \cite{Ivanenko:1991wt} can take the value in the region [$-(3n^2-1)$, $3n^2-1$]. Hence the sum with respect to $M$ is restricted to the region between $m_1$ and $m_2$ defined below
\beqn
m_1=-(3n^2-1)-\min\GM{k_{\mu'}(s'),k_{\nu'}(s'+\hat{\mu'}),
                      -k_{\mu'}(s'+\hat{\nu'}),-k_{\nu'}(s')},\nn\\
m_2=(3n^2-1)-\max\GM{k_{\mu'}(s'),k_{\nu'}(s'+\hat{\mu'}),
                      -k_{\mu'}(s'+\hat{\nu'}),-k_{\nu'}(s')}.
\eeqn
Finally we find $\hat{\cal O}_a[k]$ is rewritten by
\beqn
\hat{\cal O}_a[k]=\frac{\sum_{M=m_1}^{m_2}
                            {\cal O}_a[\bar{k}]
             \exp\BM{-\sum_i F(i) \tilde{S}_i [\bar{k}]}}
     {\sum_{M=m_1}^{m_2}
             \exp\BM{-\sum_i F(i) \tilde{S}_i [\bar{k}]}},\label{eqn:sw4}
\eeqn
Here
\beqn
\bar{k}_\mu \equiv k_\mu (s)
               +M(\delta_{s,s'}\delta_{\mu,\mu'}
                 +\delta_{s,s'+\hat{\mu}'}\delta_{\mu,\nu'}
                 -\delta_{s,s'+\hat{\nu}'}\delta_{\mu,\mu'}
                 -\delta_{s,s'}\delta_{\mu,\nu'}).
\eeqn
Then
\beqn
\lefteqn{
\BS{{\prod_{s,\mu}}{\sum_{k_\mu(s)=-\infty}^{\infty}}}
\BS{{\prod_{s}}\delta_{\partial'_\mu k_\mu(s),0}}
         {\cal O}_a[k]\exp\BM{-\sum_i F(i) S_i [k]}}\nn\\
&=&\BS{{\prod_{s,\mu}}\sum_{k_\mu(s)=-\infty}^{\infty}}
    \BS{{\prod_{s}}\delta_{\partial'_\mu k_\mu(s),0}}
        \hat{\cal O}_a[k]
        \exp\BM{-\sum_i F(i) S_i [k]}.
\eeqn
The final expression is the following
\beqn
\langle {\cal O}_a[k] \rangle=\langle \hat{\cal O}_a[k] \rangle.
\label{eqn:sw5}
\eeqn

As an arbitrary operator $O_a(k)$, we adopt $S_a(k)$ in the monopole action. When we consider here only quadratic monopole interactions,
we can get 
\beqn
S_i(\hat{k},\{k\}')=a_i^{(2)}M^2+a_i^{(1)}M+S_i(k).
\eeqn
Then Eq.(\ref{eqn:sw5}) is reduced to
\beqn
\langle\frac{\sum_M(a_i^{(2)}M^2+a_i^{(1)}M)\exp[-(\sum_jg_ja_j^{(2)})M^2
-(\sum_jg_ja_j^{(1)})M]}{\exp[-(\sum_jg_ja_j^{(2)})M^2
-(\sum_jg_ja_j^{(1)})M]}\rangle = 0
\label{eq:sw55}
\eeqn

Using this identity (\ref{eq:sw55}), 
we can estimate the monopole action $S[k]$
iteratively. For that purpose, we define an operator $\overline{\cal O}_a[k]$
where the coupling constants are replaced by a trial set $\{\tilde{F}_i\}$
in Eq.(\ref{eqn:sw4}):
\beqn
\overline{\cal O}_a[k] \equiv\frac{\sum_{M=m_1}^{m_2}
                            {\cal O}_a[\bar{k}]
             \exp\BM{-\sum_i \tilde{F}_i \tilde{S}_i [\bar{k}]}}
     {\sum_{M=m_1}^{m_2}
             \exp\BM{-\sum_i \tilde{F}_i \tilde{S}_i [\bar{k}]}}.
\label{eqn:sw6}
\eeqn
If $F(i)$ are not eqaul to $\tilde{F}_i$ for all $i$, we expand $\langle{\cal O}_a-\overline{\cal O}_a\rangle$ upto the first order of \{$F(i)-\tilde{F}_i$\} and get
\beqn
\langle{\cal O}_a-\overline{\cal O}_a\rangle=
\sum_b
\langle\overline{\cal O}_a \overline{S}_b-
       \overline{{\cal O}_a S_b}\rangle
(g_b-\tilde{F}_b).
\label{eqn:sw7}
\eeqn
Practically, we take a set of trial coupling constants $\{\tilde{F}_a\}$ and 
evaluate the expectation value $\langle{\cal O}_a-\overline{\cal O}_a\rangle$
using the thermalized monopole vacua. If 
$\langle{\cal O}_a-\overline{\cal O}_a\rangle$ become zero for all $a$, then $\{\tilde{F}_a\}$ can be regarded as the real coupling constants.
Otherwise, we solve the equation (\ref{eqn:sw7}) numerically and adopt the solution $\{g_a\}$ as a new trial set of coupling constants.
This is the way to get the effective monopole action iteratively.

Eq.(\ref{eqn:sw7}) can be expressed as
\beqn
&&\langle\overline{a_i^{(2)}M^2+a_i^{(1)}M}\rangle\\
&=&\sum_j\{\langle\overline{(a_i^{(2)}M^2+a_i^{(1)}M)}\ \overline{(a_j^{(2)}M^2+a_j^{(1)}M)}\rangle\\
&&-\langle\overline{(a_i^{(2)}M^2+a_i^{(1)}M)(a_j^{(2)}M^2+a_j^{(1)}M)}\rangle\}(g_j-\tilde{F}_j)
\rangle
\label{eq:sw8}
\eeqn
\end{widetext}

\begin{table*}
\caption{The quadratic interactions used for the modified Swendsen method. Color index $a$ of the monopole current $k_{\mu}^a$ is omitted.}
\label{tbl:appquad} 
\begin{tabular}{cllcll}
{\it coupling $\mbra{F(i)}$} &  distance& $ \ \ \ \ \ \ \ \ $  {\it type} $ \ \ \ \ \ \ \ \ $ &
{\it coupling $\mbra{F(i)}$} &  distance& $ \ \ \ \ \ \ \ \ $  {\it type} \\ 
\hline
$F(1)$    & (0,0,0,0) & $k_\mu(s)$ &
$F(15)$ & (2,1,1,0) & $k_\mu(s+2\hat{\mu}+\hat{\nu}+\hat{\rho})$ \\
$F(2)$    & (1,0,0,0) & $k_\mu(s+\hat{\mu})$ &
$F(16)$ & (1,2,1,0) & $k_\mu(s+\hat{\mu}+2\hat{\nu}+\hat{\rho})$ \\
$F(3)$    & (0,1,0,0) & $k_\mu(s+\hat{\nu})$ &
$F(17)$ & (0,2,1,1) & $k_\mu(s+2\hat{\nu}+\hat{\rho}+\hat{\sigma})$ \\
$F(4)$    & (1,1,0,0) & $k_\mu(s+\hat{\mu}+\hat{\nu})$ &
$F(18)$ & (2,1,1,1) & $k_\mu(s+2\hat{\mu}+\hat{\nu}+\hat{\rho}+\hat{\sigma})$ \\
$F(5)$    & (0,1,1,0) & $k_\mu(s+\hat{\nu}+\hat{\rho})$ &
$F(19)$ & (1,2,1,1) & $k_\mu(s+\hat{\mu}+2\hat{\nu}+\hat{\rho}+\hat{\sigma})$ \\
$F(6)$    & (1,1,1,0) & $k_\mu(s+\hat{\mu}+\hat{\nu}+\hat{\rho})$ &
$F(20)$ & (2,2,0,0) & $k_\mu(s+2\hat{\mu}+2\hat{\nu})$ \\
$F(7)$    & (0,1,1,1) & $k_\mu(s+\hat{\nu}+\hat{\rho}+\hat{\sigma})$ &
$F(21)$ & (0,2,2,0) & $k_\mu(s+2\hat{\nu}+2\hat{\rho})$ \\
$F(8)$    & (2,0,0,0) & $k_\mu(s+2\hat{\mu})$ &
$F(22)$ & (3,0,0,0) & $k_\mu(s+3\hat{\mu})$ \\
$F(9)$    & (1,1,1,1) & $k_\mu(s+\hat{\mu}+\hat{\nu}+\hat{\rho}+\hat{\sigma})$ &
$F(23)$ & (0,3,0,0) & $k_\mu(s+3\hat{\nu})$ \\
$F(10)$ & (0,2,0,0) & $k_\mu(s+2\hat{\nu})$ &
$F(24)$ & (2,2,1,0) & $k_\mu(s+2\hat{\mu}+2\hat{\nu}+\hat{\rho})$ \\
$F(11)$ & (2,1,0,0) & $k_\mu(s+2\hat{\mu}+\hat{\nu})$ &
$F(25)$ & (1,2,2,0) & $k_\mu(s+\hat{\mu}+2\hat{\nu}+2\hat{\rho})$ \\ 
$F(12)$ & (1,2,0,0) & $k_\mu(s+\hat{\mu}+2\hat{\nu})$ &
$F(26)$ & (0,2,2,1) & $k_\mu(s+2\hat{\nu}+2\hat{\rho}+\hat{\sigma})$ \\
$F(13)$ & (0,2,1,0) & $k_\mu(s+2\hat{\nu}+\hat{\rho})$ &
$F(27)$ & (2,1,1,0) & $k_\rho(s+2\hat{\mu}+2\hat{\nu}+\hat{\rho})$ \\
$F(14)$ & (2,1,0,0) & $k_\nu(s+2\hat{\mu}+\hat{\nu})$ &
         &           & \\
\end{tabular}
\end{table*}

\begin{table}
\caption{The higher order interactions used for
the modified Swendsen method. }
\label{tbl:higher} 
\begin{tabular}{cll}
{\it coupling } & distance  & $ \ \ \ \ \ \ \ \ $  {\it type} \\ 
\hline
4-point         & (0,0,0,0) &
$S^{(4)}= \sum_{s}\sum_a\sbra{\sum_{\mu=-4}^4 (k_\mu^a)^2(s)}^2$ \\
6-point         & (0,0,0,0) &
$S^{(6)}= \sum_{s}\sum_a\sbra{\sum_{\mu=-4}^4 (k_\mu^a)^2(s)}^3$ \\
\end{tabular}
\end{table}
\begin{table}[htbp]
\caption{An example of the results with the monopole action with $10$ quadratic+four-point + six-point interactions having color mixing. 
The case for $\beta=3.3$ and $n=3$ blocking in MCG gauge is shown on $48^4$ lattice. Here $F(12)$ and $F(14)$ are coupling constants with color-mixed terms.}
\label{tbl:1022}
\begin{center}
\begin{tabular}{|l|r|r|}
\hline 
&\textrm{quadratic}& \textrm{error}\\
\hline	
F(  1)=	&	2.13E+00	&	4.99E-03	\\
F(  2)=	&	3.89E-01	&	1.92E-03	\\
F(  3)=	&	3.15E-01	&	1.62E-03	\\
F(  4)=	&	1.15E-01	&	7.00E-04	\\
F(  5)=	&	8.62E-02	&	1.26E-03	\\
F(  6)=	&	2.36E-02	&	1.81E-03	\\
F(  7)=	&	2.68E-02	&	4.14E-04	\\
F(  8)=	&	1.73E-02	&	7.12E-04	\\
F(  9)=	&	4.19E-02	&	5.90E-04	\\
F(  10)=	&	2.80E-02	&	1.06E-03	\\
\hline
&\textrm{four-point} &\textrm{error}\\
 \hline
F( 11)=	&	-6.87E-02	&	4.41E-04	\\
F( 12)=	&	-3.18E-02	&	1.13E-04	\\
\hline 
&\textrm{six-point}& \textrm{error}\\
\hline
F (13)=	&	2.81E-03	&	3.17E-05	\\
F( 14)=	&	4.62E-05	&	3.75E-04	\\
\hline
\end{tabular}
\end{center}
\end{table}
\begin{table*}[htbp]
 \caption{Comparison of the monopole actions: An example of  $n=4$ and $\beta=3.2$ ($b=1.052 ~(\sigma_{phys}^{-1/2})$) on $48^4$ lattice in MCG gauge.}
 \label{tbl:comparison}
 \begin{center}
  \begin{tabular}{||c|c|c||c|c||c|c||c|c||}
\hline\hline																						
		&	$S^2_{10}$	&	error	&	$S^2_{27}$	&	error	&	$S^2_{10}+S^4+S^6$	&	error	&	$S^2_{27}+S^4+S^6$	&	error			\\
\hline\hline																						
	F(1)	&	9.02E-01	&	4.13E-04	&	9.22E-01	&	8.45E-05	&	1.49E+00	&	1.16E-02	&	1.56E+00	&	7.06E-03			\\
\hline																						
	F(2)	&	2.96E-01	&	2.41E-03	&	3.20E-01	&	9.50E-05	&	2.47E-01	&	5.99E-04	&	2.74E-01	&	1.10E-04				\\
\hline																						
	F(3)	&	2.11E-01	&	1.37E-03	&	2.50E-01	&	1.05E-05	&	1.91E-01	&	1.25E-03	&	2.32E-01	&	7.58E-04					\\
\hline																						
	F(4)	&	7.75E-02	&	1.15E-03	&	9.96E-02	&	1.23E-04	&	6.74E-02	&	8.83E-04	&	9.30E-02	&	1.73E-03					\\
\hline																						
	F(5)	&	5.79E-02	&	1.24E-03	&	9.11E-02	&	1.22E-04	&	5.01E-02	&	1.26E-03	&	8.59E-02	&	1.60E-03					\\
\hline																						
	F(6)	&	2.85E-02	&	2.85E-04	&	5.18E-02	&	1.12E-04	&	1.65E-02	&	5.70E-04	&	4.76E-02	&	9.68E-04					\\
\hline																						
	F(7)	&	2.02E-02	&	8.86E-04	&	4.01E-02	&	9.13E-07	&	1.32E-02	&	2.94E-04	&	3.81E-02	&	3.17E-04					\\
\hline																						
	F(8)	&	1.64E-02	&	2.05E-03	&	2.54E-02	&	1.99E-06	&	1.01E-02	&	7.24E-05	&	2.33E-02	&	1.28E-04					\\
\hline																						
	F(9)	&	1.13E-02	&	4.67E-04	&	4.70E-02	&	2.02E-06	&	2.52E-02	&	1.87E-04	&	4.35E-02	&	3.61E-04				\\
\hline																						
	F(10)	&	1.49E-02	&	1.02E-03	&	4.71E-02	&	1.75E-05	&	1.76E-02	&	4.74E-04	&	4.42E-02	&	4.73E-04					\\
\hline																						
	F(11)	&		&		&	2.34E-02	&	1.42E-04	&	-4.29E-02	&6.27E-04		&	2.28E-02	&	1.28E-03				\\
\hline																						
	F(12)	&		&		&	2.34E-02	&	2.91E-05	&	1.15E-03	&1.75E-05	&	2.15E-02	&	2.29E-04				\\
\hline																						
	F(13)	&		&		&	2.13E-02	&	1.52E-05	&		&		&	2.01E-02	&	1.48E-06					\\
\hline																						
	F(14)	&		&		&	3.18E-05	&	4.21E-05	&		&		&	-5.07E-04	&	4.50E-04					\\
\hline																						
	F(15)	&		&		&	1.17E-02	&	1.69E-04	&		&		&	1.21E-02	&	1.57E-03					\\
\hline																						
	F(16)	&		&		&	1.19E-02	&	9.78E-06	&		&		&	1.08E-02	&	4.79E-05					\\
\hline																						
	F(17)	&		&		&	1.28E-02	&	2.80E-05	&		&		&	1.23E-02	&	3.23E-04					\\
\hline																						
	F(18)	&		&		&	6.18E-03	&	1.47E-04	&		&		&	6.95E-03	&	1.32E-03					\\
\hline																						
	F(19)	&		&		&	6.34E-03	&	2.25E-05	&		&		&	5.94E-03	&	2.45E-04					\\
\hline																						
	F(20)	&		&		&	6.84E-03	&	3.77E-05	&		&		&	6.83E-03	&	4.15E-04					\\
\hline																						
	F(21)	&		&		&	4.63E-03	&	1.15E-05	&		&		&	4.44E-03	&	2.86E-04					\\
\hline																						
	F(22)	&		&		&	5.71E-03	&	1.22E-04	&		&		&	4.66E-03	&	9.76E-04					\\
\hline																						
	F(23)	&		&		&	1.08E-03	&	4.76E-06	&		&		&	1.10E-03	&	3.54E-05					\\
\hline																						
	F(24)	&		&		&	1.91E-03	&	7.34E-05	&		&		&	2.31E-03	&	6.97E-04					\\
\hline																						
	F(25)	&		&		&	2.98E-03	&	8.67E-05	&		&		&	2.08E-03	&	7.48E-04					\\
\hline																						
	F(26)	&		&		&	2.88E-03	&	7.03E-06	&		&		&	2.75E-03	&	2.71E-05					\\
\hline																						
	F(27)	&		&		&	1.16E-03	&	9.94E-05	&		&		&	5.12E-04	&	8.02E-04					\\
\hline																						
	F(28)	&		&		&		&		&		&		&	-4.55E-02	&	4.38E-04				\\
\hline																						
	F(29)	&		&		&		&		&		&		&	1.23E-03	&	1.22E-05					\\
\hline\hline																						
  \end{tabular}
 \end{center}
\end{table*}

\begin{figure}[htb]
\caption{Comparison of the coupling constants of the self and two nearest-neighbor interactions between the actions composed of  27 ($NF2=27$) and 10 ($NF2=10$) quadratic interactions alone. The data are taken on $48^4$ in MCG}
\label{figA_F123}
  \begin{minipage}[b]{0.9\linewidth}
    \centering
    \includegraphics[width=8cm,height=6.cm]{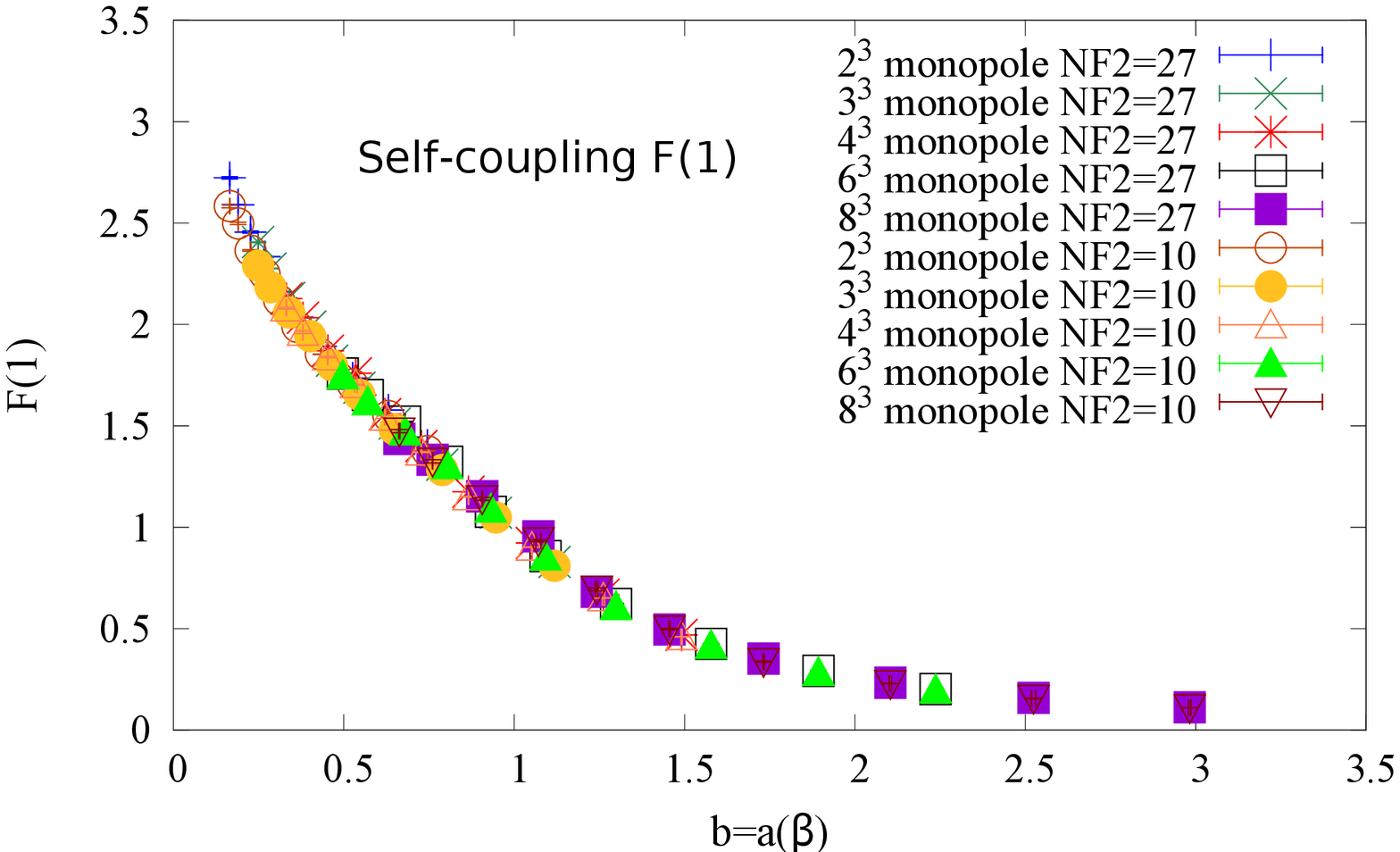}
  \end{minipage}
  \begin{minipage}[b]{0.9\linewidth}
    \centering
    \includegraphics[width=8cm,height=6.cm]{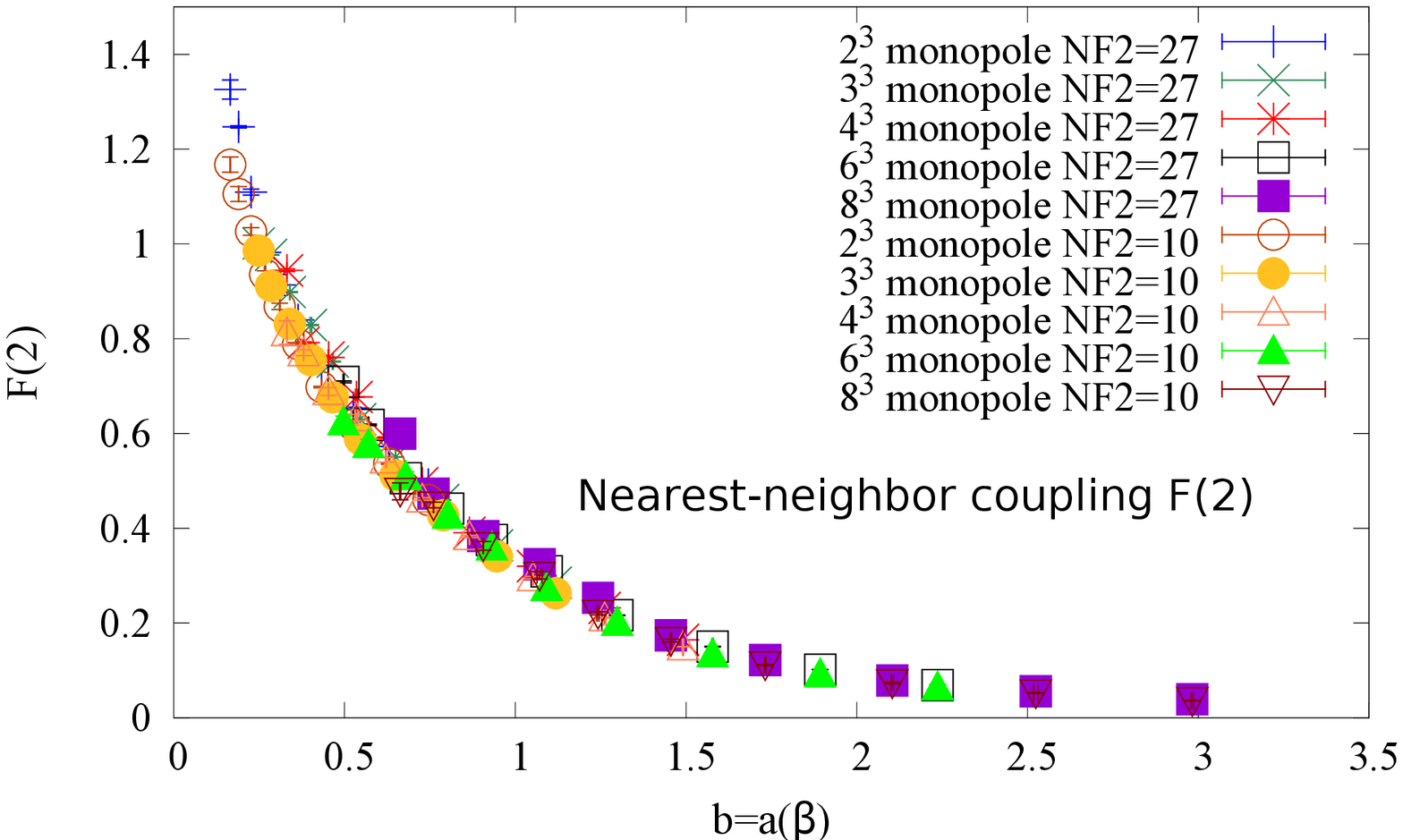}
  \end{minipage}
  \begin{minipage}[b]{0.9\linewidth}
    \centering
    \includegraphics[width=8cm,height=6.cm]{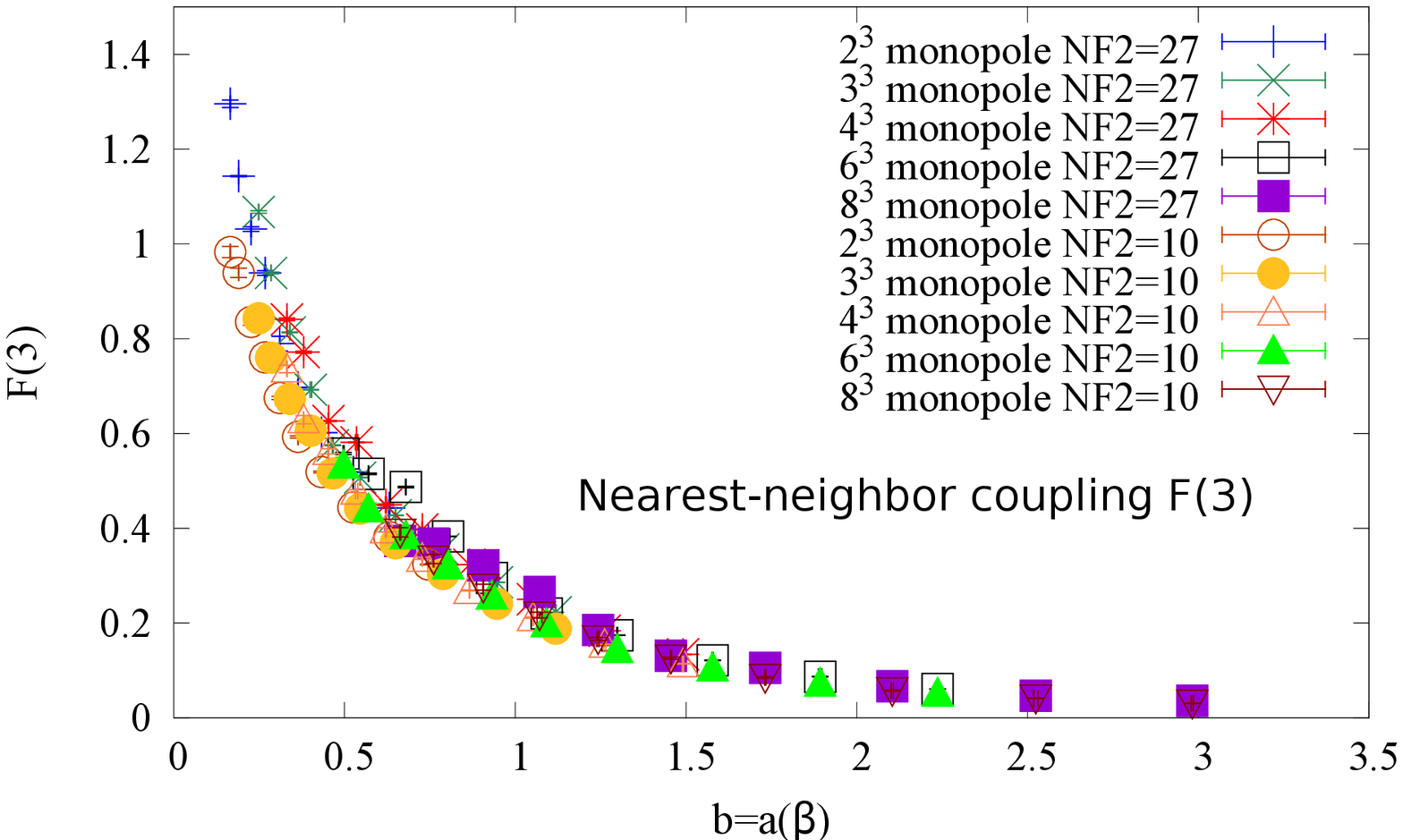}
  \end{minipage}
\end{figure}
\begin{figure}[htb]
\caption{Comparison of the coupling constants of the  two next nearest-neighbor interactions between the actions composed of  27 ($NF2=27$) and 10 ($NF2=10$) quadratic interactions alone.}
\label{figA_F45}
  \begin{minipage}[b]{0.9\linewidth}
    \centering
    \includegraphics[width=8cm,height=6.cm]{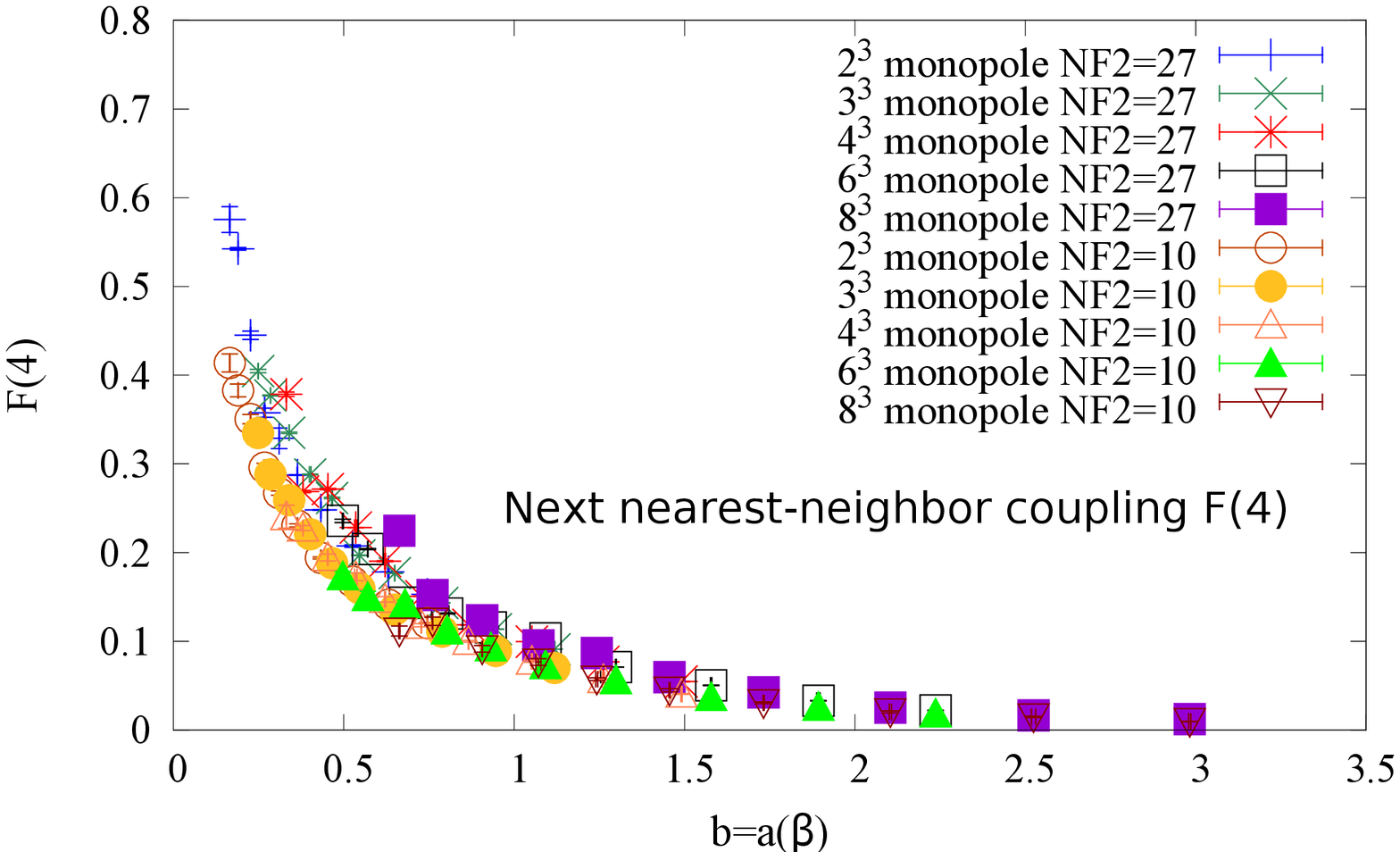}
  \end{minipage}
  \begin{minipage}[b]{0.9\linewidth}
    \centering
    \includegraphics[width=8cm,height=6.cm]{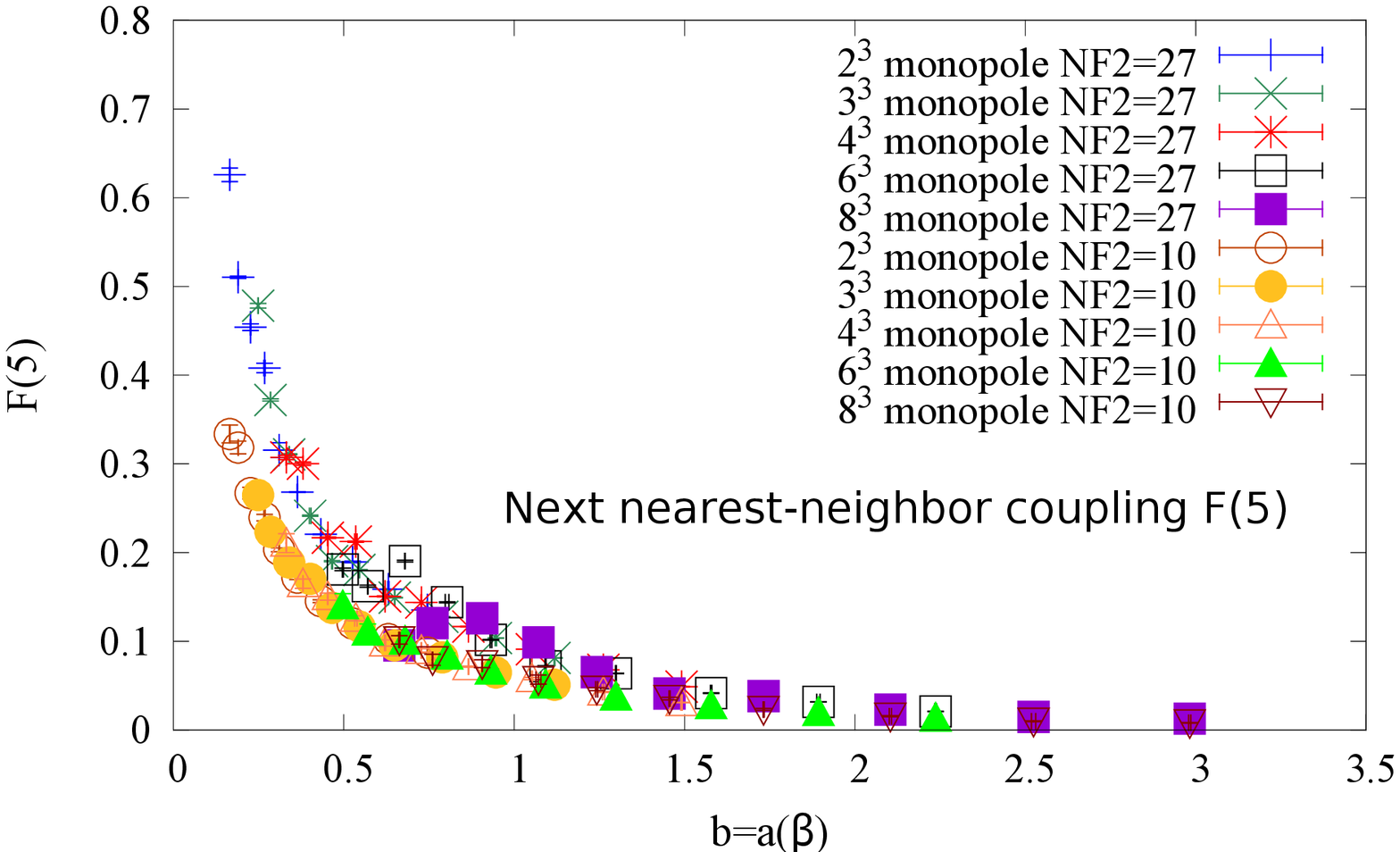}
  \end{minipage}
\end{figure}
\begin{figure}[htb]
\caption{Comarison of the coupling constants of the self and two nearest-neighbor interactions versus $b=na(\beta)$ between the actions of $10$ two-point interactions with and without higher interactions
on $48^4$ in MCG}
\label{compF123}
  \begin{minipage}[b]{0.9\linewidth}
    \centering
    \includegraphics[width=8cm,height=6.cm]{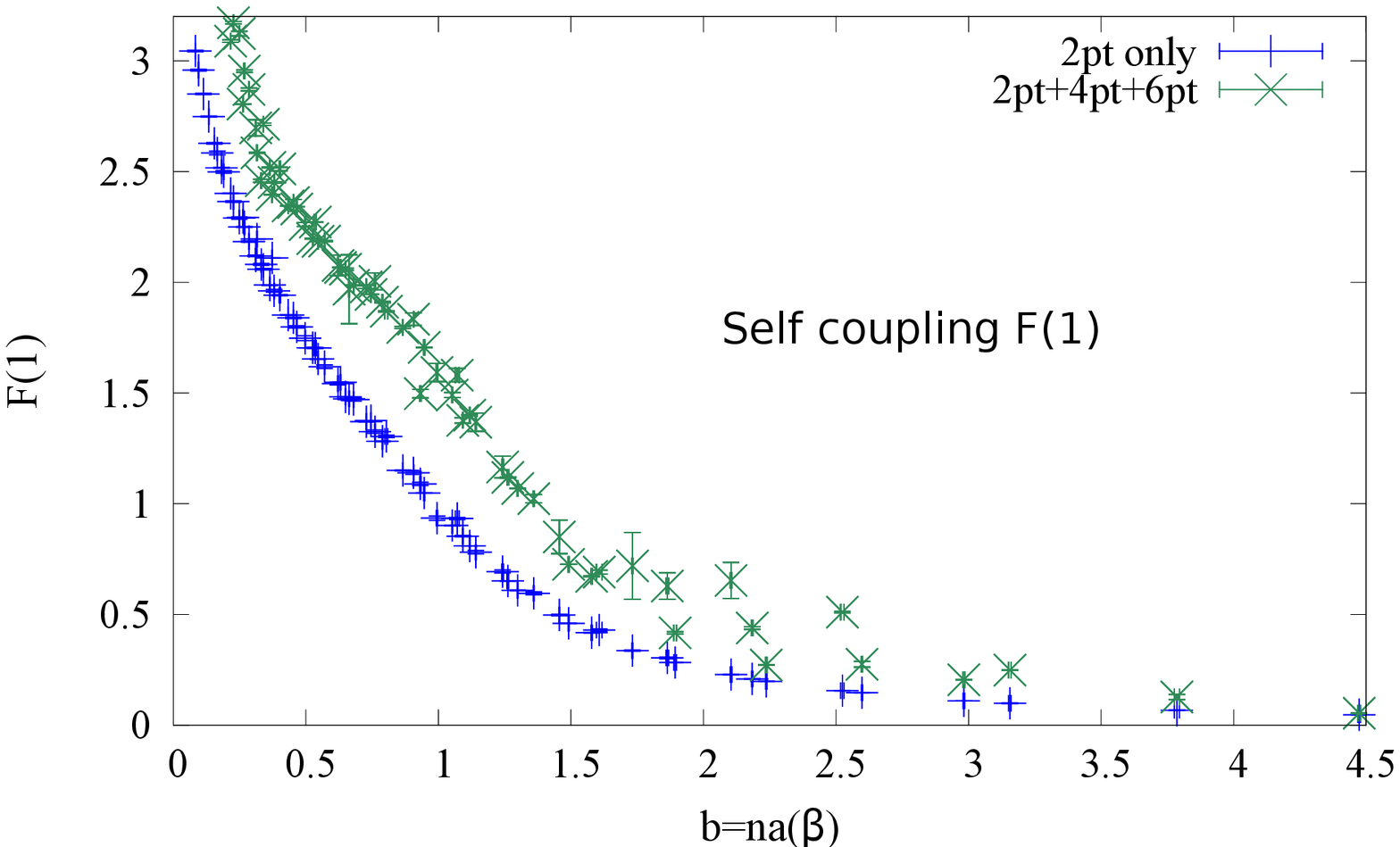}
  \end{minipage}
  \begin{minipage}[b]{0.9\linewidth}
    \centering
    \includegraphics[width=8cm,height=6.cm]{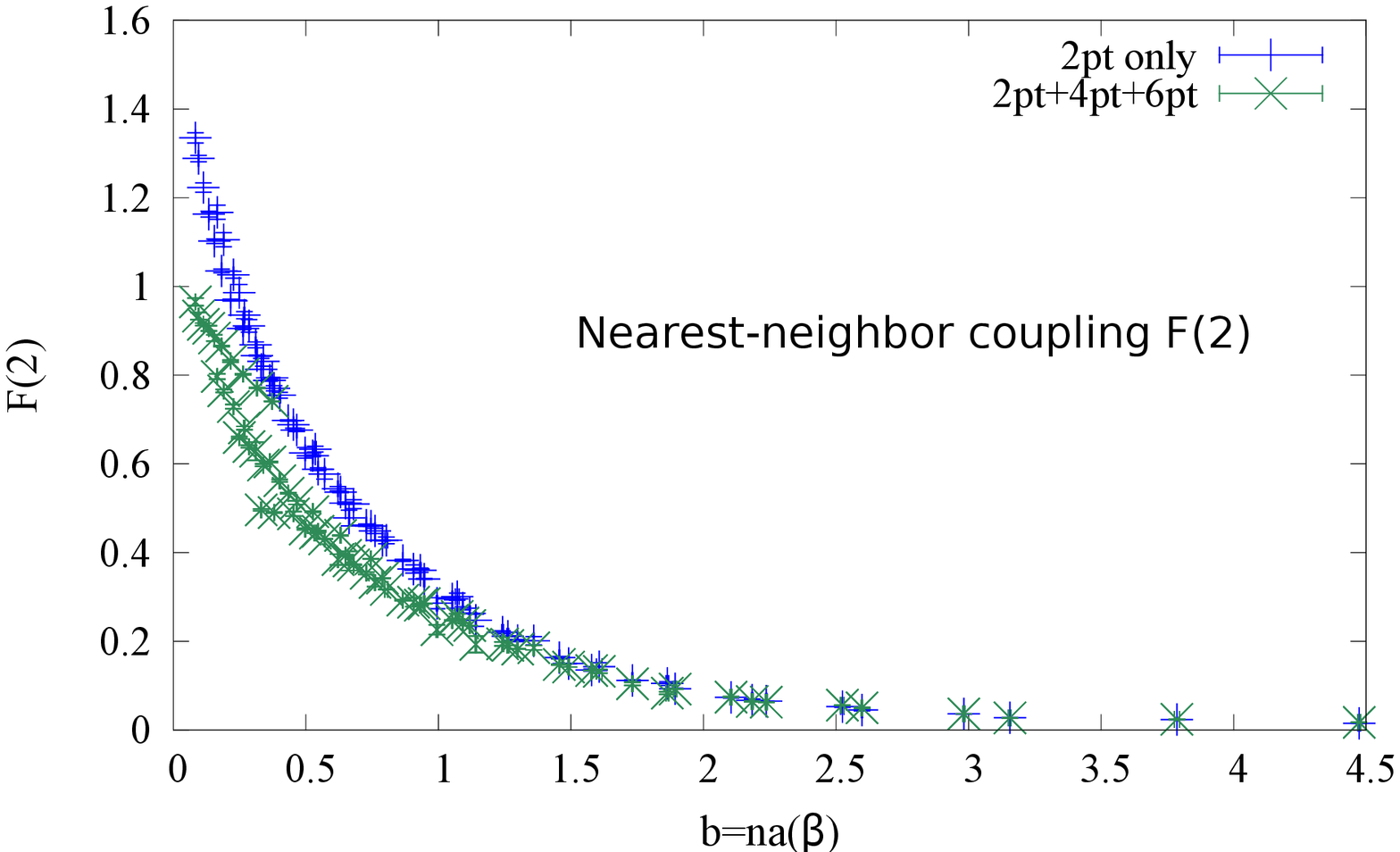}
  \end{minipage}
  \begin{minipage}[b]{0.9\linewidth}
    \centering
    \includegraphics[width=8cm,height=6.cm]{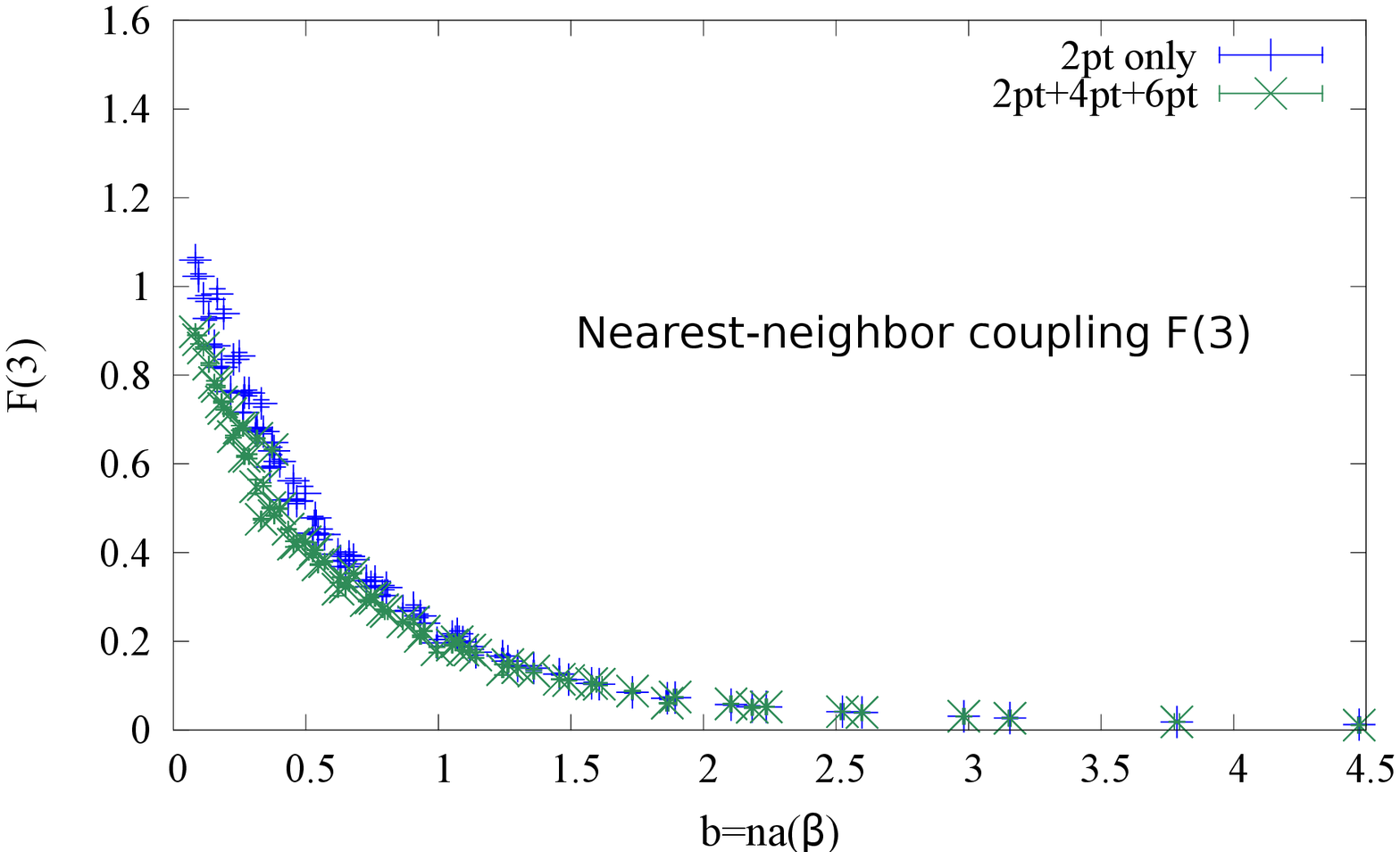}
  \end{minipage}
\end{figure}
\begin{figure}[htb]
\caption{Comparison of the coupling constants of two next to nearest-neighbor interactions versus $b=na(\beta)$ between the actions of $10$ two-point interactions with and without higher interactions on $48^4$ in MCG}
\label{compF45}
  \begin{minipage}[b]{0.9\linewidth}
    \centering
    \includegraphics[width=8cm,height=6.cm]{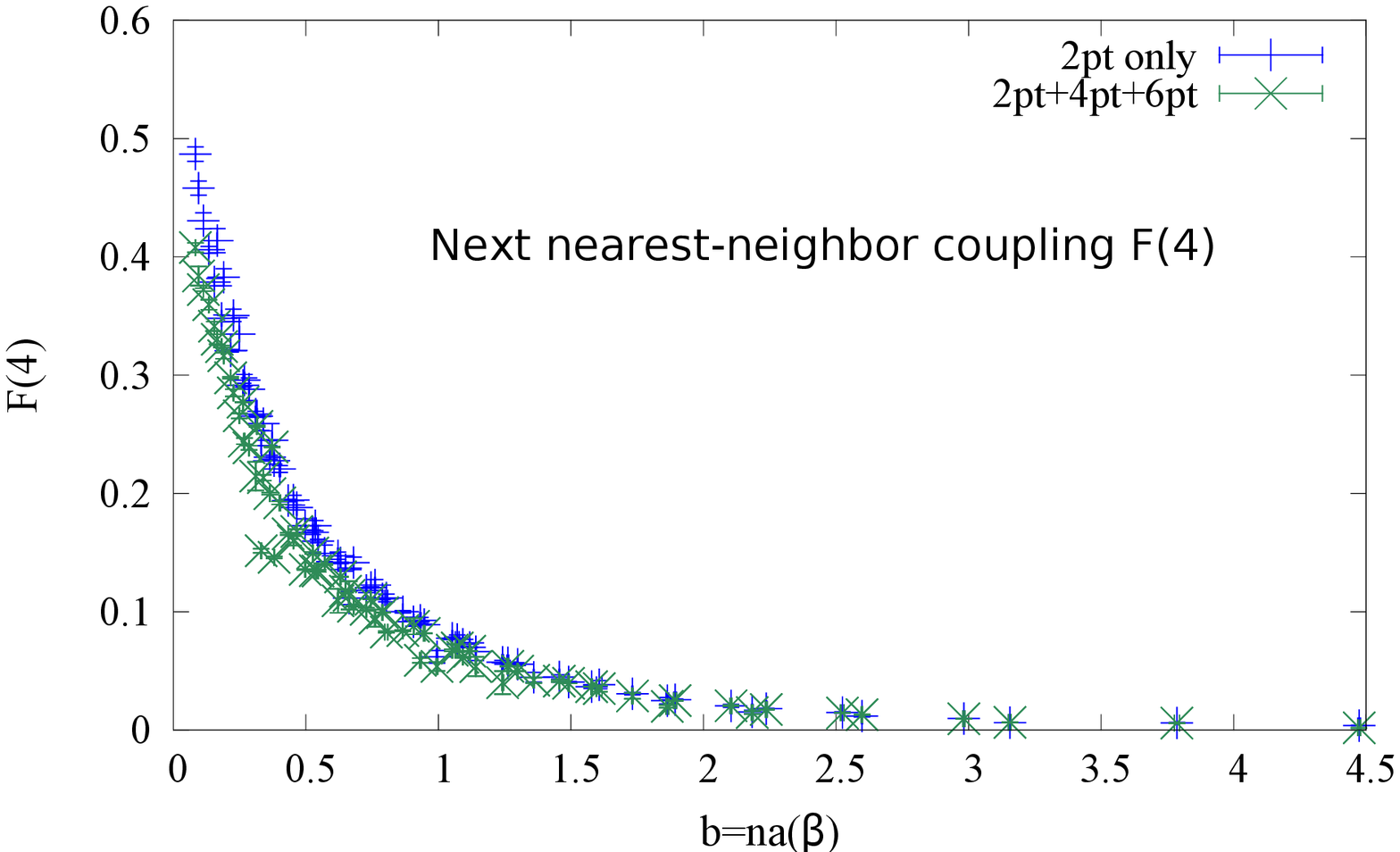}
  \end{minipage}
  \begin{minipage}[b]{0.9\linewidth}
    \centering
    \includegraphics[width=8cm,height=6.cm]{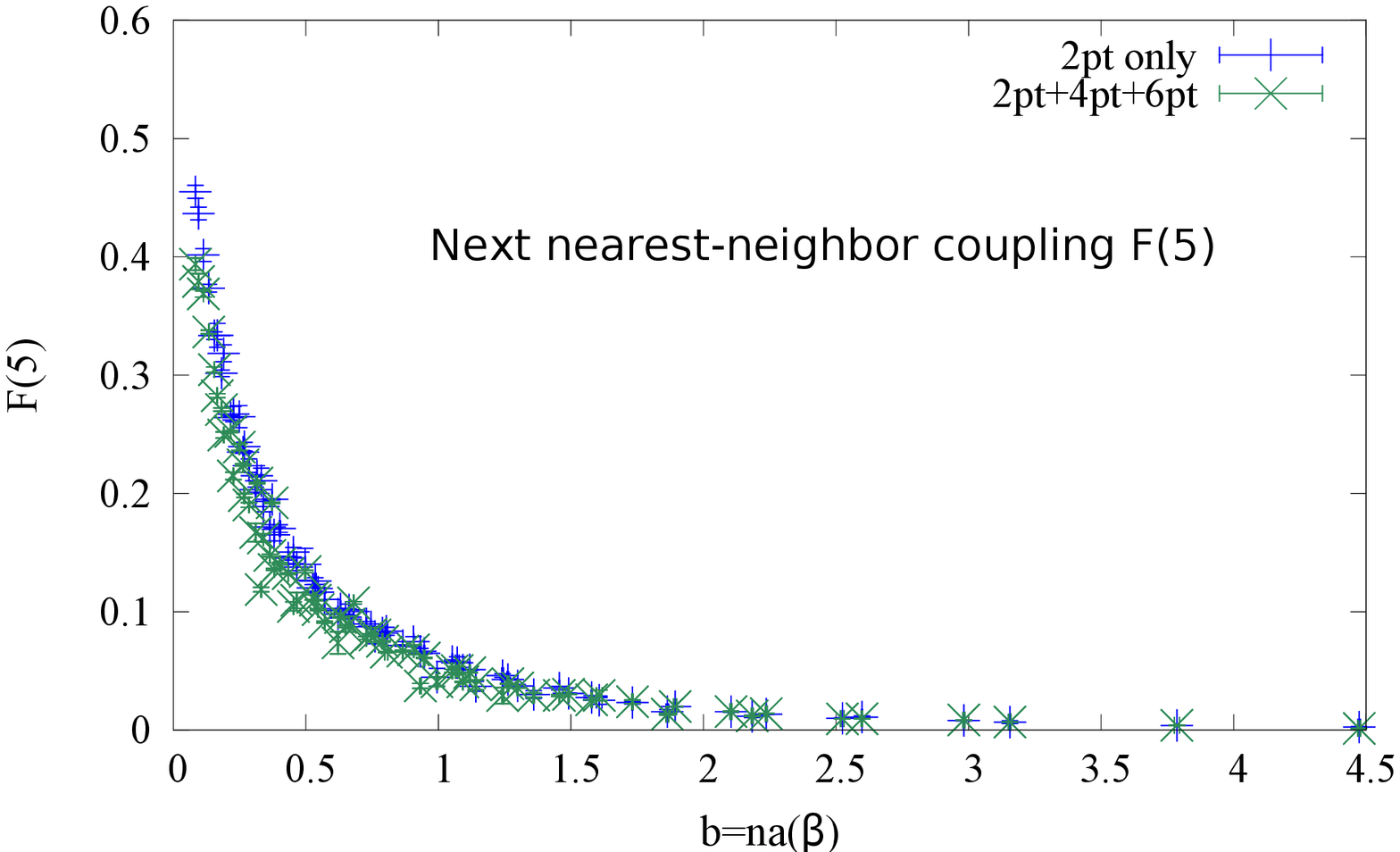}
  \end{minipage}
\end{figure}
\begin{figure}[htb]
\caption{The coupling constants of four- and six-point interactions versus $b=na(\beta)$ in the action of $10$ two-point interactions with higher interactions on $48^4$ in MCG}
\label{F11_12}
  \begin{minipage}[b]{0.9\linewidth}
    \centering
    \includegraphics[width=8cm,height=6.cm]{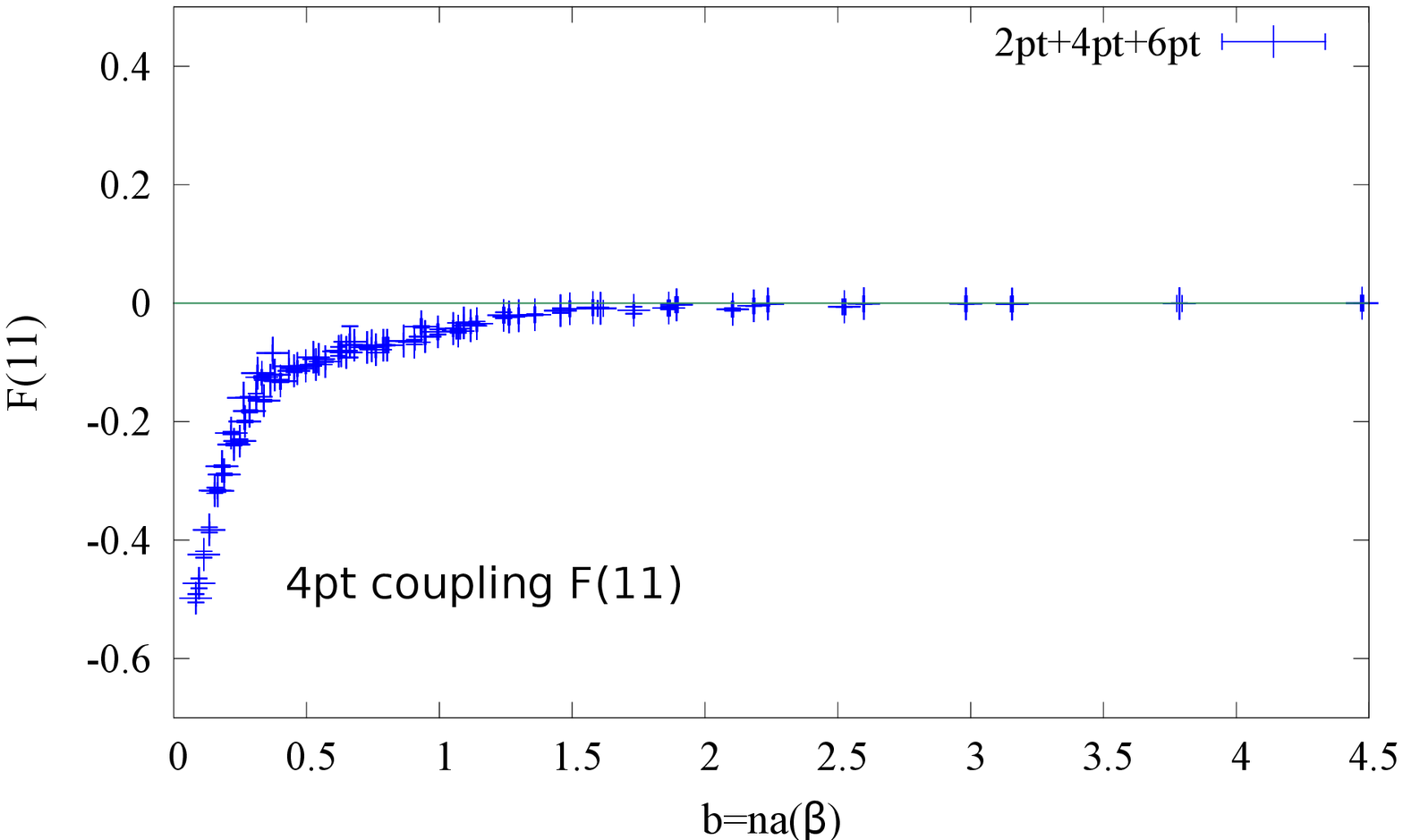}
  \end{minipage}
  \begin{minipage}[b]{0.9\linewidth}
    \centering
    \includegraphics[width=8cm,height=6.cm]{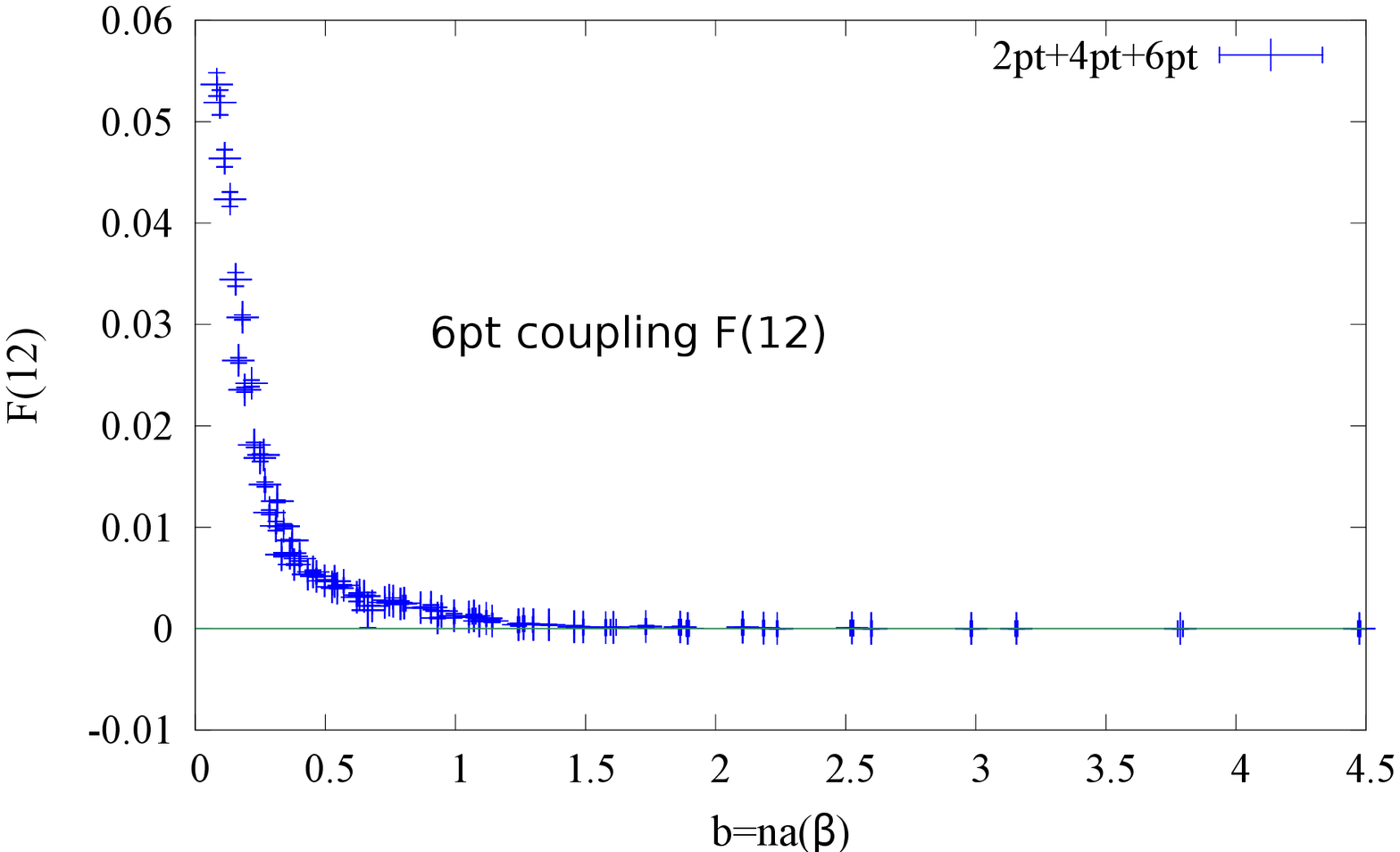}
  \end{minipage}
\end{figure}

\section{The form of the effective monopole action\label{APD:action}} 
As the form of the effective monopole action, we assume that 
only local and short-ranged interactions  
 are dominant.

The quadratic interactions for each color $a$ used for the modified Swendsen method
are shown in Table~\ref{tbl:appquad}.
Only the partner of the current multiplied by $k_{\mu}^a(s)$ are listed.
All terms in which the relation of the two currents is equivalent should
be added to satisfy translation and rotation invariances. 

To check the dominance of quadratic interactions, we include the following four-point and six-point interactions among monopoles of the same 
color component listed in Table~\ref{tbl:higher}. The six-point interaction is included, since the coefficient of the four-point interaction is found to be negative numerically.

  In the case of four and six-point interactions, there may exist color-mixing interactions via interactions with the gauge fields. 
We discuss the following color-mixed interactions as a simple example:
\begin{eqnarray*}
S^{(4)}_{dc}(k)&=&\sum_{s}\sbra{\sum_{\mu=-4}^4 \sum_{a\neq b}(k_\mu^a(s))^2(k_\mu^b(s))^2} \\
S^{(6)}_{dc}(k)&=&\sum_{s}\sbra{\sum_{\mu=-4}^4(k_\mu^1(s))^2\sum_{\mu=-4}^4(k_\mu^2(s))^2\sum_{\mu=-4}^4(k_\mu^3(s))^2}
\end{eqnarray*}
\section{Comparison  of the effective monopole actions from numerical analyses}\label{APD:comparison}
Various combinations of monopole interactions are tested numerically.
\begin{enumerate}
  \item Color mixing effects are checked first by adopting 
\begin{eqnarray*}
S&=& \sum_{i=1}^{10}F(i)S^{(2)}_i(k)+F(11)S^{(4)}(k)+F(12)S^{(4)}_{dc}(k)\\
&&+F(13)S^{(6)}(k)+F(14)S^{(6)}_{dc}(k),
\end{eqnarray*}
where the first $10$ quadratic interactions $S^{(2)}(k)$ alone in Table~\ref{tbl:appquad} are used for simplicity. 

As a whole, the convergence is rather difficult. When the convergence condition is relaxed, we get the convergent results for $n\ge3$. An example is shown in Table~\ref{tbl:1022} for $n=3$ and $\beta=3.3$.
Since the coupling constants of the color-mixed interactions $F(12)$ and $F(14)$ are suppressed in comparison with those without no color mixing and stable convergence is not obtained for all cases, we did not consider any color mixing in the extensive studies done in this paper. The form of effective monopole action having no stable convergence is not a good choice
for the application of the renormalization group study. 

  \item  Under the condition of no color-mixing, we study four cases of effective monopole actions:\\
(1) $ 27$ quadratic interactions in Table~\ref{tbl:appquad} plus higher interactions in Table~\ref{tbl:higher}.\\
(2) First $10$ quadratic interactions with lattice distance $R\le 2$ plus higher interactions in Table~\ref{tbl:higher}.\\
(3) $ 27$ quadratic interactions in Table~\ref{tbl:appquad} alone.\\
(4) First $10$ quadratic interactions with lattice distance $R\le 2$ in Table~\ref{tbl:appquad} alone.\\
An example for $\beta=3.2$ and $n=4$ blocking is shown in Table~\ref{tbl:comparison}. The comparison can be done only for $n<8$ due to boundary effects, since the reduced lattice volume in $n=8$ is $6^4$ and $4^4$ in $n=12$ blocking. Similar behaviors 
are found for all $n<8$ and all $\beta$.
\begin{itemize}
  \item The coupling constants of four- and six-point interactions are very small, but they have  non-negligible efffects on the most important quadratic self interaction $F(1)$ as seen from the data in the second and the fourth rows in Table~\ref{tbl:comparison}. Big effects are not seen for other couplings up to $F(21)$. 
  \item The coupling constant $F(28)$ of the four-point interaction is negative, whereas that of the six-point interaction $F(29)$ is positive. This is similar to the results observed previously in MA gauge~\cite{Chernodub:2000ax}.
  \item The first and the second rows in Table~\ref{tbl:comparison} show the comparison of both quadratic actions in (3) and (4). The most important self and the nearest-neighbor interactions are much the same. The couplings of the first $5$ quadratic interactions are compared in 
Fig.\ref{figA_F123} and Fig.\ref{figA_F45}. 
\item
The differences of the cases (2) and (4) with and without higher interactions are shown in Fig.\ref{compF123} and Fig.\ref{compF45}. All data satisfy the scaling but the differences are not negligible especially in the self coupling case.
The coupling constants of higher interactions in the case (2) are plotted in Fig.\ref{F11_12}.
Also scaling is seen beautifully.  
\item
In the main part of this paper, we focus on the most simple case (4), i.e., the action composed of first $10$ quadratic interactions alone, since then even $n=12$ could be studied in the renormalization group flow and the comparison between numerical data and analytic results from the blocking from the continuum is easy. Namely we will study the projection on to the coupling constant plane composed of the quadratic $10$ interactions of the renormalized action.
\end{itemize}
\end{enumerate}

\section{Evaluation of the self-coupling term $D_{ii}^{-1}(0)$\label{dinv}}
The 10 quadratic interactions of $D_{\mu\nu}(s,s')$ are explicitly written from Table~\ref{tbl:appquad} for each color component as
\begin{eqnarray}
D_{\mu\nu}(s,s')&=& \sum_i F(i)(S_i)_{\mu\nu}(s,s'),
\end{eqnarray}
where $F(i)$ are coupling constants and the operators $S_i$ are shown as follows:
\begin{eqnarray*}
S_1&=&\delta_{s',s}\delta_{\mu,\nu}\\
S_2&=&\frac{1}{2}\left[\delta_{s',s+\mu}+ \delta_{s',s-\mu}\right]\delta_{\mu,\nu}\\
S_3&=&\frac{1}{2}\sum_{\alpha(\neq\mu)}\left[\delta_{s',s+\alpha}+ \delta_{s',s-\alpha}\right]\delta_{\mu,\nu}\\
S_4&=&\frac{1}{4}\sum_{\alpha(\neq\mu)}[\delta_{s',s+\mu+\alpha}+ \delta_{s',s+\mu-\alpha}\\
&&+\delta_{s',s-\mu+\alpha}+ \delta_{s',s-\mu-\alpha}]\delta_{\mu,\nu}
\end{eqnarray*}
\begin{eqnarray*}
S_5&=&\frac{1}{4}\sum_{\alpha\neq\beta(\neq\mu)}[\delta_{s',s+\alpha+\beta
}+ \delta_{s',s+\alpha-\beta}\\
&&+\delta_{s',s-\alpha+\beta}+ \delta_{s',s-\alpha-\beta}]\delta_{\mu,\nu}\\
S_8&=&\frac{1}{2}\left[\delta_{s',s+2\mu}+ \delta_{s',s-2\mu}\right]\delta_{\mu,\nu}\\
S_{10}&=&\frac{1}{2}\sum_{\alpha(\neq\mu)}\left[\delta_{s',s+2\alpha}+ \delta_{s',s-2\alpha}\right]\delta_{\mu,\nu}.
\end{eqnarray*}
Here irrelevant terms $S_6, S_7, S_9$ are not written explicitly.
As shown in Table~\ref{tbl:comparison}, the self-coupling $F(1)$ is much larger than other coupling constants.  Hence the inverse propagator $D_{\mu\nu}^{-1}(s,s')$ can be evaluated by the expansion with respect to $F(1)$.
It is easy to see the self-coupling term contribution to the inverse proprgator comes only from the quadratic terms of $S_i$ in the expansion. Considering the numerical data showing $F(1)\gg F(2)\sim F(3)\gg F(4)\sim F(5)\gg \textrm{higher terms}$, the relevent non-negligible operators are $S_2^2,\ S_3^2,\ S_4^2,\ S_5^2,\ S_2^4,\ S_2^2S_3^2,\ S_3^4$. These operators are evaluated explicitly  as
\begin{eqnarray*}
S_2^2&=&\frac{1}{2}S_1+\frac{1}{2}S_8, \\ 
S_3^2&=&\frac{3}{2}S_1+S_8+\frac{1}{2}S_{10},\\
S_4^2&=&\frac{3}{4}S_1+\frac{1}{2}S_5+\frac{3}{4}S_8+\ldots,\\
S_5^2&=&\frac{3}{4}S_1+\frac{3}{4}S_{10}+\ldots,\\
\ S_2^4&=&\frac{3}{8}S_1+\ldots\\
S_2^2S_3^2&=&\frac{3}{4}S_1+\ldots,\\
S_3^4&=&\frac{25}{8}S_1+\ldots.
\end{eqnarray*}

Hence we get
\begin{eqnarray}
D_{ii}^{-1}(0)&=&\frac{1}{F(1)}+\frac{F(2)^2}{2F(1)^3}+\frac{3F(3)^2}{2F(1)^3}
+\frac{3F(4)^2}{4F(1)^3}\nn\\
&+&\frac{3F(5)^2}{4F(1)^3} 
+\frac{3F(2)^4}{8F(1)^5}+\frac{9F(2)^2F(3)^2}{2F(1)^5}
\nn\\
&+&\frac{25F(3)^4}{8F(1)^5}+\ldots.\label{D_in}
\end{eqnarray}

\end{document}